\pdfoutput=1
\documentclass[12pt,a4paper]{article}

\usepackage{ifthen} 
\newboolean{pdflatex}
\setboolean{pdflatex}{true} 

\newboolean{articletitles}
\setboolean{articletitles}{true} 

\newboolean{uprightparticles}
\setboolean{uprightparticles}{false} 

\def\paperauthors{LHCb collaboration} 
\def\paperasciititle{Measurement of lepton universality parameters in B+ -> K+l+l- and B0 -> K*0l+l- decays} 
\def\papertitle{Measurement of lepton universality parameters in \BuToKll and \BdToKstll decays} 

\def\paperkeywords{{High Energy Physics}, {LHCb}} 
\def\papercopyright{\the\year\ CERN for the benefit of the LHCb collaboration} 
\def\paperlicence{CC BY 4.0 licence}
\def\paperlicenceurl{https://creativecommons.org/licenses/by/4.0/}

\usepackage[top=1in, bottom=1.25in, left=1in, right=1in]{geometry}

\usepackage{microtype}
\usepackage{lineno}  
\usepackage{xspace} 
\usepackage{caption} 

\usepackage{graphicx}  
\usepackage{color}
\usepackage{colortbl}
\graphicspath{{./figs/}} 

\usepackage{amsmath} 
\usepackage{amssymb}
\usepackage{amsfonts}
\usepackage{upgreek} 

\newcommand*\patchAmsMathEnvironmentForLineno[1]{%
\expandafter\let\csname old#1\expandafter\endcsname\csname #1\endcsname
\expandafter\let\csname oldend#1\expandafter\endcsname\csname
end#1\endcsname
 \renewenvironment{#1}%
   {\linenomath\csname old#1\endcsname}%
   {\csname oldend#1\endcsname\endlinenomath}%
}
\newcommand*\patchBothAmsMathEnvironmentsForLineno[1]{%
  \patchAmsMathEnvironmentForLineno{#1}%
  \patchAmsMathEnvironmentForLineno{#1*}%
}
\AtBeginDocument{%
\patchBothAmsMathEnvironmentsForLineno{equation}%
\patchBothAmsMathEnvironmentsForLineno{align}%
\patchBothAmsMathEnvironmentsForLineno{flalign}%
\patchBothAmsMathEnvironmentsForLineno{alignat}%
\patchBothAmsMathEnvironmentsForLineno{gather}%
\patchBothAmsMathEnvironmentsForLineno{multline}%
\patchBothAmsMathEnvironmentsForLineno{eqnarray}%
}

\usepackage{hyperxmp}

\usepackage[pdftex,
            pdfauthor={\paperauthors},
            pdftitle={\paperasciititle},
            pdfkeywords={\paperkeywords},
            pdfcopyright={Copyright (C) \papercopyright},
            pdflicenseurl={\paperlicenceurl}]{hyperref}

\usepackage[colorinlistoftodos,textsize=scriptsize]{todonotes}

\usepackage[bottom,flushmargin,hang,multiple]{footmisc}

\usepackage[all]{hypcap} 

\usepackage{xspace} 
\usepackage{upgreek}

\def\lhcb   {\mbox{LHCb}\xspace}
\def\atlas  {\mbox{ATLAS}\xspace}
\def\cms    {\mbox{CMS}\xspace}

\def\MagUp {\mbox{\em Mag\kern -0.05em Up}\xspace}

\def\lone   {L0\xspace}
\def\hlt    {HLT\xspace}

\ifthenelse{\boolean{uprightparticles}}%
{

 \def\Peta        {\ensuremath{\upeta}\xspace}

 \def\Pmu         {\ensuremath{\upmu}\xspace}                 
 \def\Pnu         {\ensuremath{\upnu}\xspace}                 
                  
 \def\Ppi         {\ensuremath{\uppi}\xspace}

 \def\Pphi        {\ensuremath{\upphi}\xspace}

 \def\Ppsi        {\ensuremath{\uppsi}\xspace}                 
 \def\Pomega      {\ensuremath{\upomega}\xspace}                 

 \def\PDelta      {\ensuremath{\Delta}\xspace}                 
 \def\PXi         {\ensuremath{\Xi}\xspace}                 
 \def\PLambda     {\ensuremath{\Lambda}\xspace}                 
 \def\PSigma      {\ensuremath{\Sigma}\xspace}                 
 \def\POmega      {\ensuremath{\Omega}\xspace}                 
 \def\PUpsilon    {\ensuremath{\Upsilon}\xspace}
 \let\oldPi\Pi
 \def\PPi         {\ensuremath{\oldPi}\xspace}

 \def\PB      {\ensuremath{\mathrm{B}}\xspace}                 
                  
 \def\PD      {\ensuremath{\mathrm{D}}\xspace}

 \def\PJ      {\ensuremath{\mathrm{J}}\xspace}                 
 \def\PK      {\ensuremath{\mathrm{K}}\xspace}

 \def\PW      {\ensuremath{\mathrm{W}}\xspace}

 \def\PZ      {\ensuremath{\mathrm{Z}}\xspace}                 
                  
 \def\Pb      {\ensuremath{\mathrm{b}}\xspace}                 
 \def\Pc      {\ensuremath{\mathrm{c}}\xspace}                 
                  
 \def\Pe      {\ensuremath{\mathrm{e}}\xspace}

 \def\Pi      {\ensuremath{\mathrm{i}}\xspace}

 \def\Pp      {\ensuremath{\mathrm{p}}\xspace}

 \def\Ps      {\ensuremath{\mathrm{s}}\xspace}

 \def\thebaroffset{0.0em}
}
{

 \def\Peta        {\ensuremath{\eta}\xspace}

 \def\Pmu         {\ensuremath{\mu}\xspace}                 
 \def\Pnu         {\ensuremath{\nu}\xspace}                 
                  
 \def\Ppi         {\ensuremath{\pi}\xspace}

 \def\Pphi        {\ensuremath{\phi}\xspace}

 \def\Ppsi        {\ensuremath{\psi}\xspace}                 
 \def\Pomega      {\ensuremath{\omega}\xspace}                 
 \mathchardef\PDelta="7101
 \mathchardef\PXi="7104
 \mathchardef\PLambda="7103
 \mathchardef\PSigma="7106
 \mathchardef\POmega="710A
 \mathchardef\PUpsilon="7107
 \mathchardef\PPi="7105
                  
 \def\PB      {\ensuremath{B}\xspace}                 
                  
 \def\PD      {\ensuremath{D}\xspace}

 \def\PJ      {\ensuremath{J}\xspace}                 
 \def\PK      {\ensuremath{K}\xspace}

 \def\PW      {\ensuremath{W}\xspace}

 \def\PZ      {\ensuremath{Z}\xspace}                 
                  
 \def\Pb      {\ensuremath{b}\xspace}                 
 \def\Pc      {\ensuremath{c}\xspace}                 
                  
 \def\Pe      {\ensuremath{e}\xspace}

 \def\Pi      {\ensuremath{i}\xspace}

 \def\Pp      {\ensuremath{p}\xspace}

 \def\Ps      {\ensuremath{s}\xspace}

 \def\thebaroffset{0.18em}
}
\newcommand{\offsetoverline}[2][\thebaroffset]{\kern #1\overline{\kern -#1 #2}}%

\makeatletter
\ifcase \@ptsize \relax
  \newcommand{\miniscule}{\@setfontsize\miniscule{4}{5}}
\or
  \newcommand{\miniscule}{\@setfontsize\miniscule{5}{6}}
\or
  \newcommand{\miniscule}{\@setfontsize\miniscule{5}{6}}
\fi
\makeatother

\DeclareRobustCommand{\optbar}[1]{\shortstack{{\miniscule (\rule[.5ex]{1.25em}{.18mm})}
  \\ [-.7ex] $#1$}}

\def\en         {{\ensuremath{\Pe^-}}\xspace}   
\def\ep         {{\ensuremath{\Pe^+}}\xspace}

\def\epem       {{\ensuremath{\Pe^+\Pe^-}}\xspace}

\def\mup        {{\ensuremath{\Pmu^+}}\xspace}
\def\mun        {{\ensuremath{\Pmu^-}}\xspace} 

\def\mumu       {{\ensuremath{\Pmu^+\Pmu^-}}\xspace}

\def\ellm       {{\ensuremath{\ell^-}}\xspace}
\def\ellp       {{\ensuremath{\ell^+}}\xspace}

\def\ellell     {\ensuremath{\ell^+ \ell^-}\xspace}

\def\neu        {{\ensuremath{\Pnu}}\xspace}
\def\neub       {{\ensuremath{\overline{\Pnu}}}\xspace}
\def\neue       {{\ensuremath{\neu_e}}\xspace}
\def\neueb      {{\ensuremath{\neub_e}}\xspace}
\def\neul       {{\ensuremath{\neu_\ell}}\xspace}
\def\neulb      {{\ensuremath{\neub_\ell}}\xspace}

\def\W      {{\ensuremath{\PW}}\xspace}

\def\Z      {{\ensuremath{\PZ}}\xspace}

\def\squark    {{\ensuremath{\Ps}}\xspace}

\def\cquark    {{\ensuremath{\Pc}}\xspace}
\def\cquarkbar {{\ensuremath{\overline \cquark}}\xspace}
\def\ccbar     {{\ensuremath{\cquark\cquarkbar}}\xspace}
\def\bquark    {{\ensuremath{\Pb}}\xspace}
\def\bquarkbar {{\ensuremath{\overline \bquark}}\xspace}
\def\bbbar     {{\ensuremath{\bquark\bquarkbar}}\xspace}

\def\pion   {{\ensuremath{\Ppi}}\xspace}
\def\piz    {{\ensuremath{\pion^0}}\xspace}
\def\pip    {{\ensuremath{\pion^+}}\xspace}
\def\pim    {{\ensuremath{\pion^-}}\xspace}

\def\kaon    {{\ensuremath{\PK}}\xspace}
\def\Kbar    {{\ensuremath{\offsetoverline{\PK}}}\xspace}

\def\KorKbar {\kern \thebaroffset\optbar{\kern -\thebaroffset \PK}{}\xspace}

\def\Kp      {{\ensuremath{\kaon^+}}\xspace}
\def\Km      {{\ensuremath{\kaon^-}}\xspace}

\def\KS      {{\ensuremath{\kaon^0_{\mathrm{S}}}}\xspace}

\def\Kstarz  {{\ensuremath{\kaon^{*0}}}\xspace}
\def\Kstarzb {{\ensuremath{\Kbar{}^{*0}}}\xspace}
\def\Kstar   {{\ensuremath{\kaon^*}}\xspace}

\def\Kstarp  {{\ensuremath{\kaon^{*+}}}\xspace}

\newcommand{\etaz}{\ensuremath{\Peta}\xspace}
\newcommand{\etapr}{\ensuremath{\Peta^{\prime}}\xspace}
\newcommand{\phiz}{\ensuremath{\Pphi}\xspace}
\newcommand{\omegaz}{\ensuremath{\Pomega}\xspace}

\def\Dbar    {{\ensuremath{\offsetoverline{\PD}}}\xspace}
\def\D       {{\ensuremath{\PD}}\xspace}

\def\DorDbar {\kern \thebaroffset\optbar{\kern -\thebaroffset \PD}\xspace}
\def\Dz      {{\ensuremath{\D^0}}\xspace}
\def\Dzb     {{\ensuremath{\Dbar{}^0}}\xspace}
\def\Dp      {{\ensuremath{\D^+}}\xspace}
\def\Dm      {{\ensuremath{\D^-}}\xspace}

\def\DpDm    {\ensuremath{\Dp {\kern -0.16em \Dm}}\xspace}

\def\Dstarp  {{\ensuremath{\D^{*+}}}\xspace}

\def\B       {{\ensuremath{\PB}}\xspace}

\def\BorBbar {\kern \thebaroffset\optbar{\kern -\thebaroffset \PB}\xspace}
\def\Bz      {{\ensuremath{\B^0}}\xspace}

\def\Bd      {{\ensuremath{\B^0}}\xspace}

\def\BdorBdbar {\kern \thebaroffset\optbar{\kern -\thebaroffset \Bd}\xspace}
\def\Bu      {{\ensuremath{\B^+}}\xspace}

\def\Bp      {{\ensuremath{\Bu}}\xspace}

\def\Bs      {{\ensuremath{\B^0_\squark}}\xspace}

\def\BsorBsbar {\kern \thebaroffset\optbar{\kern -\thebaroffset \Bs}\xspace}

\def\jpsi     {{\ensuremath{{\PJ\mskip -3mu/\mskip -2mu\Ppsi}}}\xspace}
\def\psitwos  {{\ensuremath{\Ppsi{(2S)}}}\xspace}

\def\Y#1S{\ensuremath{\PUpsilon{(#1S)}}\xspace}

\def\proton      {{\ensuremath{\Pp}}\xspace}

\def\Lz          {{\ensuremath{\PLambda}}\xspace}

\def\LorLbar     {\kern \thebaroffset\optbar{\kern -\thebaroffset \PLambda}\xspace}

\def\Lb           {{\ensuremath{\Lz^0_\bquark}}\xspace}

\newcommand{\decay}[2]{\ensuremath{#1\!\to #2}\xspace} 

\def\to                 {\ensuremath{\rightarrow}\xspace}

\def\order   {{\ensuremath{\mathcal{O}}}\xspace}

\def\qsq       {{\ensuremath{q^2}}\xspace}

\newcommand{\etot}{{\ensuremath{\varepsilon_{\mathrm{ tot}}}}\xspace}

\def\BdToKstmm    {\decay{\Bd}{\Kstarz\mup\mun}}

\def\bsll     {\decay{\bquark}{\squark \ell^+ \ell^-}}

\def\AT#1     {\ensuremath{A_{\mathrm{T}}^{#1}}\xspace}           

\def\C#1      {\ensuremath{\mathcal{C}_{#1}}\xspace}                       
\def\Cp#1     {\ensuremath{\mathcal{C}_{#1}^{'}}\xspace}                    
\def\Ceff#1   {\ensuremath{\mathcal{C}_{#1}^{\mathrm{(eff)}}}\xspace}        
\def\Cpeff#1  {\ensuremath{\mathcal{C}_{#1}^{'\mathrm{(eff)}}}\xspace}       
\def\Ope#1    {\ensuremath{\mathcal{O}_{#1}}\xspace}                       
\def\Opep#1   {\ensuremath{\mathcal{O}_{#1}^{'}}\xspace}                    

\newcommand{\nospaceunit}[1]{\ensuremath{\text{#1}}}       
\newcommand{\aunit}[1]{\ensuremath{\text{\,#1}}}       

\newcommand{\tev}{\aunit{Te\kern -0.1em V}\xspace}
\newcommand{\gev}{\aunit{Ge\kern -0.1em V}\xspace}
\newcommand{\mev}{\aunit{Me\kern -0.1em V}\xspace}
\newcommand{\kev}{\aunit{ke\kern -0.1em V}\xspace}
\newcommand{\ev}{\aunit{e\kern -0.1em V}\xspace}
 
\newcommand{\mevc}{\ensuremath{\aunit{Me\kern -0.1em V\!/}c}\xspace}
\newcommand{\gevc}{\ensuremath{\aunit{Ge\kern -0.1em V\!/}c}\xspace}
\newcommand{\mevcc}{\ensuremath{\aunit{Me\kern -0.1em V\!/}c^2}\xspace}
\newcommand{\gevcc}{\ensuremath{\aunit{Ge\kern -0.1em V\!/}c^2}\xspace}
\newcommand{\gevgevcccc}{\ensuremath{\gev^2\!/c^4}\xspace} 

\def\mum  {\ensuremath{\,\upmu\nospaceunit{m}}\xspace}

\def\mbarn{\aunit{mb}\xspace}

\def\fb   {\ensuremath{\aunit{fb}}\xspace}
\def\invfb   {\ensuremath{\fb^{-1}}\xspace}

\newcommand{\stat}{\aunit{(stat)}\xspace}
\newcommand{\syst}{\aunit{(syst)}\xspace}

\def\order{{\ensuremath{\mathcal{O}}}\xspace}
\newcommand{\chisq}{\ensuremath{\chi^2}\xspace}

\newcommand{\chisqip}{\ensuremath{\chi^2_{\text{IP}}}\xspace}

\def\deriv {\ensuremath{\mathrm{d}}}

\def\gsim{{~\raise.15em\hbox{$>$}\kern-.85em
          \lower.35em\hbox{$\sim$}~}\xspace}
\def\lsim{{~\raise.15em\hbox{$<$}\kern-.85em
          \lower.35em\hbox{$\sim$}~}\xspace}

\def\pt         {\ensuremath{p_{\mathrm{T}}}\xspace}

\def\ptot       {\ensuremath{p}\xspace}
\def\et         {\ensuremath{E_{\mathrm{T}}}\xspace}

\def\evtgen     {\mbox{\textsc{EvtGen}}\xspace}

\def\geant      {\mbox{\textsc{Geant4}}\xspace}

\def\photos     {\mbox{\textsc{Photos}}\xspace}

\def\pythia     {\mbox{\textsc{Pythia}}\xspace}

\def\roofit     {\mbox{\textsc{RooFit}}\xspace}
\def\root       {\mbox{\textsc{Root}}\xspace}

\def\tell1  {TELL1\xspace}
\def\ukl1   {UKL1\xspace}

\newcommand{\eg}{\mbox{\itshape e.g.}\xspace}
\newcommand{\ie}{\mbox{\itshape i.e.}\xspace}

\newcommand{\vs}{\mbox{\itshape vs.}\xspace}

\newcommand{\lhcborcid}[1]{\href{https://orcid.org/#1}{\hspace*{0.1em}\raisebox{-0.45ex}{\includegraphics[width=1em]{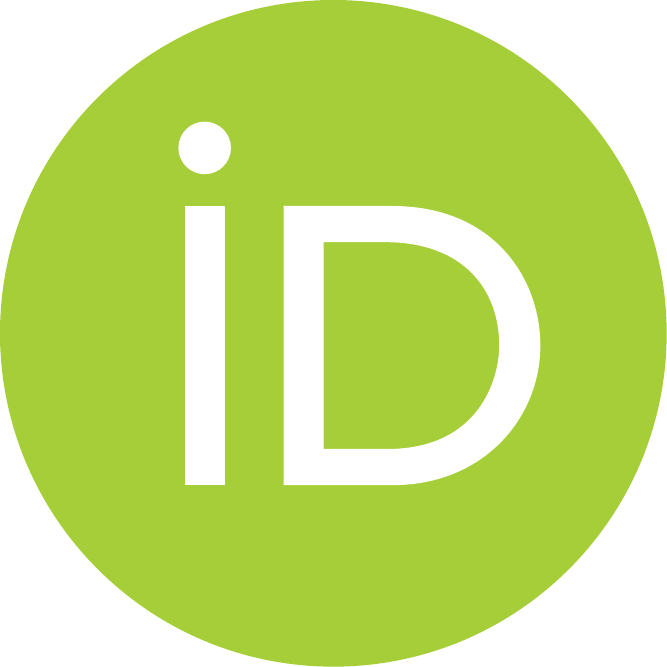}}}}

\usepackage{cite} 
\usepackage{mciteplus}

\def\RK         {\ensuremath{R_K}\xspace}
\def\RKst       {\ensuremath{R_{\Kstar}}\xspace}
\def\RKorKst       {\ensuremath{R_{K,\Kstar}}\xspace}

\def\NRKorKst       {\ensuremath{R_{(K,\Kstar)}}\xspace}

\def\RJPsK     {\ensuremath{r_{\jpsi}^{\kaon}}\xspace}
\def\RJPsKst   {\ensuremath{r_{\jpsi}^{\Kstar}}\xspace}

\def\RPsiK{\ensuremath{R_{\psitwos}^{\kaon}}\xspace}
\def\RPsiKst{\ensuremath{R_{\psitwos}^{\Kstar}}\xspace}

\def\lqsq{low-\qsq\xspace}
\def\cqsq{central-\qsq\xspace}

\def\runone{\textsc{Run\,1}\xspace}
\def\runonetable{\textsc{Run\,1\phantom{p1}}\xspace}
\def\runtwo{\textsc{Run\,2}\xspace}
\def\runtwopo{\textsc{Run\,2p1}\xspace}
\def\runtwopt{\textsc{Run\,2p2}\xspace}

\def\ll{\ensuremath{\ellp\ellm}\xspace}

\def\pp{\ensuremath{\proton\proton}\xspace}

\def\tistable{$\textrm{TIS\phantom{O}}_\textrm{\phantom{inc}}$\xspace}
\def\tostable{$\textrm{TOS\phantom{I}}_\textrm{\phantom{inc}}$\xspace}
\def\tosinctable{$\textrm{TOS}_\textrm{inc}\textrm{\phantom{I}}$\xspace}
\def\tosinc{$\textrm{TOS}_\textrm{inc}$\xspace}

\def\tis{\textrm{TIS}\xspace}
\def\tos{\textrm{TOS}\xspace}

\def\sig{\textrm{S}\xspace}
\def\bkg{\textrm{B}\xspace}

\def\nsig{\ensuremath{N_{\sig}}\xspace}
\def\nbkg{\ensuremath{N_{\bkg}}\xspace}

\def\hepml {\mbox{hep\_ml}\xspace}

\def\wpid{\ensuremath{w_{\textrm{PID}}}\xspace}
\def\wtrk{\ensuremath{w_{\textrm{TRK}}}\xspace}

\def\wmco{\ensuremath{w_{\textrm{Mult\&Kin}}}\xspace}

\def\wlo{\ensuremath{w_{\textrm{\lone}}}\xspace}
\def\whlt{\ensuremath{w_{\textrm{\hlt}}}\xspace}
\def\wmcreco{\ensuremath{w_{\textrm{Reco}}}\xspace}

\def\etot{\ensuremath{\varepsilon_{\textrm{tot}}}\xspace}
\def\egeo{\ensuremath{\varepsilon_{\textrm{geo}}}\xspace}

\def\mll{\ensuremath{m(\ll)}\xspace}

\def\BToKee{\decay{\B^{(+,0)}}{\kaon^{(+,*0)}\epem}}
\def\BToKmm{\decay{\B^{(+,0)}}{\kaon^{(+,*0)}\mumu}}

\def\NBToKJPsll{\decay{\B^{(+,0)}}{\kaon^{(+,*0)}\jpsi(\decay{}{\ellell})}}

\def\NBToKJPsee{\decay{\B^{(+,0)}}{\kaon^{(+,*0)}\jpsi(\decay{}{\epem})}}

\def\NBToKJPsmm{\decay{\B^{(+,0)}}{\kaon^{(+,*0)}\jpsi(\decay{}{\mumu})}}

\def\BdToKstll{\decay{\Bd}{\Kstarz \ll}}
\def\BdToKstmm{\decay{\Bd}{\Kstarz \mumu}}
\def\BdToKstee{\decay{\Bd}{\Kstarz \epem}}

\def\BdToKstG{\decay{\Bd}{\Kstarz \gamma}}

\def\BdToKstJPs{\decay{\Bd}{\Kstarz \jpsi}}

\def\BdToKstPsill{\decay{\Bd}{\Kstarz \psitwos(\to\ll)}}

\def\BuToKll{\decay{\Bu}{\Kp \ll}}
\def\BuToKmm{\decay{\Bu}{\Kp \mumu}}
\def\BuToKmm{\decay{\Bu}{\Kp \mumu}}

\def\BuToKee{\decay{\Bu}{\Kp \epem}}

\def\BuToKJPs{\decay{\Bu}{\Kp \jpsi}}

\def\BuToKJPsee{\decay{\Bu}{\Kp \jpsi(\epem)}}

\def\NBuToKJPsee{\decay{\Bu}{\Kp \jpsi(\decay{}{\epem})}}

\def\BuToPiJPsll{\decay{\Bu}{\pip \jpsi(\decay{}{\ll})}}

\def\BuToKPiPiee{\decay{\Bu}{\Kp\pip\pim \epem}}

\def\BuToKPiPiJPsmm{\decay{\Bu}{\Kp\pip\pim \jpsi(\decay{}{\mumu})}}

\def\BsToPhiJPsll{\decay{\Bs}{\phi(1020) \jpsi(\decay{}{\ll})}}

\def\BsToKstJPsll{\decay{\Bs}{\Kstarzb \jpsi(\decay{}{\ll})}}

\def\BsToKstPsill{\decay{\Bs}{\Kstarzb \psitwos(\decay{}{\ll})}}

\def\BToXJPsll{\decay{\B}{X \jpsi(\decay{}{\ll})}}
\def\BToXJPsee{\decay{\B}{X \jpsi(\decay{}{\epem})}}
\def\BToXJPsmm{\decay{\B}{X \jpsi(\decay{}{\mumu})}}
\def\BToXPsill{\decay{\B}{X \psitwos(\decay{}{\ll})}}
\def\BToXPsiee{\decay{\B}{X \psitwos(\decay{}{\epem})}}
\def\BToXPsimm{\decay{\B}{X \psitwos(\decay{}{\mumu})}}

\def\LbTopKJPsll{\decay{\Lb}{p\kaon \jpsi(\decay{}{\ll})}}

\def\LbTopKPsill{\decay{\Lb}{p\kaon \psitwos(\decay{}{\ll})}}

\def\lqsq{\ensuremath{\text{low-}\qsq}\xspace}
\def\cqsq{\ensuremath{\text{central-}\qsq}\xspace}

\def\BToJPseeX{\decay{\B}{\jpsi(\decay{}{\epem})X}}

\def\flavio        {\mbox{\texttt{flavio}}\xspace}

\def\LbTopKJPsll{\decay{\Lb}{p\Km \jpsi(\decay{}{\ll})}}
\def\LbTopKPsill{\decay{\Lb}{p\Km \psi(2S)(\decay{}{\ll})}}

\def\RJPsKorKst   {\ensuremath{r_{\jpsi}^{K,\Kstar}}\xspace}

\def\RPsiKorKst{\ensuremath{R_{\psitwos}^{K,\Kstar}}\xspace}

\def\DstDpi{ \decay{\Dstarp}{\Dz(\decay{}{\Km \pip}) \pip }}

\def\JPsmumu{\decay{\jpsi}{\mumu}}

\def\MpsitwosPDG{\ensuremath{M_{\psitwos}^{\mathrm{PDG}}}\xspace}
\def\MjpsiPDG{\ensuremath{M_{\jpsi}^{\mathrm{PDG}}}\xspace}

\def\mcorr{\ensuremath{m_{\mathrm{corr}}}\xspace}
\def\mHOP{\mcorr}
\def\Kpll{\ensuremath{\Kp\ellell}\xspace}
\def\mreco{\ensuremath{m^{\mathrm{reco}}}\xspace}
\def\mtrue{\ensuremath{m^{\mathrm{true}}}\xspace}
\def\msmeared{\ensuremath{m^{\mathrm{Res}}}\xspace}

\def\DelPID{\ensuremath{\Delta_{\mathrm{PID}}}\xspace}
\def\dECAL{\ensuremath{\mathrm{d}_{\mathrm{ECAL}}}\xspace}

\def\wsmear{\ensuremath{w_{\textrm{Res}}}\xspace}

\usepackage{comment}
\usepackage{enumitem}
\usepackage{cleveref}
\usepackage{multirow}
\usepackage{placeins}
\usepackage{tikz}
\usetikzlibrary{trees,arrows,automata,shapes}
\usetikzlibrary{decorations.pathmorphing}
\usetikzlibrary{decorations.markings}
\usetikzlibrary{shapes.symbols}
\usetikzlibrary{shapes.geometric}
\usetikzlibrary{shapes.arrows}
\usetikzlibrary{positioning}
\tikzset
{
   vector/.style = {decorate, decoration={snake,amplitude=2pt, segment length=5pt}},
   fermion/.style = {postaction={decorate}, decoration={markings,mark=at position .55 with {\arrow{>}}}},
   fermionbar/.style = {postaction={decorate}, decoration={markings,mark=at position .55 with {\arrow{<}}}},
   gluon/.style = {decorate, decoration={coil,amplitude=3pt, segment length=5pt}},
   scalar/.style = {dashed, postaction={decorate}, decoration={markings,mark=at position .55 with {\arrow{>}}}},
   scalarbar/.style = {dashed, postaction={decorate}, decoration={markings,mark=at position .55 with {\arrow{<}}}}
}
\tikzstyle{information text}=[draw,rounded corners,inner sep=1ex]
\usepackage{longtable} 

\usepackage[title]{appendix} 

\begin{document}

\renewcommand{\thefootnote}{\fnsymbol{footnote}}
\setcounter{footnote}{1}


\begin{titlepage}
\pagenumbering{roman}

\vspace*{-1.5cm}
\centerline{\large EUROPEAN ORGANIZATION FOR NUCLEAR RESEARCH (CERN)}
\vspace*{1.5cm}
\noindent
\begin{tabular*}{\linewidth}{lc@{\extracolsep{\fill}}r@{\extracolsep{0pt}}}
\ifthenelse{\boolean{pdflatex}}
{\vspace*{-1.5cm}\mbox{\!\!\!\includegraphics[width=.14\textwidth]{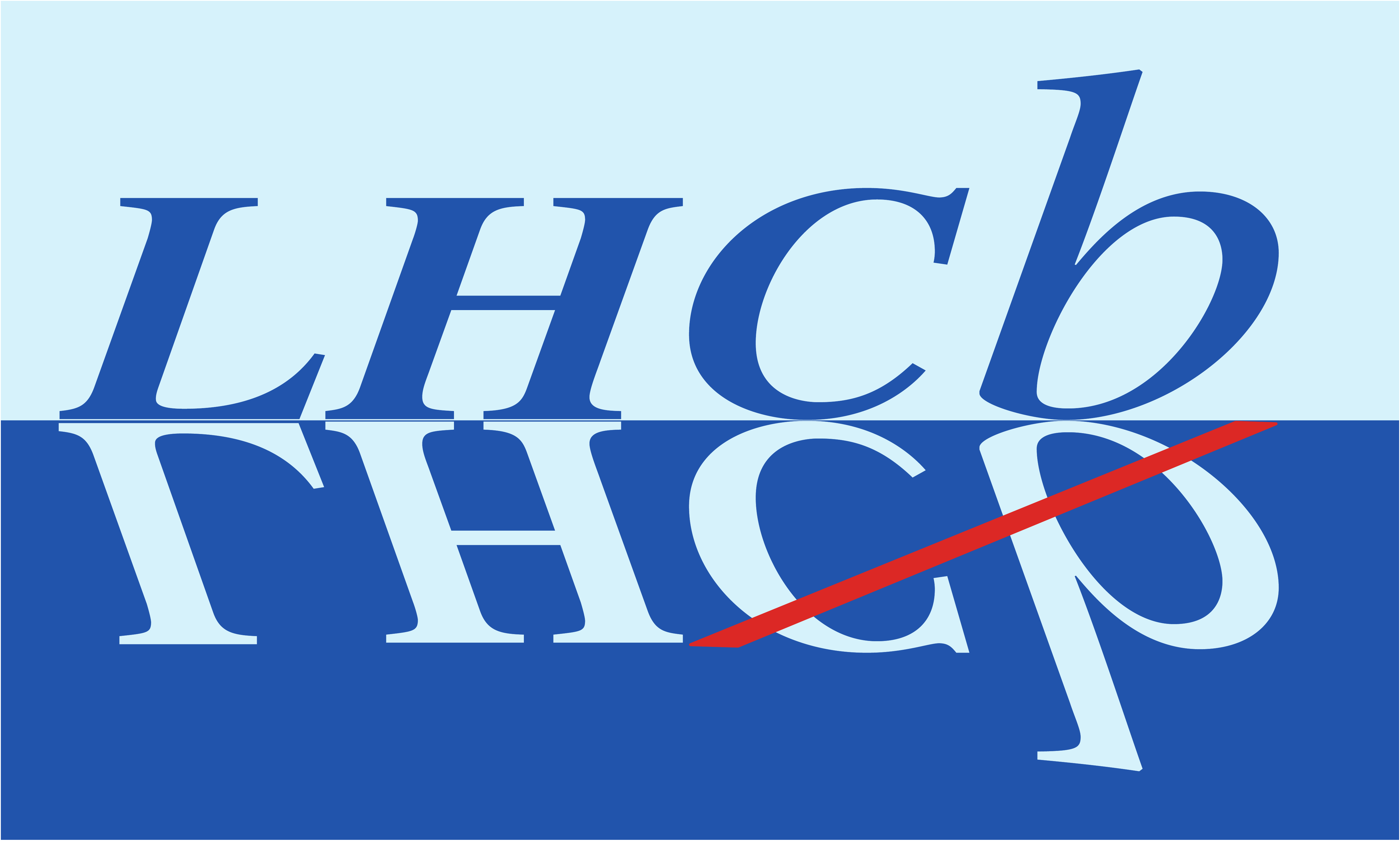}} & &}%
{\vspace*{-1.2cm}\mbox{\!\!\!\includegraphics[width=.12\textwidth]{figs/lhcb-logo.eps}} & &}%
\\
 & & CERN-EP-2022-278 \\  
 & & LHCb-PAPER-2022-045 \\  
 & & November 7, 2023 \\ 
 & & \\
\end{tabular*}

\vspace*{4.0cm}

{\normalfont\bfseries\boldmath\huge
\begin{center}
  \papertitle 
\end{center}
}

\vspace*{1.0cm}

\begin{center}
\paperauthors\footnote{Authors are listed at the end of this paper.}
\end{center}

\vspace{\fill}

\begin{abstract}
   A simultaneous analysis of the  \mbox{\BuToKll} and \mbox{\BdToKstll} decays is performed to test muon-electron universality in two ranges of the square of the dilepton invariant mass, \qsq.
   The measurement uses a sample of beauty meson decays produced in proton-proton collisions collected with the \lhcb detector between 2011 and 2018, corresponding to an integrated luminosity of 9\invfb. 
   A sequence of multivariate selections and strict particle identification requirements produce a higher signal purity and a  better statistical sensitivity 
   per unit luminosity than previous \lhcb lepton universality tests using the same decay modes. Residual backgrounds due to misidentified hadronic decays are studied using data and included in the fit model.
   Each of the four lepton universality measurements reported is either the first in the given \qsq interval or supersedes previous \lhcb measurements. The results are compatible with the predictions of the Standard Model.
  
\end{abstract}

\vspace*{2.0cm}

\begin{center}
  Published in Phys.~Rev.~D 108 (2023) 032002
\end{center}

\vspace{\fill}

{\footnotesize 
\centerline{\copyright~\papercopyright. \href{\paperlicenceurl}{\paperlicence}.}}
\vspace*{2mm}

\end{titlepage}


\newpage
\setcounter{page}{2}
\mbox{~}


\renewcommand{\thefootnote}{\arabic{footnote}}
\setcounter{footnote}{0}

\cleardoublepage


\pagestyle{plain} 
\setcounter{page}{1}
\pagenumbering{arabic}


\section{Introduction}
\label{sec:Introduction}
In the Standard Model (SM) of particle physics, gauge bosons have identical couplings with each of the 
three families of leptons, a phenomenon known as lepton universality (LU). The decay rates of SM hadrons 
to final states involving leptons are therefore independent of the  lepton family, with differences arising 
purely from lepton mass effects rather than from any intrinsic differences in couplings.
The validity of LU has been demonstrated at the percent level in \W boson decays
and at the per mille level in \Z boson decays~\cite{UA1:1988rck,CDF:1991mse,UA2:1992udr, CDF:1992jvc,D0:1995gzy,D0:1999bqi,ALEPH:2005ab,ALEPH:2013dgf,ATLAS:2016nqi,LHCb-PAPER-2016-024,LHCb-PAPER-2018-016}.

Interactions that violate LU arise naturally in extensions to the SM, because there is no fundamental principle requiring beyond the SM (BSM) particles to have the same couplings as their SM counterparts. However, to date 
there is no direct evidence for the existence of BSM particles, with particularly stringent limits on their 
couplings to SM processes and masses being set by the \atlas and \cms experiments at the LHC, see \eg Refs.~\cite{CMS:2021far,ATLAS:2021jyv}. Beyond the SM particles that are too heavy to be produced directly at 
the LHC can still participate in SM decays as virtual particles in higher-order contributions, altering decay 
rates and other observables with respect to the corresponding SM expectations. 

Measurements of rare, ``nonresonant'' semileptonic \bsll decays, where $\ell$ represents either an electron or a muon, are particularly sensitive 
probes of LU because the theoretical uncertainties on ratios of decay rates can be controlled at the percent level~\cite{Bordone:2016gaq,Isidori:2020acz,Isidori:2022bzw}. 
As a consequence, measurements of LU in these processes are powerful null tests of the SM that can 
probe the existence of BSM particles at energy scales  up to $\mathcal{O}(50\tev)$~\cite{LHCb-PII-Physics} 
with current  data, depending on the assumed nature of BSM couplings to SM particles.

While there had been longstanding theoretical interest in these processes~\cite{Wang:2003je,Hiller:2003js}, the
experimental interest increased significantly following LHCb's first test~\cite{LHCb-PAPER-2014-024} of LU in \mbox{\BuToKll}
decays,\footnote{The inclusion of charge-conjugate processes is implied throughout, unless stated otherwise.} 
which was consistent with the value predicted by the SM at the  $2.5\sigma$ level. Comparable levels of consistency 
were seen in measurements of \mbox{\BdToKstll}~\cite{LHCb-PAPER-2017-013}, \decay{\Lb}{\proton\Km\ellell}~\cite{LHCb-PAPER-2019-040}, \decay{\Bd}{\KS\ellell} and \decay{\Bu}{\Kstarp\ellell} \cite{LHCb-PAPER-2021-038} decays. The most recent \lhcb measurement using \BuToKll decays~\cite{LHCb-PAPER-2021-004} 
resulted in evidence of LU breaking with a significance of $3.1\sigma$ and, with a combined statistical and 
systematic uncertainty of approximately 5\%, is the  most precise such measurement to date.
If the current experimental central value were to be confirmed, there is consensus that the  deviation could not be 
explained through underestimated theoretical uncertainties of the SM prediction:
establishing LU breaking in \bsll decays would constitute an unambiguous sign of physics beyond 
the Standard Model. It is therefore vital to improve the experimental precision and consider potential correlations 
among \bsll LU measurements.

This paper presents the first simultaneous test of muon-electron LU using nonresonant \mbox{\BuToKll} and \mbox{\BdToKstll} decays.
A more concise description of this test is reported in a companion article~\cite{LHCb-PAPER-2022-046}.
Here, \Kstarz  represents a $\Kstar(892)^0$ meson, which is reconstructed in the $\Kp\pim$ 
final state by selecting candidates within 100\mevcc of its known mass \cite{PDG2020}. The  relative decay rates to 
muon and electron final states, integrated over a region of the square of the dilepton invariant mass (\qsq), 
$q^2_a<\qsq <q^2_b$, are used to construct the observables \RK and \RKst in terms of the decay rates $\Gamma$: 
\begin{align}
   \RKorKst(q^2_a,q^2_b) = &
     \frac
         {\displaystyle\int _{q^2_a}^{q^2_b}
            \frac{\deriv\Gamma(\BToKmm)}{\deriv\qsq}\deriv\qsq}
         {\displaystyle\int _{q^2_a}^{q^2_b}
            \frac{\deriv\Gamma(\BToKee)}{\deriv\qsq}\deriv\qsq } \;\;\;.
\end{align}
These observables are measured in two \qsq intervals: $0.1<\qsq<1.1$\gevgevcccc (\lqsq);  $1.1<\qsq<6.0$\gevgevcccc (\cqsq).
All proton-proton collision data recorded by the \lhcb detector between 2011 and 2018 are used, corresponding to integrated 
luminosities of 1.0, 2.0, and $6.0\invfb$ at center-of-mass energies of 7, 8 and 13\tev, respectively.

While the \BdToKstmm and \BuToKmm  muon-mode signal decays are experimentally independent of one another, 
this is not the case for the electron-mode signal decays due to their poorer mass resolution: partially reconstructed  \BdToKstee decays  
represent a significant background to the \BuToKee  decay. The simultaneous measurement introduced here 
allows this background to be determined directly from the observed yields of the signal \BdToKstee decay.

The processes \BdToKstJPs and \BuToKJPs (``resonant modes''), with \decay{\jpsi}{\ellell}, share the same 
final state as the signal modes and therefore dominate in \qsq regions corresponding to the square of the \jpsi meson mass.
The large resonant mode samples serve as a normalization channel for the signal decays  and allow  determination of correction factors, which  
account for imperfect modeling of the \lhcb detector. 
The corrections obtained from the \Bp (\Bz) channel are applied to the \decay{\Bd}{\Kstarz\ellell} (\decay{\Bu}{\Kp\ellell}) decay and the two sets are shown to be interchangeable.

This analysis is performed at a higher purity level than previous \lhcb tests of LU, due to both stricter particle
identification (PID) criteria and dedicated multivariate selections to reject misidentified and partially reconstructed backgrounds.
The trigger strategy is also optimized to improve the signal purity and to minimize the differences in trigger efficiency 
between electrons and muons. Finally, data are used to estimate residual backgrounds that
survive all these criteria and allow them to be modeled in the analysis. Taken together, these choices lead to both a better
statistical sensitivity per unit integrated luminosity and a more accurate estimate of systematic uncertainties.

This paper is structured as follows. First, the \lhcb detector is described in Sec.~\ref{sec:detector}.
Subsequently, the phenomenology of \bsll decays in the context of LU tests is briefly discussed in Sec.~\ref{sec:pheno}, 
and the analysis strategy is outlined in Sec.~\ref{sec:method}. The 
event selection and modeling of backgrounds is discussed in Sec.~\ref{sec:selection}, followed 
by a description of how the simulation is calibrated and used to calculate the efficiencies 
in Sec.~\ref{sec:effs}. The simultaneous fit to the \Bd and \Bu invariant-mass distributions is 
described in Sec.~\ref{sec:simfit}, and the cross-checks performed to validate the robustness of 
the analysis procedure are documented in Sec.~\ref{sec:crosschecks}. Systematic
uncertainties are discussed in Sec.~\ref{sec:systematics}, results are detailed in 
Sec.~\ref{sec:results} and summarized in Sec.~\ref{sec:conclusion}.

\section{LHCb detector and simulation}
\label{sec:detector}

The \lhcb detector~\cite{LHCb-DP-2008-001,LHCb-DP-2014-002} is a single-arm forward
spectrometer covering the \mbox{pseudorapidity} range $2<\eta <5$,
designed for the study of particles containing \bquark or \cquark
quarks. The detector includes a high-precision charged-particle reconstruction (tracking) 
system consisting of a silicon-strip vertex detector surrounding the $pp$
interaction region~\cite{LHCb-DP-2014-001}, a large-area silicon-strip detector~(TT) located
upstream of a dipole magnet with a bending power of about
$4{\mathrm{\,Tm}}$, and three stations of silicon-strip detectors and straw
drift tubes~\cite{LHCb-DP-2013-003,LHCb-DP-2017-001}
placed downstream of the magnet.
The tracking system provides a measurement of the momentum, \ptot, of charged particles with
a relative uncertainty that varies from 0.5\% at low momentum to 1.0\% at 200\gevc.
The minimum distance of a track to a primary vertex (PV), the impact parameter (IP), 
is measured with a resolution of $(15+29/\pt)\mum$,
where \pt is the component of the momentum transverse to the beam, in\,\gevc.
Different types of charged hadrons are distinguished from one another using information
from two ring-imaging Cherenkov detectors~\cite{LHCb-DP-2012-003}. 
Photons, electrons and hadrons are identified by a calorimeter system consisting of
scintillating-pad and preshower detectors, an electromagnetic and a hadronic calorimeter.
Information from these detectors is combined to build global log-likelihoods corresponding to various mass hypotheses for each particle in the event. 
The electromagnetic calorimeter (ECAL) consists of three regions with square cells of side length $40.4$~mm, $60.6$~mm or $121.2$~mm, with the smaller sizes closer to the beam.
The calorimeter system is used to reconstruct photons with at least 75\mev energy transverse to the beam \cite{LHCb-DP-2020-001}. The transverse energy is estimated as $\et = E \sin\theta$, where $E$ is the measured energy deposit in a given ECAL cell, and $\theta$ is the angle between the beam direction and a line from the PV to the center of that cell~\cite{LHCb-DP-2020-001}.
Photons are associated with 
reconstructed electron trajectories to take into account potential bremsstrahlung energy losses incurred while passing through the \lhcb detector.
Muons are identified by a system composed of alternating layers of iron and multiwire
proportional chambers~\cite{LHCb-DP-2012-002}.

The real-time selection of LHC $pp$ interactions is performed by a trigger~\cite{LHCb-DP-2012-004}, 
which consists of a hardware stage (\lone), based on information from the calorimeter and muon
systems, followed by a software stage (\hlt), which applies a full event
reconstruction. At the hardware trigger stage, events are required to
have a muon with high \pt, or a hadron or an electron with high transverse energy in the calorimeters.
In addition, the hardware trigger rejects events having  too many hits in the scintillating-pad
detector, since large occupancy events have large backgrounds, which reduces the  
reconstruction and PID performance.
The software trigger requires a two- or three-body secondary vertex with significant displacement from any primary $pp$ interaction vertex. At least one charged particle
must have significant transverse momentum and be inconsistent with originating from a PV.
A multivariate algorithm~\cite{BBDT,LHCb-PROC-2015-018} based on kinematic, geometric and lepton identification criteria is used for the identification of secondary vertices consistent with the decay of a \bquark hadron.

Simulation is used to model the effects of the detector acceptance, resolution and the imposed selection requirements. In the simulation, $pp$ collisions are generated using \pythia~\cite{Sjostrand:2007gs,*Sjostrand:2006za} with a specific \lhcb configuration~\cite{LHCb-PROC-2010-056}. Decays of unstable particles are described by \evtgen~\cite{Lange:2001uf}, in which final-state radiation is generated using \photos~\cite{davidson2015photos}. The interaction of the generated particles with the detector, and its response, are implemented using the \geant toolkit~\cite{Allison:2006ve, *Agostinelli:2002hh} as described in Ref.~\cite{LHCb-PROC-2011-006}. As the cross-section for $\ccbar$ production~\cite{LHCb-PAPER-2015-041} exceeds 1\mbarn
in the LHCb acceptance, abundant samples of charm hadron and charmonia decays have been collected using a
tag-and-probe approach~\cite{LHCb-DP-2018-001} for all data-taking periods. 
These are used to calibrate the simulated hadron and muon track reconstruction and PID 
performance to ensure that they describe  data in the kinematic and geometric ranges of interest to this analysis. Electron reconstruction and identification efficiencies are calibrated using tag-and-probe samples of inclusive 
\BToJPseeX decays, as discussed further in Sec.~\ref{sec:effs}.
\section{Phenomenology of LU in \texorpdfstring{\boldmath \bsll}{b->sl+l-} decays}
\label{sec:pheno}
The \bsll decay rate has a strong \qsq dependence due to the various contributing processes. 
Discrepancies 
between the true and reconstructed \qsq distributions arise due to the resolution and efficiency of the detector. These effects are modeled and taken into account in the analysis as discussed in Sections~\ref{sec:method}--\ref{sec:simfit}. The remainder of this section
will discuss the \bsll phenomenology in terms of the true \qsq.
\begin{figure}[ht]
\includegraphics[width=0.3\textwidth]{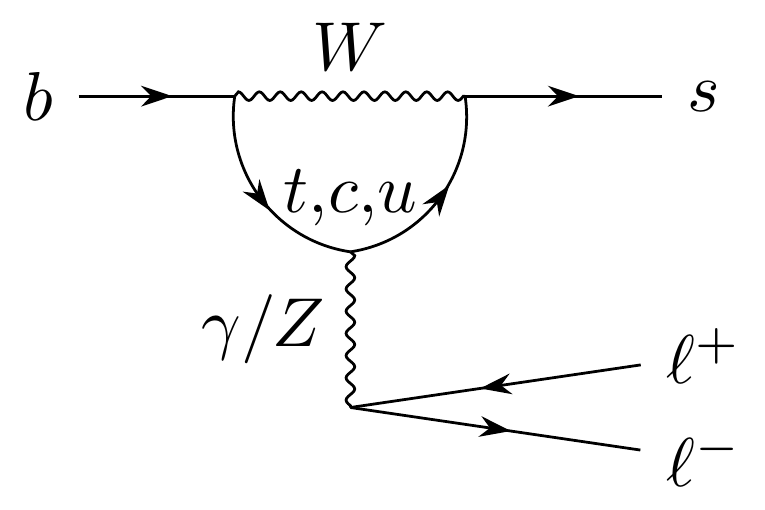}
\includegraphics[width=0.3\textwidth]{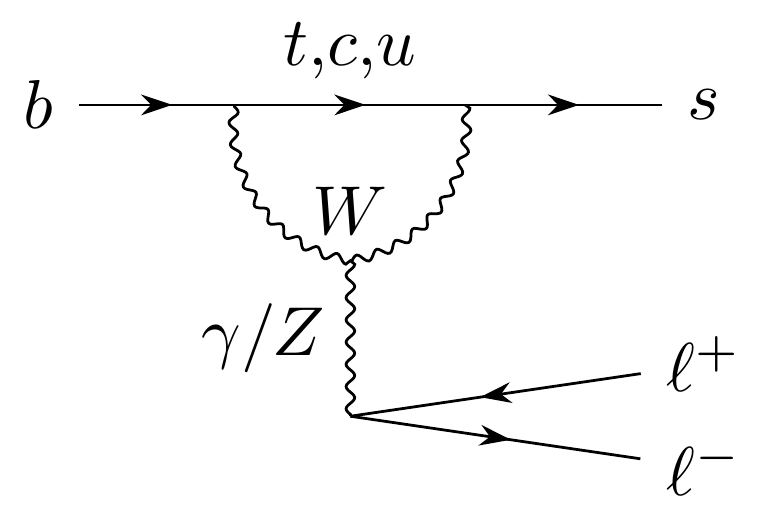}
\includegraphics[width=0.3\textwidth]{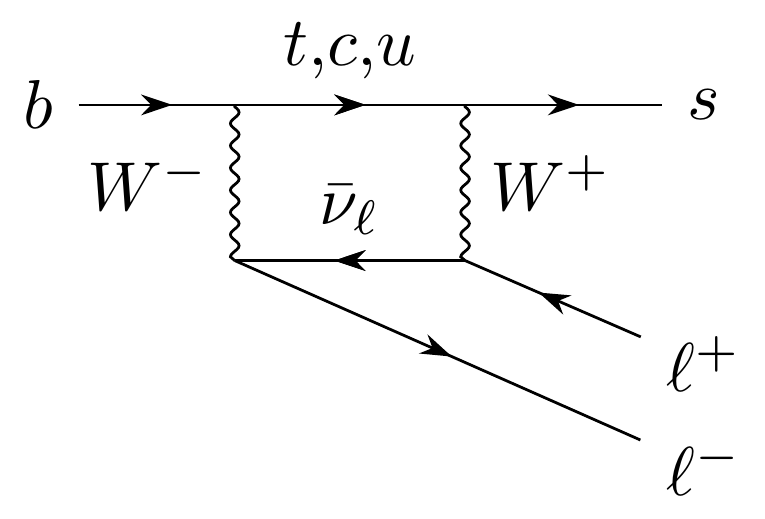}
\caption{Leading-order Feynman diagrams for \bsll transitions in the SM.\label{fig:smdiagrams}}
\end{figure}

The SM forbids flavor-changing neutral current (FCNC) processes at tree level, and so they proceed via amplitudes involving electroweak loop (penguin and box) Feynman diagrams. 
The SM description of \bsll decays is often expressed in terms of an effective
field theory (EFT) ansatz that factorizes the heavy, short-distance (perturbative) physics from the light, long-distance (non-perturbative) effects~\cite{Buchalla:1995vs}. While theoretical predictions of non-local effects have substantial associated uncertainties, these are confined to the hadronic part 
of \bsll decays. 
Within the EFT approach, a set of Wilson coefficients 
encodes the effective coupling strengths of local operators. Muon-electron universality therefore implies that the muon and electron Wilson coefficients are equal in \bsll decays.

The leading-order FCNC SM diagrams for \bsll decays are shown in Fig.~\ref{fig:smdiagrams}. They result in differential branching fractions, integrated over given \qsq regions, of $\order(10^{-7})$, \eg Ref.\cite{LHCb-PAPER-2014-006}. 
In the vicinity of the photon pole, the \BdToKstee decay branching fraction is dominated by the lepton-universal
electromagnetic penguin operator \C7, and the electron-muon mass difference induces significant  
LU-breaking effects.
Additional SM diagrams play a role in regions of \qsq near hadronic resonances that can decay to
dileptons.  In these regions corresponding to light meson resonances such as the \etaz, $\rho(770)$, $\omegaz(782)$, 
$\etapr(958)$ and $\phiz(1020)$, the resonant decay proceeds primarily through gluonic FCNC $b\to (s,d)$ transitions. The branching fractions of the decays of these light resonances to dileptons are  $\order(10^{-4})$ or smaller. As a result, the diagrams in Fig.~\ref{fig:smdiagrams} dominate the \qsq region of this analysis.
In \qsq regions corresponding to the \jpsi 
and \psitwos charmonium resonances, decays are dominated by tree-level \decay{\bquark}{\ccbar s}
processes. These have  branching fractions of ${\cal O}(10^{-3})$,  which are orders of magnitude larger than the FCNC contribution. 
As LU has
been established to hold to within 0.4\% in \jpsi meson decays~\cite{BESIII:2013csc,PDG2022}, contributions from charmonium resonances are considered lepton-flavor universal. The resonant charmonium decays are therefore used in this analysis as both calibration and normalization counterparts to the FCNC signals.

Calculations of decay rates to inclusive muon and electron final states in the SM are affected by sizable form-factor uncertainties, as well as uncertainties due to the contributions from non-resonant \ccbar loop diagrams. 
As mentioned above, these uncertainties cancel in the ratio outside the photon pole region~\cite{Wang:2003je,Hiller:2003js} and 
the leading source of uncertainty in the SM predictions is from the modeling of radiative effects in \photos~\cite{davidson2015photos}.

\begin{figure}[t]
\begin{center}
\begin{tikzpicture}[scale=0.65]
    \draw [fermion, line width = 0.7pt] (0.,0) node[above]{$b$} -- (3.5,0);
    \draw [fermion, line width = 0.7pt] (3.5,0) -- (7,0) node[above]{$s$};
    \draw [vector, line width = 0.7pt] (3.0,0) -- (4.0,-2.5) node[midway,left]{$Z'~$};
    \draw [fermion,    line width = 0.7pt] (4.0,-2.5) -- (6.6,-3) node[right]{$\ell^-$};
    \draw [fermionbar, line width = 0.7pt] (4.0,-2.5) -- (6.6,-2) node[right]{$\ell^+$};

    \draw [fermion, line width = 0.7pt] (10.,0) node[above]{$b$} -- (13.5,0);
    \draw [fermion, line width = 0.7pt] (13.5,0) -- (17,0) node[above]{$\ell^-$};
    \draw [very thick,double, line width = 0.7pt] (13.0,0) -- (14.0,-2.5) node[midway,left]{$LQ~$};
    \draw [fermion,    line width = 0.7pt] (14.0,-2.5) -- (16.6,-3) node[right]{$s$};
    \draw [fermionbar, line width = 0.7pt] (14.0,-2.5) -- (16.6,-2) node[right]{$\ell^+$};
\end{tikzpicture}
\end{center}
\caption{Examples of Feynman diagrams for \bsll decays beyond the SM. Potential contributions from new heavy $Z^\prime$ gauge bosons are shown on the left, contributions from leptoquarks (LQ) on the right.\label{fig:npdiagrams}}
\end{figure}
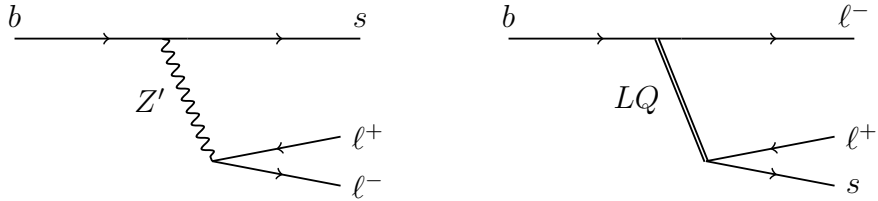

The tensions with the SM prediction in previous tests of LU in \bsll\ decays, combined with tensions of similar size in 
angular analyses and branching fraction measurements of \decay{b}{s\mumu} decays~\cite{LHCb-PAPER-2014-006,LHCb-PAPER-2015-009,LHCb-PAPER-2015-023,LHCb-PAPER-2016-012, LHCb-PAPER-2021-014,LHCb-PAPER-2021-022,LHCb-PAPER-2015-051,LHCb-PAPER-2020-002,LHCb-PAPER-2020-041,ATLAS:2018gqc,CMS:2015bcy,CMS:2017rzx}, have led to many proposed BSM explanations, see \eg Refs.~\cite{Hiller:2014yaa,Gripaios:2014tna,deMedeirosVarzielas:2015yxm,Barbieri:2016las,Altmannshofer:2014cfa,Crivellin:2015mga,Celis:2015ara,Falkowski:2015zwa}. Models involving $\Z'$ bosons and leptoquarks, 
illustrated in Fig.~\ref{fig:npdiagrams}, are particularly popular in the literature. 
 New particles that couple to the SM sector and break LU will influence the rates of many SM processes other than \decay{b}{s\mumu} decays. 
 The conventional way to confront BSM models with these constraints is through global EFT fits
in which the hypothetical BSM particles modify the Wilson coefficients from their SM values. 

Taken by themselves, measurements of relative muon-electron decay rates do not determine whether LU-violating effects arise from anomalous couplings to muons, electrons, or both. Due to the coherent pattern of deviations from the SM predictions that is observed in angular analyses and branching fractions of \decay{b}{s\mumu} decays~\cite{LHCb-PAPER-2014-006,LHCb-PAPER-2015-009,LHCb-PAPER-2015-023,LHCb-PAPER-2016-012, LHCb-PAPER-2021-014,LHCb-PAPER-2021-022,LHCb-PAPER-2015-051,LHCb-PAPER-2020-002,LHCb-PAPER-2020-041,ATLAS:2018gqc,CMS:2015bcy,CMS:2017rzx}, most models proposed  
introduce a shift of the muonic vector- and axial-vector couplings denoted by the Wilson coefficients \C9  and \C10, respectively.
\begin{figure}[t]
\centering
\includegraphics[width=\textwidth]{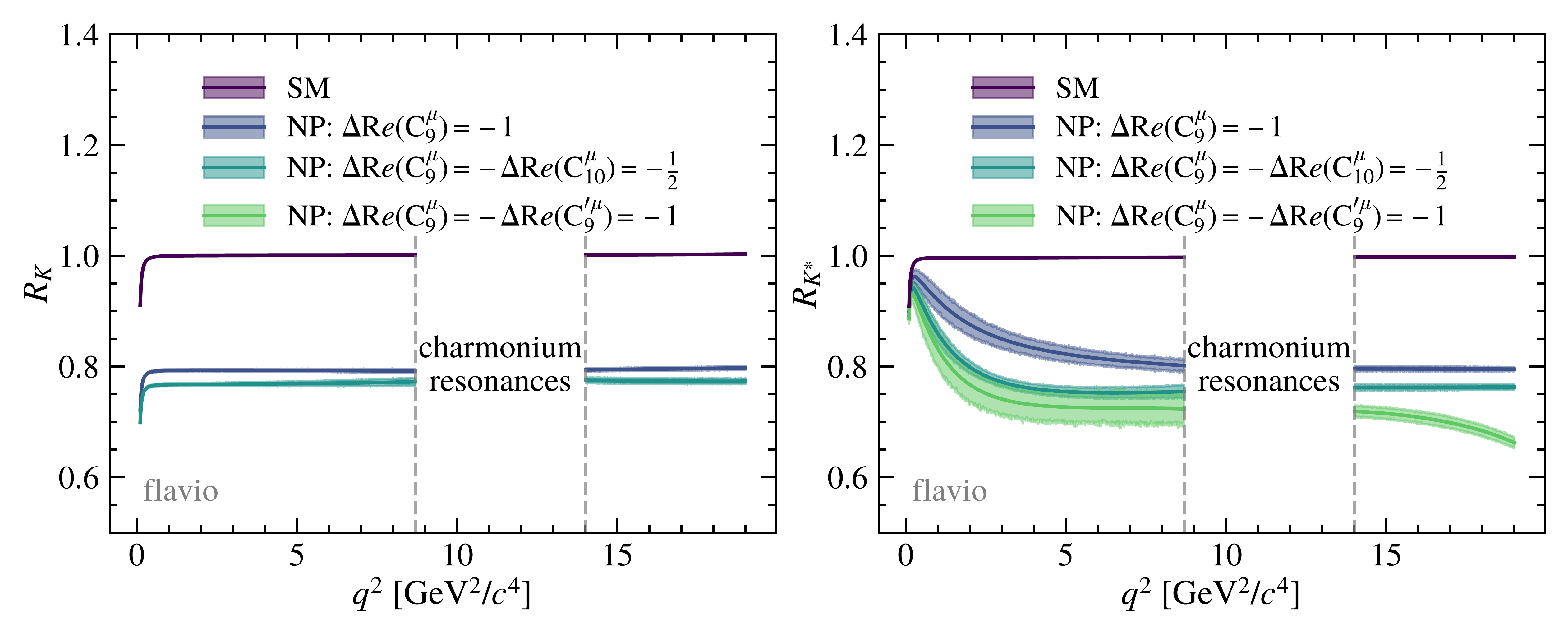}
\caption{Variation of (left) \RK and (right) \RKst as a function of \qsq within the SM obtained using the \flavio software package~\cite{Straub:2018kue}, taking into account potential heavy BSM contributions to the Wilson coefficients. The contributions from \ccbar resonances are subtracted in both cases. For \RK, the SM prediction overlaps with the BSM scenario  $\Delta {\cal R}e(C_9^\mu)=-\Delta {\cal R}e(C_9^{\prime\mu})=-1$.}
\label{fig:rxvsq2preds}
\end{figure}
The impact of modifying the muonic \C9 and \C10 Wilson coefficients on the \RK and \RKst LU ratios
is illustrated in Fig.~\ref{fig:rxvsq2preds}. 
The strikingly different \qsq behavior between the predicted values of \RK and \RKst would allow precise measurements to resolve the contributions from the different Wilson coefficients.

\section{Analysis strategy}
\label{sec:method}
The fundamental approach of this analysis is to treat the measurements of \RK and \RKst as null
tests of the Standard Model. The analysis is designed to maximize the signal significance at the expected SM decay rates,  and achieves a higher signal purity than previous \lhcb analyses of  
these decay modes. The treatment of decays with different final states is also made as coherent as possible, including at the triggering stage.
A multivariate selection based on decay kinematics, geometric features and displacement from the associated PV is 
used to reject combinatorial background. In addition, PID requirements and two dedicated selections, defined later, are designed to suppress
backgrounds from other partially reconstructed beauty hadron decays, as well as to improve purity for electron
signals in the region below the \bquark hadron masses.
Three data-taking periods based on common center-of-mass energies and trigger thresholds are defined and used throughout this analysis: \runone (2011--2012), \runtwopo
(2015--2016), and \runtwopt (2017--2018). Given that many aspects of the analysis depend on the treatment
of electron bremsstrahlung, three further categories are defined based on whether the dielectron
system has zero, one, or at least two associated bremsstrahlung photons.

The definition of the \cqsq region from 1.1 to 6.0\gevgevcccc is the same as in previous
LHCb analyses of these decay modes~\cite{LHCb-PAPER-2017-013,LHCb-PAPER-2021-004}. The lower limit excludes the light meson resonances,
while the upper limit minimizes background contamination from resonant \decay{\jpsi}{\epem} decays that 
 can undergo bremsstrahlung emission,  resulting in a reconstructed dilepton invariant mass well below the known \jpsi mass. 
The definition
of the \lqsq region is changed from that used in the previous \lhcb measurement of \RKst \cite{LHCb-PAPER-2017-013},  where it extended down to the dimuon mass threshold of $0.045$\gevgevcccc to increase the signal yield, leading to substantial contamination from
the photon pole in the electron mode. This lepton mass effect induces significant LU breaking also within the SM,
with an expected \RKst value of $\sim\! 0.9$. The current analysis defines the \lqsq region from 0.1 to 1.1\gevgevcccc, excluding most of the photon pole and leading to an expected \RKst value
of $\sim\! 0.98$ within the SM~\cite{Bordone:2016gaq}, 
close to unity as also expected for the \cqsq region. 
The same definition of the \lqsq region is used for \RK.

As in previous \lhcb  analyses of LU, the \RK and \RKst ratios are measured by forming
double ratios of efficiency corrected yields in the nonresonant and resonant modes,
\begin{align}
\NRKorKst & \equiv 
\frac{\tfrac{\cal{N}}{\varepsilon}(\BToKmm)}
     {\tfrac{\cal{N}}{\varepsilon}(\NBToKJPsmm)}
\bigg{/} 
\frac{\tfrac{\cal{N}}{\varepsilon}(\BToKee)}
     {\tfrac{\cal{N}}{\varepsilon}(\NBToKJPsee)} \, ,
     \label{eq:doubleratiorx}
\end{align}
where $\tfrac{\cal{N}}{\varepsilon}(X)$ represents the efficiency corrected yield for process $X$.
Potential systematic uncertainties arising from differences in the 
detection efficiencies for muons and electron largely cancel in the double ratios, apart from those induced by kinematic differences between the signal and resonant modes. The single ratios of efficiency corrected yields in the resonant \jpsi modes,
\begin{align}
\label{eq:rawrjpsik}
\RJPsKorKst & \equiv
\frac{\tfrac{\cal{N}}{\varepsilon}(\NBToKJPsmm)}
     {\tfrac{\cal{N}}{\varepsilon}(\NBToKJPsee)}
\end{align}

\noindent are used extensively to perform  cross-checks of the analysis procedure, as described in 
Section~\ref{sec:crosschecks}. Additional cross-checks are performed using two  
double ratios, \RPsiK and \RPsiKst, which are defined in direct analogy with 
Eq.~\ref{eq:doubleratiorx}, substituting the signal modes with the resonant \psitwos\ decays to $\ep\en$ and $\mumu$. 

A simultaneous fit to the reconstructed $\B^{0,+}$ candidate mass distributions in the signal modes and resonant \jpsi modes is used to determine \RK and \RKst within the low and  \cqsq ranges.
This approach allows the $4\times 4$ covariance matrix of statistical and systematic uncertainties to be determined so that they can be incorporated into global fits or alternative interpretations.
Partially reconstructed \BdToKstee decays, where the pion from the \decay{\Kstarz}{\Kp\pim} 
decay chain is not selected, represent a significant background in the invariant-mass
spectrum of \BuToKee decays.  
The simultaneous fit allows the yield of the partially reconstructed \Kstarz\epem  background, as well as contributions from the isospin-related decay \decay{\Bu}{\Kstarp\ep\en}, to 
be constrained by the fully reconstructed \BdToKstee signal and known detector efficiencies.
This improves the sensitivity of the fit, and for the first time also ensures that the background yield in the \Kp\epem spectrum is consistent with the measured value of \RKst.

Trigger decisions are associated with particles reconstructed offline.  Requirements can be made on whether the decision
was due to the reconstructed signal candidate (triggered on signal or \tos);  or independent of the signal candidates and due to other particles produced in the \pp collision (triggered independent of signal or \tis); or a combination of both.
This analysis divides the events into mutually exclusive categories based on the \lone trigger decision, similar to previous LU tests. 

The \lone trigger makes decisions based on  kinematic information from the muon and 
calorimeter systems, with associated quantities having lower resolution and reconstruction efficiency than their offline counterparts.
The \lone trigger has a significant fraction of \tis events which can be used for the analysis. 
As the \lone hadron trigger can have a different performance for $\Kp$ and \Kstarz final states
due to overlapping clusters in the hadronic calorimeter, 
events exclusively selected by it are excluded from this analysis for both muons and electrons, leading to a negligible loss of efficiency for \BdToKstee decays and up to 14\% for \BuToKee decays.
In the case of muon signals, over 90\% of events 
selected by the \lone trigger are \tos, and only around 25\% are \tis (the excess in the sum over 100\% is due to some events being both \tos and \tis). 
Due to larger background rates, the \lone electron trigger has more stringent requirements
and a lower signal efficiency than the muon trigger. As a result, the \tos fraction 
is only around 60\%, while the \tis fraction is around 50\%. 
Even though the fraction of muon \tis is small, 
the overall detector efficiency for muons is much larger 
than for electrons. Therefore, the absolute yield of muon \tis is still 
larger than that of electron signals in either the \tis or \tos categories.

In order to define mutually exclusive samples, the primary trigger category is chosen to be \tis for both muon and electron final states.
Events that are \tos on the \lone muon or electron trigger, whilst being not \tis (\ie they are not in the primary trigger category), are placed into the
secondary trigger category for muon and electron final states, respectively. 
This approach has several advantages compared to that used in previous \lhcb LU analyses where the \lone hadron trigger on the \Kp and \Kstarz candidates was used and preference to \tos category was given. Firstly, it increases the muon signal yields and gives 
two almost equally populated trigger categories for electron signals. 
Secondly, although trigger decisions due to the signal
candidate are directly correlated with kinematic quantities, trigger decisions due to the rest of the
\pp collision only modify the signal kinematics indirectly; this occurs through correlations between the signal and
other particles produced in the same \pp  collision. The \tis category therefore not only minimizes
efficiency differences between the muon and electron signals, but also minimizes the impact of
differences in the signal kinematics between data and simulation.

The \hlt selects events based on tracking information,  with loose lepton identification requirements also applied. It is therefore sufficiently well aligned with the offline selection not to require any special treatment beyond
the choice of appropriate trigger paths for the electron and muon modes described earlier.
Only a few percent of events are \tis at the \hlt stage. These \hlt-\tis events are crucial
for calibrating the \tos trigger performance in data as described in Sec.~\ref{sec:effs}, 
but are not otherwise used in the analysis (unless they are also \tos).

\section{Event selection and background}
\label{sec:selection}
The reconstruction of \BuToKll and \BdToKstll
candidates requires a dilepton system, which consists of a pair of oppositely charged particles, identified as either electrons or muons and required to originate from a common vertex.
Muons and electrons are required to have \pt greater than 800 and 500\mevc, respectively, and to have momentum greater than 3\gevc.
All tracks used in this analysis are required to 
satisfy track quality requirements, using their \chisq as determined by a Kalman filter and the output of a neural network trained to distinguish between genuine and fake tracks~\cite{DeCian:2255039}. A dedicated algorithm associates reconstructed bremsstrahlung photons to tracks identified as electrons; when a given photon is associated with both electron tracks, it is attached to one chosen randomly. The bremsstrahlung energy loss recovery procedure is used to improve the electron momentum resolution by  searching for photon clusters that are not already associated with particle tracks in the event.
This takes place 
within regions in the electromagnetic calorimeters into which electron tracks segments reconstructed upstream of the magnet have been extrapolated. 
Lepton tracks and dilepton candidates are 
required to satisfy criteria on transverse momenta, displacement from the PV and, for dilepton candidates, their vertex fit quality.
A similar approach is used to reconstruct \decay{\Kstarz}{\Kp\pim} candidates.
The \B candidates are subsequently formed by combining the dilepton candidates with either a
charged particle  identified as a \Kp, or with the \Kstarz candidates for
which the invariant mass of the \kaon\pion system is required to be within 100\mevcc of the known \Kstarz 
mass~\cite{PDG2022}. The \B candidates need to satisfy minimal criteria 
on their transverse momentum, displacement from the PV and vertex fit quality. 
The fit of the \B candidate is performed using the decay tree fitter~\cite{Hulsbergen:2005pu} algorithm. 
In addition, the \B-candidate momentum vector is required to be consistent with the vector connecting the \B candidate's production and decay vertices (the displacement vector). 

Minimum requirements on the angles between final-state particle trajectories 
ensure that the \B candidates are not constructed from duplicated tracks using the same track segment in the vertex detector.
The criteria applied in this reconstruction and preselection are identical for the signal and resonant
control modes, and are aligned as much as possible between the \Kp and \Kstarz final states.
Finally, the \B candidates are divided into regions based on their reconstructed dilepton \qsq:
\begin{align*}
    &\textrm{\lqsq region:}& 0.1 < \qsq < 1.1\gevgevcccc \,,\\
    &\textrm{\cqsq region:}& 1.1 < \qsq < 6.0\gevgevcccc \,,\\
    &\textrm{electron \jpsi region:}& 6 < \qsq < 11\gevgevcccc \,,\\
    &\textrm{muon \jpsi region:}& |\mll - \MjpsiPDG| <100\mevcc \,,\\
    &\textrm{electron \psitwos region:}& 11 < \qsq < 15\gevgevcccc \,,\\
    &\textrm{muon \psitwos region:}& |\mll - \MpsitwosPDG| <100\mevcc \,, 
\end{align*}
where \MjpsiPDG and \MpsitwosPDG are the known masses of the \jpsi and \psitwos mesons~\cite{PDG2022}, respectively.
The low- and \cqsq regions are identical for muons and electrons, whereas the resonant regions
are significantly broader for electrons due to their poorer dilepton mass resolution.

Particle identification requirements are used to suppress backgrounds. Two families of variables are used: the difference in log-likelihood between the given charged-species hypothesis and the pion hypothesis (named DLL), and the output of artificial neural networks trained to identify each charged-particle species (normalized between 0 and 1 and named ProbNN) \cite{LHCb-PUB-2016-020,LHCb-DP-2014-002,LHCb-DP-2018-001}.
The multivariate approach uses information from
all subdetectors to compute the compatibility of each track with a given particle hypothesis.
Muons and electrons are required to satisfy stringent compatibility criteria with their assigned
particle hypothesis. Kaons and pions must satisfy both a minimal compatibility requirement with their assigned particle hypothesis and be incompatible with an alternative hypothesis.
The alternative hypotheses considered are protons in the case of kaon candidates, and protons and kaons in the case of pion candidates. 
Kaons and electrons are required to satisfy minimal criteria with respect to the pion hypothesis.

As PID requirements are calibrated using data from control samples as discussed in Section~\ref{sec:effs},  further kinematic and geometric fiducial requirements are necessary to align the selection of tracks in the candidates with those in the control samples. 
Wherever possible the same requirements are applied
to the resonant control modes. This reduces potential systematic uncertainties associated with the 
determination of relative selection efficiencies in the \RK and \RKst double ratios. 
While PID criteria factorize for most particle species, an electron-positron pair can have correlated PID
 efficiencies due to overlapping clusters in the ECAL. Therefore, as discussed in Sec.~\ref{sec:effs_pid}, a fiducial requirement is used to remove such candidates from the analysis. 
Although  the preselection and PID requirements achieve  acceptable purity for the resonant control modes, further selection requirements are essential to improve the purity of the signal modes. 

The remaining backgrounds are divided into four groups: random combinations of particles originating
from multiple physical sources (combinatorial); backgrounds having missing energy in which all particles 
originate from a single physical process (partially reconstructed); individual backgrounds that are vetoed with specific criteria or taken into account in the invariant mass fit (exclusive); and
residual backgrounds from hadrons misidentified as electrons, with or without missing energy, that
must be taken into account in the invariant mass fit (misidentified). 
With the application of all criteria, less than one percent of events 
have multiple candidates; in such cases a single reconstructed candidate is chosen randomly.

\subsection{Combinatorial and partially reconstructed backgrounds}
A multivariate classifier~\cite{Likhomanenko:2015hca,2017arXiv170609516P} is trained to distinguish between 
$\decay{B^{(+,0)}}{K^{(+,*0)}\ellp\ellm}$
decays and combinatorial background. The training
 uses simulated signal events, and data with reconstructed 
\B meson invariant masses above 5400 (5600)\mevcc as a  
proxy for the muon (electron) combinatorial background. Background events
are combined for the low- and \cqsq regions to increase the size of the training
samples. The full set of preselection and PID requirements are applied to the 
 data before training, for which 
 the same number of signal and background events are used. 
Separate classifiers are trained for the \runone, \runtwopo, and \runtwopt data-taking periods.
Ten different classifiers are trained for each period, using a \textit{k-fold} cross-validation 
approach to avoid biases \cite{Blum1999Jul} in which each of the ten classifiers is trained leaving out a different 
10\% of the data sample. 
The list of classifier inputs is reduced by repeating the training,  
excluding inputs sequentially and retaining only those whose inclusion increases the area under the ROC (Receiver Operating Characteristic) curve by  at least 1\%.
The same inputs are used for all three run periods.

The response of the multivariate classifier is verified to have no significant correlation with the \B candidate mass.
The final set of inputs is based on the following features of the candidates:
\begin{description}[rightmargin=10mm]
    \item[\boldmath{\B}] transverse momentum, vertex fit quality, displacement from the PV, compatibility of momentum and displacement vectors;
    \item[\boldmath{\ellell}] transverse momentum, vertex fit quality, displacement from the PV;
    \item[\boldmath{\Kp,\,\Kstarz}] transverse momentum, displacement from the PV;
    \item[Leptons] minimum and maximum transverse momentum and displacement of the two leptons from the PV;
\item[\boldmath{\decay{\Kstarz}{\Kp\pim}} final state hadrons]
    minimum and maximum transverse momentum and displacement of the two hadrons.
\end{description}

Partially reconstructed backgrounds are particularly important for the electron final states as 
 bremsstrahlung leads to missing energy even in the case of correctly reconstructed candidates, 
introducing a significant overlap of signal and backgrounds.
A dedicated classifier is therefore trained for the electron modes to distinguish between 
$\decay{B^{(+,0)}}{K^{(+,*0)}\epem}$ 
signal decays
and partially reconstructed backgrounds. In this case, a phase space simulation of \BuToKPiPiee is used as proxy for partially reconstructed background in \BdToKstee decays, while simulated \BdToKstee decays serve as a background proxy for \BuToKee decays. The training follows the same procedure as used for the combinatorial classifier.
In addition to observables  that describe the kinematic and geometric properties of the decays, isolation variables, such as the track multiplicity and the vertex quality obtained adding extra tracks from the underlying event to the reconstructed vertex, are evaluated. Only tracks from the underlying event contained within a cone defined by $\sqrt{(\eta-\eta_{B})^2 + (\phi-\phi_{B})^2)} < 0.5$ are considered, where $\eta_{B}$ and  $\phi_{B}$  are the  pseudorapidity and the azimuthal angle (given in radians) relative to the beam direction of the reconstructed \B candidate, respectively. Such variables contribute significantly to the rejection of candidates originating from partially reconstructed decay processes.
These isolation variables
consider the multiplicity of particles other than the \B candidate within this cone, the
scalar sum of their transverse momenta and the fraction of transverse momentum within the cone
attributed to the \B candidate. A further set of isolation variables is computed by
sequentially adding other tracks in the event to the \B candidate vertex 
and computing the mass of this new candidate vertex. The obtained vertex \chisq is used to define which new candidate vertex is most similar to that of the original \B candidate vertex. The \chisq and
invariant mass of this vertex are retained for use in the classifier to reject partially reconstructed backgrounds.

The classifiers developed to reduce combinatorial and partially reconstructed backgrounds are optimized using the expected signal significance $\nsig/\sqrt{\nsig + \nbkg}$ as a figure of merit, where \nsig and \nbkg represent the expected numbers of signal and background events within signal intervals defined as $\pm 50\mevcc$ around the known \B meson mass~\cite{PDG2022} for muon modes and, to account for bremsstrahlung, 5150--5350\mevcc for electron modes. Here \nsig is obtained from simulated samples of resonant \jpsi decays, normalized to the measured yields in data with no selections applied on the classifiers. It is scaled by the SM expectation for the ratios of nonresonant and resonant branching fractions, computed using \texttt{flavio} package~\cite{Straub:2018kue}, as well as the ratio of efficiencies between the nonresonant and resonant modes at the respective working points of the classifiers. 
The expected number of combinatorial background events in the signal window, \nbkg, is obtained from simplified fits to samples of data candidates passing
the preselection and PID  requirements.

A one-dimensional optimization of
the combinatorial classifier response is performed for muon signals, while a two-dimensional
optimization of the combinatorial and partially reconstructed classifier response is performed
for the electron signals.
The classifiers for the low- and \cqsq regions in each run period are optimized separately. 
It is verified that the classifiers do not sculpt the reconstructed \B meson mass lineshape and \qsq spectrum, 
and the optimal working points are located on broad plateaus of signal significance 
in all cases. Analogous optimizations are performed for the \jpsi and \psitwos resonant control
modes, with appropriate adjustments to take into account their different backgrounds. 
A single set of combinatorial and partially reconstructed classifier response criteria is chosen
for all electron signals and resonant muon modes, while muon signals and resonant electron modes
are selected using a different set of classifier response criteria for each run period.

\begin{figure}[t]
\centering
\includegraphics[width=1.0\linewidth]{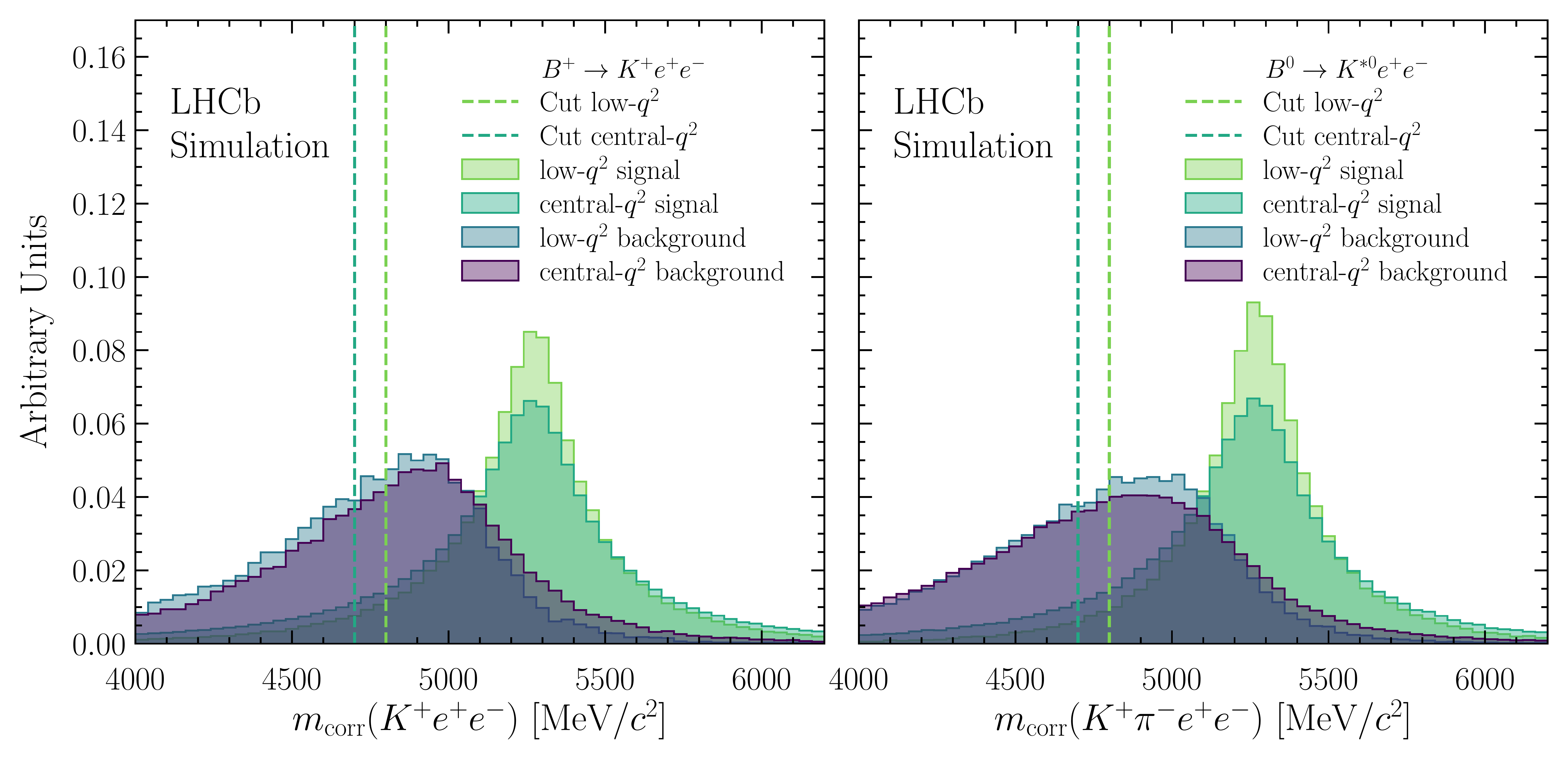}
\caption{Distribution of \mHOP for the simulated (left) \BuToKee and (right) \BdToKstee candidates after applying the nominal analysis criteria to the response of the combinatorial and partially reconstructed multivariate classifiers. 
The distributions of the signal in the low- and \cqsq regions and the partially reconstructed background are shown (unit normalizations).
The \Bu and \Bd partially reconstructed backgrounds are taken from simulated \BdToKstee and  \BuToKPiPiee decays, respectively. The vertical lines show the selection requirements applied for the two signal regions.}
\label{fig:HOPDistribution}
\end{figure}

Partially reconstructed backgrounds in electron modes are further suppressed by 
using the ratio of the hadronic and dielectron momentum components transverse to the \B direction of flight to correct the momentum of the dielectron pair \cite{LHCb-PAPER-2017-013}.
In the approximation that the dielectron direction is not modified significantly, this ratio is expected to be unity unless electrons have lost energy due to bremsstrahlung that is not recovered.
 The invariant mass calculated using the corrected dielectron momentum, \mHOP, has significant power to distinguish between signals and backgrounds that satisfy the nominal combinatorial and partially reconstructed classifier criteria, as illustrated in Fig.~\ref{fig:HOPDistribution}. 
 The \mHOP criteria are optimized in a similar manner to the multivariate classifiers and are applied after them  to reduce further combinatorial and partially reconstructed backgrounds. 
 Since the \mHOP criteria sculpt the combinatorial background, potential biases introduced by them are considered as a source of systematic uncertainty.

\subsection{Exclusive backgrounds}
Dedicated simulated event samples are used to study backgrounds which remain after all previously
described selection criteria have been applied. Specific vetoes are used to reduce many of these
backgrounds to a negligible level. To ensure high efficiency for the signal, stronger PID  requirements are imposed in the mass interval close to a resonance rather than applying a veto on invariant mass only.
It is necessary to evaluate these backgrounds and vetoes separately 
for the \BuToKll and \decay{\Bz}{\Kstarz \ellell} decay modes.

\begin{figure}[t]
\centering
\begin{minipage}{0.49\linewidth}
\centering
\includegraphics[width=\linewidth]{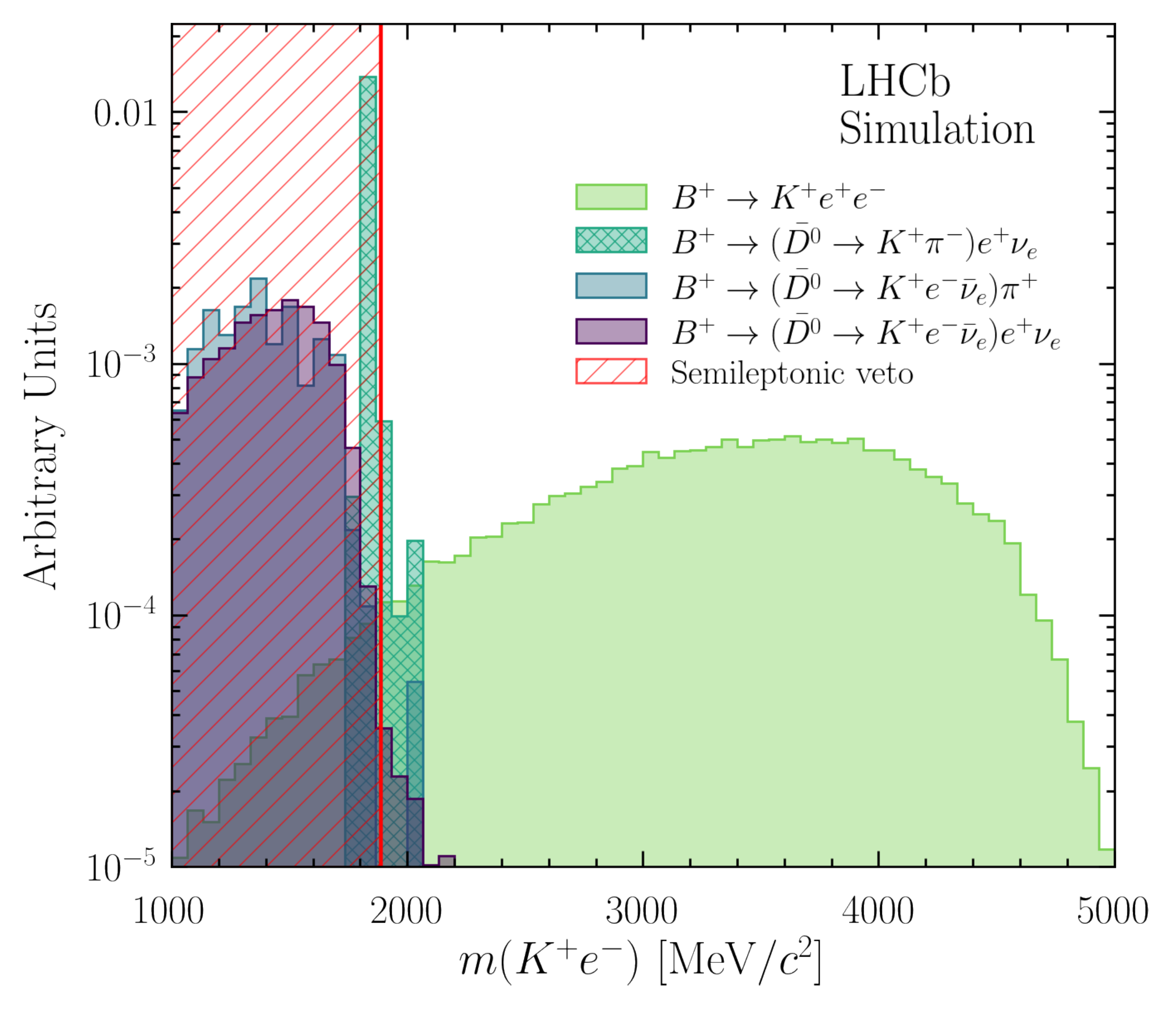}
\end{minipage}
\begin{minipage}{0.49\linewidth}
\centering
\includegraphics[width=\linewidth]{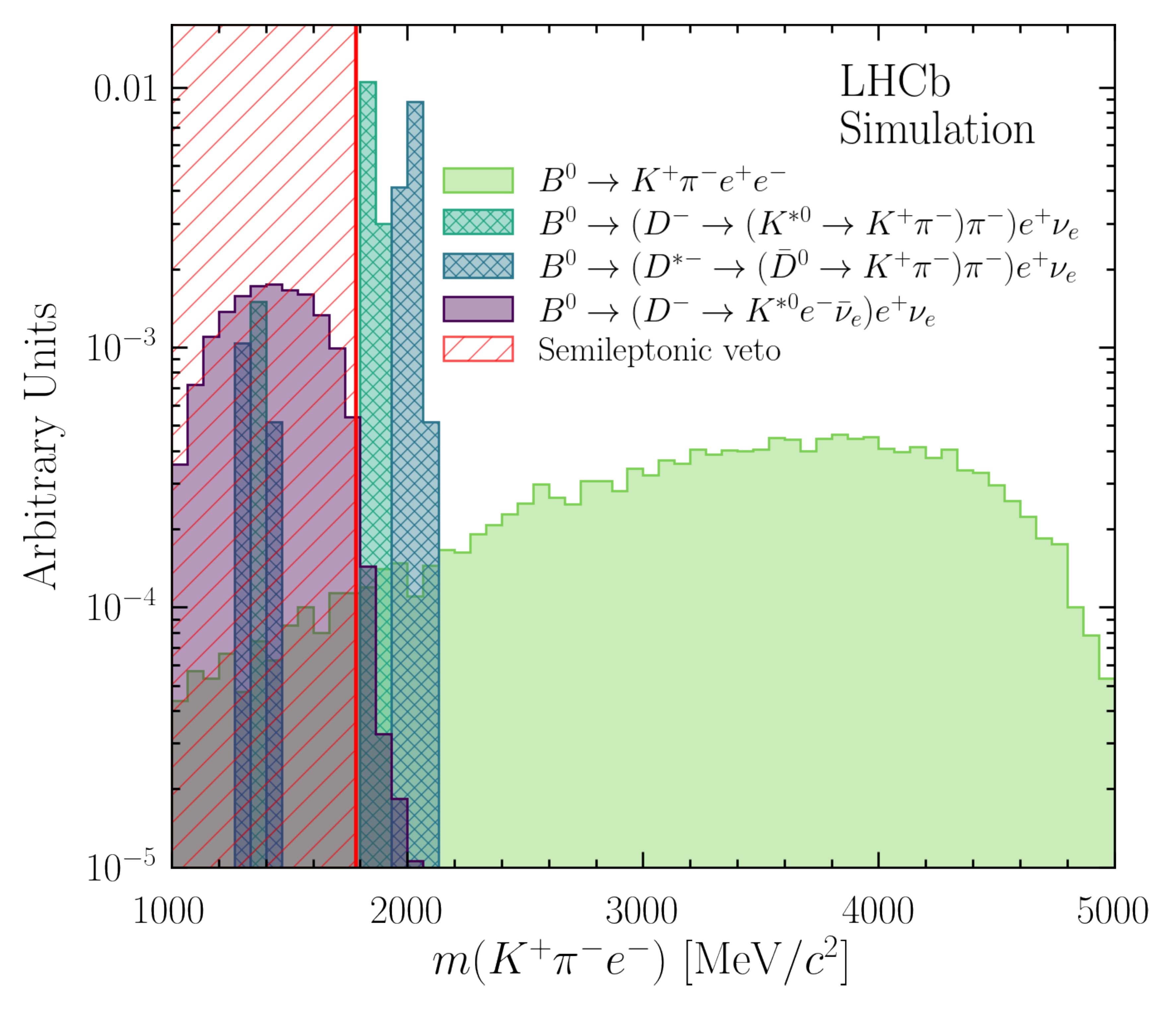}
\end{minipage}
\caption{Simulated distributions of (left) $m(\Kp\en)$  for \Bp candidates and (right)  $m(\Kp\pim\en)$ for \Bd candidates. Signal and various semileptonic cascade backgrounds are shown. The full selection is applied except for the semileptonic background vetos. The hatched areas show decay modes that are also vetoed, recomputing $m(\Kp\en)$ and $m(\Kp\pim\en)$ while assigning the pion mass hypothesis to the electron and not accounting for bremsstrahlung corrections.}
\label{fig:SemiLepBAckgrounds}
\end{figure}

For \BuToKll decays, the residual backgrounds accounted for in the fits are summarized in Table~\ref{tab:BkgforfitsRK} and additional selection criteria are applied to suppress background contributions from: 
\begin{description}[rightmargin=10mm]
    \item[\boldmath{\decay{\Bu} {(\decay{\Dzb} {\Kp\pim}) \ellp\neul}}:]
    This decay has one pion misidentified as a charged lepton.  If the invariant mass of the kaon and oppositely charged lepton, computed assigning the pion mass hypothesis to the lepton, differs from the known \Dz mass~\cite{PDG2022} by less than 40\mevcc, the charged lepton must satisfy tighter PID requirements.
    This background affects all \qsq regions.
    
    \item[\boldmath{\decay{\Bu} {(\decay{\Dzb} {\Kp\ellm\neulb})} \ellp\neul}:]
    This decay has two additional neutrinos compared to the signal mode resulting in significant missing energy. To suppress this background, the invariant mass of the kaon and the lepton with opposite charge to the kaon is required to be greater than 1780\mevcc as illustrated in Fig.~\ref{fig:SemiLepBAckgrounds}.
    This background affects the low- and \cqsq regions.
    
    \item[Hadron-lepton swap:] This background involves a double misidentification which may cause a resonant mode candidate to be misidentified as signal since the overall invariant mass of the \Kpll system still peaks in the vicinity of the \Bp meson mass while the reconstructed dilepton mass is mistakenly different from the charmonium mass. In the muon mode, 
    where the invariant mass of the system formed by the kaon (under the muon mass hypothesis) and the oppositely charged muon differ by less than 60\mevcc from the known masses of the \jpsi and \psitwos mesons,
    the muon is required to satisfy stringent PID criteria.
    In the electron mode,
    the \Kpll invariant mass is recomputed swapping the kaon and same-charge electron mass hypotheses and constraining the invariant mass
    of the dilepton system to the \jpsi or \psitwos masses. Where this  \Kpll mass differs by less than 60\mevcc from the known \Bp mass, the electron is required to satisfy stringent electron identification criteria.
    This background affects all \qsq regions.
    
    \item[\boldmath{\decay{\Bp} {\psitwos(\decay{} {\jpsi X) \Kp}}}:] The invariant mass of the reconstructed 
    \Bp candidate is required to be at least 200 \mevcc greater than the \Bp meson mass when the dilepton mass is constrained to the known \psitwos meson mass.
    This background affects the \jpsi region.
\end{description}

\begin{table}[t]
	\centering
	\caption{Exclusive backgrounds modeled in the \BuToKll invariant mass fits, along with the \qsq region of interest and the mode(s) for which the background is relevant.}
	\label{tab:BkgforfitsRK}
	\renewcommand\arraystretch{1.3}
	\begin{tabular}{l|c|c}
		\textbf{Decay mode} & \textbf{\qsq region} & \textbf{Relevant mode(s)} \\
		\hline
		\BuToPiJPsll  & \jpsi & electron and muon\\
		\BsToKstJPsll & \jpsi & electron and muon\\
		\BdToKstPsill & \psitwos & electron and muon\\
		$B^{+,0}\to (K\pi)^{+,0} \ell\ell$    & low/central & electron\\
	\end{tabular}
\end{table}

For \BdToKstll decays, the residual backgrounds accounted for in the fits are summarized in Table~\ref{tab:BkgforfitsRkst} and additional selection criteria are applied to suppress background contributions from:
\begin{description}[rightmargin=10mm]
    \item[\boldmath{\decay{\Bs}{\phi(1020)\ellell}}:]
    This decay has one kaon misidentified as a pion. Where the \Kp\pim invariant mass, recomputed under the \Kp\Km mass hypothesis, is less than 1040\mevcc, the pion is required to satisfy stringent
    PID  criteria.
    This background affects all \qsq regions and can only be fully vetoed
    in the low- and central-\qsq regions.
    In the resonant modes a non-negligible amount of this background remains after the veto and is modeled in the fits.
    
    \item[\boldmath{\decay{\Bd}{(\decay{\Dzb}{\Kp\pim}) \pim\ellp\neul}}:]
    This decay has one pion misidentified as a charged lepton and one neutrino compared to the signal mode.
    Where the invariant mass of the kaon and oppositely charged
    lepton, computed by assigning the pion mass hypothesis to the lepton, differs by less than 30\mevcc from the known \Dz meson mass, the lepton is required to satisfy stringent
    PID criteria.
    This background affects all \qsq regions;
   
    \item[{\boldmath\decay{\Bd} {(\decay{\Dm}{(\decay{\Kstarz}{\Kp\pim})\pim}) \ellp\neul}}:]
    If the invariant mass of the \Kp\pim system and the
    lepton with opposite charge to the kaon (computed under the pion mass hypothesis) 
    differs by less than 30\mevcc from the known \Dm meson mass, the lepton is required to satisfy stringent PID  criteria.
    This background affects all \qsq regions.
    
    \item[\boldmath{\decay{\Bd} {(\decay{\Dm}{(\decay{\Kstarz}{\Kp\pim})\ellm\neulb}) \ellp\neul} }:]
    This decay differs from the signal mode by having two additional neutrinos in the final state.
    The invariant mass of the \Kp\pim system and the
    lepton with opposite charge with respect to the kaon is required to be greater than 1780\mevcc as illustrated in Fig.~\ref{fig:SemiLepBAckgrounds}.
    This background affects the low- and \cqsq regions.
    
    \item[\boldmath\decay{\Bp}{\Kp \ellell}:]
    This decay, with the addition of a random pion from the underlying event can constitute a background for the \BdToKstll candidates.
    This background is suppressed applying an invariant mass requirement to the $\pim\ellell$ system, assigning the kaon mass hypothesis to the pion, and to the invariant mass of the 
    $\Kp\ellell$ system. Both the invariant masses for a given \BdToKstll candidate are required to be smaller than 5100\mevcc. 
    This background affects all \qsq regions.
    
    \item[\textbf{Hadron-lepton swap}:]
    This background has the same physical origin as, and is treated
    analogously to, its counterpart in the \Bp decay.
    
    \item [\boldmath{\decay{\Bz}{\psitwos(\decay{}{\jpsi X})\Kstarz}}:]
    This background also has the same physical origin as, and is treated analogously to, its counterpart in the \Bp decay.
\end{description}

\begin{table}[t]
  \centering
  \caption{Exclusive backgrounds modeled in the \decay{\Bz}{\Kstarz\ellell} invariant mass fits, the \qsq region of interest and the mode(s) for  which the background is relevant.
  The $K-\pi$ swap backgrounds refer to cases where the mass hypotheses of the kaon and pion from a genuine \decay{\Bz}{\Kstarz\ellell} decay are swapped.}
	\label{tab:BkgforfitsRkst}
	\renewcommand\arraystretch{1.3}
	\begin{tabular}{l|c|c}
		\textbf{Decay mode}  &  \textbf{\qsq region} & \textbf{Relevant mode(s)} \\
		\hline
                \BsToKstJPsll & \jpsi & electron and muon\\
                \BsToPhiJPsll & \jpsi & electron and muon\\
                \LbTopKJPsll  & \jpsi & electron and muon\\
                \BToXJPsll    & \jpsi & electron and muon\\
                $K-\pi$ swap  & \jpsi & electron and muon\\
                \BsToKstPsill & \psitwos & electron and muon\\
                \LbTopKPsill  & \psitwos & electron and muon\\
                \BToXPsill    & \psitwos & electron and muon\\
                $K-\pi$ swap  & \psitwos & electron and muon\\
                \decay{B^{+,0}}{(K\pi\pi)^{+,0} \ellell}  & low/central & electron
	\end{tabular}
\end{table}

\begin{figure}[t]
\centering
\includegraphics[width=\linewidth]{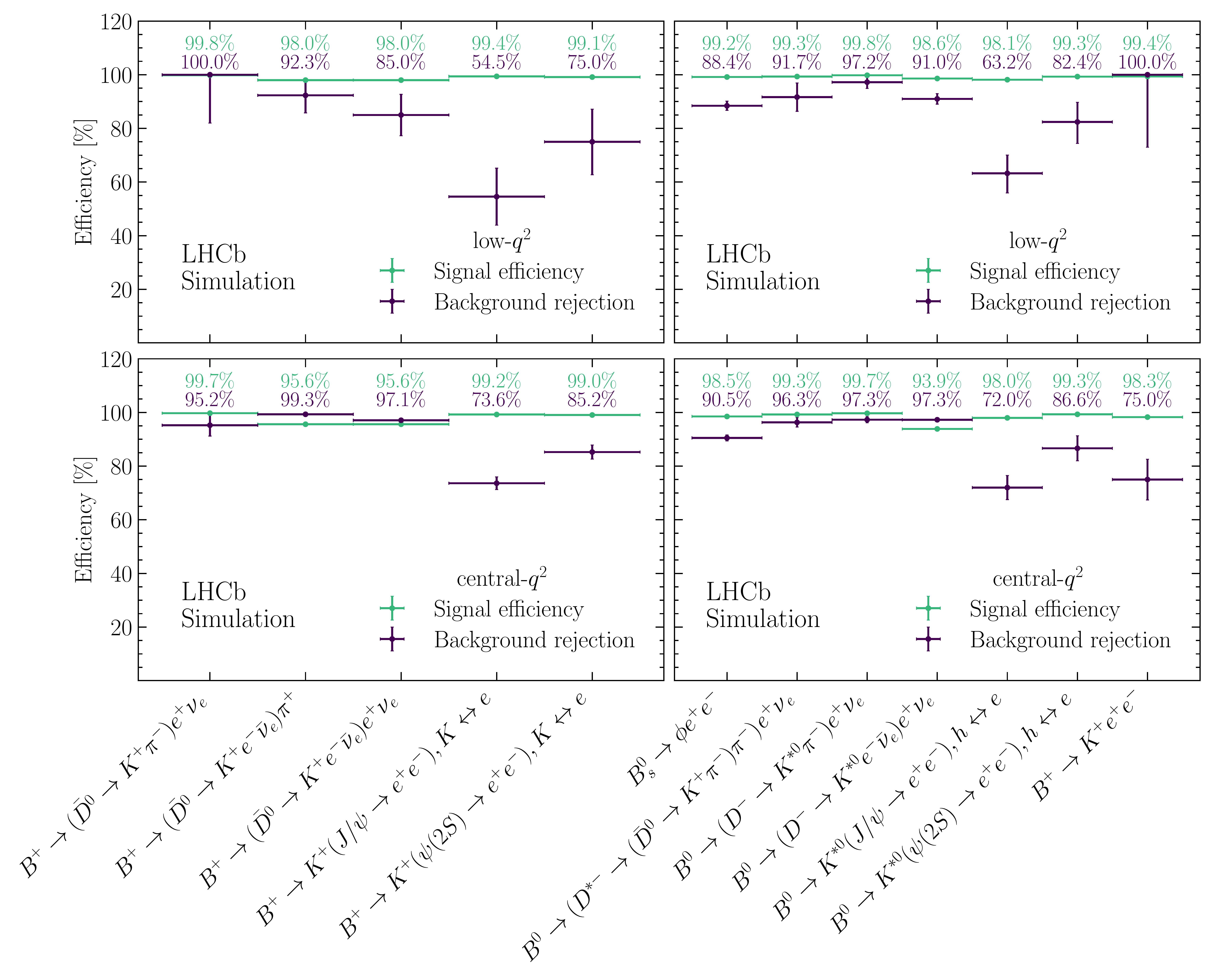}
\caption{Upper: signal efficiencies and background rejection factors for all vetos against physical backgrounds, for (left) \Bp and (right) \Bd modes, in the \lqsq region; lower: analogous plots for the \cqsq region.}
\label{fig:VetoEfficiencies}
\end{figure}

The residual contamination of exclusive backgrounds in the low- and \cqsq signal regions 
is evaluated using large samples of simulated background events (Fig.~\ref{fig:VetoEfficiencies}). Backgrounds that would form a peaking
structure in the \B invariant mass, such as \decay{\Bs}{\phi(1020) \ellell} or
\decay{\Lb}{\proton\Km \ellell}, are found to have yields at a few per mille of the
expected signal yield, and are therefore considered negligible. Due to their large branching fractions, double-semileptonic decays of the form 
\decay{\Bd}{(\decay{\Dm}{\Kstarz\en\neueb})\ep\neue} are found
to have yields of a few percent of the expected signal yield. Since the selection efficiency for these decays is very small, modeling them with dedicated 
templates in the invariant-mass fit would require prohibitively large simulated event samples to be generated. As these decays involve two neutrinos and 
significant missing energy they do not form a peaking structure near the invariant mass signal region. They are therefore not modeled explicitly but rather 
absorbed by other, larger, missing energy background components in the invariant-mass fit.

\subsection{Misidentified backgrounds}
\label{sec:misidentifiedbkgds}
After applying all selection criteria, a significant contribution from backgrounds in which one or more hadrons are misidentified as leptons, with or without additional missing energy, still remains. 
These backgrounds
have various impacts on the invariant mass fit. Fully reconstructed misidentified decays 
of the type \decay{\Bp}{\Kp h_1 h_2} and \decay{\Bd}{\Kstarz h_1 h_2}, where $h_{1,2}$ are kaons or pions,
create clear peaking structures in both the electron and muon invariant-mass fits. There are however
also numerous backgrounds specific to the electron final states which feature a combination of either 
single or double misidentification, as well as missing energy. These backgrounds create more complex structures. 

One specific example is the decay \decay{\Bd}{\Kp \pim ( \decay{\piz}{\ep\en\gamma} )}, where the 
electron from the $\piz$ decay is missed, the photon is missed or reconstructed as bremsstrahlung, 
and the negatively charged pion is misidentified as an electron. This example is similar to the backgrounds 
discussed in Ref.~\cite{Robinson:2021cws}, with a misidentified hadron substituted 
for one of the electrons. More generally, however, any decay of the type
\decay{\Bp}{\Kp \pim (\piz,\gamma) X} or \decay{\Bd}{\Kstarz \pim (\piz,\gamma) X}, 
where $X$ is any number of other final state particles, can contribute. Not all particles from such 
processes are used to reconstruct the signal, therefore such backgrounds are characterized by low invariant masses. 

Compared to previous LU measurements at \lhcb, the tighter PID requirements used for electrons reduces the expected rates for pions and kaons to be misidentified as electrons. 
Table~\ref{tab:misidrates} compares the misidentification rates at the working point used in this analysis to those from Ref.~\cite{LHCb-PAPER-2021-004}, for each of the three data taking periods considered. 
The misidentification rates are determined from data using $\decay{\Dstarp}{\Dz(\to \Km\pip)\pip}$ decays. 
It is noted that in \runone a similar  pion-to-electron misidentification rate is found from this data-driven method, while a factor two suppression is achieved for kaon-to-electron misidentification. 
For \runtwopo and \runtwopt, the pion-to-electron misidentification rates are reduced by a factor two and the kaon-to-electron misidentification rates are reduced by almost a factor of ten; \runtwopo and \runtwopt rates are found to be consistent with one another. 
Table~\ref{tab:signaleffcompare} shows the impact of the tighter PID  requirements on the overall electron mode signal efficiencies, separated by data-taking period and trigger category. 
The improved background reduction has only a small impact on the signal efficiencies: in \runone these are unchanged, while in \runtwo, they are reduced by around 10\%.

\begin{table}[th]
\caption{Single-particle misidentification rates obtained on data averaging over the kinematics of prompt $D^{\ast+}\rightarrow D^{0}(\rightarrow K^{-}\pi^{+})\pi^{+}$ decays. The misidentification rates are evaluated for the PID  criteria used in this analysis given the acceptance and kinematic requirements applied in the track final state. The misidentification rates corresponding to the PID requirements of Ref.~\cite{LHCb-PAPER-2021-004} are given in parentheses.}\label{tab:misidrates}
\centering 
\begin{tabular}{lcc}
\hline
 Sample            & $\pi \rightarrow e$   & $K \rightarrow e$    \\
\hline
 \runonetable   & $1.78 \,( 1.70)\,\%$  & $0.69 \,( 1.24)\,\%$ \\
 \runtwopo & $0.83 \,( 1.51)\,\%$  & $0.18 \,( 1.25)\,\%$ \\
 \runtwopt & $0.80 \,( 1.50)\,\%$  & $0.16 \,( 1.23)\,\%$ \\
 \hline
\end{tabular}
\end{table}

\begin{table}[th]
\caption{Overall signal efficiency for electron mode in percent. The impact of global event cuts on the efficiency determination is not included as it cancels in the \NRKorKst ratio defined in Eq.~\eqref{eq:doubleratiorx}.  The efficiency values obtained applying the same PID requirements of Ref.~\cite{LHCb-PAPER-2021-004} are given in parentheses.} \label{tab:signaleffcompare}
\centering 
\resizebox{\linewidth}{!}{
\begin{tabular}{l|cc|cc}
\hline
\multirow{2}{*}{Sample}              & \multicolumn{2}{c|}{\BuToKee} & \multicolumn{2}{c}{\BdToKstee} \\
                                       & \lqsq                       & \cqsq                      & \lqsq                         & \cqsq                         \\ \hline
 \runonetable \tistable   & $0.152$ ($0.152$)\%                              & $0.138$ ($0.140$)\%                                 & $0.054$ ($0.054$)\%                                   & $0.051$ ($0.051$)\%                                      \\
 \runonetable \tostable   & $0.126$ ($0.127$)\%                              & $0.127$ ($0.127$)\%                                  & $0.044$ ($0.044$)\%                                   & $0.044$ ($0.044$)\%                                       \\
 \runtwopo \tistable & $0.250$ ($0.273$)\%                              & $0.230$ ($0.252$)\%                                 & $0.084$ ($0.092$)\%                                   & $0.087$ ($0.095$)\%                                      \\
 \runtwopo \tostable & $0.239$ ($0.258$)\%                             & $0.228$ ($0.247$)\%                                  & $0.074$ ($0.080$)\%                                   & $0.081$ ($0.087$)\%                                       \\
 \runtwopt \tistable & $0.256$ ($0.285$)\%                             & $0.232$ ($0.260$)\%                                  & $0.086$ ($0.094$)\%                                  & $0.084$ ($0.095$)\%                                       \\
 \runtwopt \tostable & $0.228$ ($0.253$)\%                              & $0.226$ ($0.249$)\%                                  & $0.079$ ($0.086$)\%                                   & $0.078$ ($0.087$)\%                                       \\

\hline
\end{tabular}
}
\end{table}

It is essential to establish whether a significant number of misidentified background candidates pass
the full selection criteria, and whether they create distinctive invariant mass distributions that
cannot be absorbed by combinatorial or other background components. This task is complicated by the fact
that there is a very large number of such backgrounds, many of which are poorly known. Even where the
branching fractions of individual \decay{\Bp}{\Kp \pim (\piz,\gamma) X} or \decay{\Bd}{\Kstarz \pim (\piz,\gamma) X}
decays have been measured, their Dalitz structure is often unknown. 
A representative subset of these backgrounds is studied using simulation, and the expected contribution of each 
individual background found to be negligible. However, even if the contribution
of any given background is small, the contribution of all these backgrounds taken together can be large and have a shape that differs from combinatorial background. 
These considerations lead to a data-driven 
strategy for modeling the distributions of the residual misidentified backgrounds in this analysis, using control samples enriched with misidentified hadrons.
This strategy consists of inverting the stringent lepton identification requirements in the selection, 
while maintaining the preselection requirements. The resulting dataset (referred to as control region in the following) predominantly contains misidentified background  rather than signal candidates and can be used, together with standard PID  calibration samples, to estimate the residual misidentified backgrounds.

\begin{figure}[htb]
\centering
\includegraphics[width=0.98\linewidth]{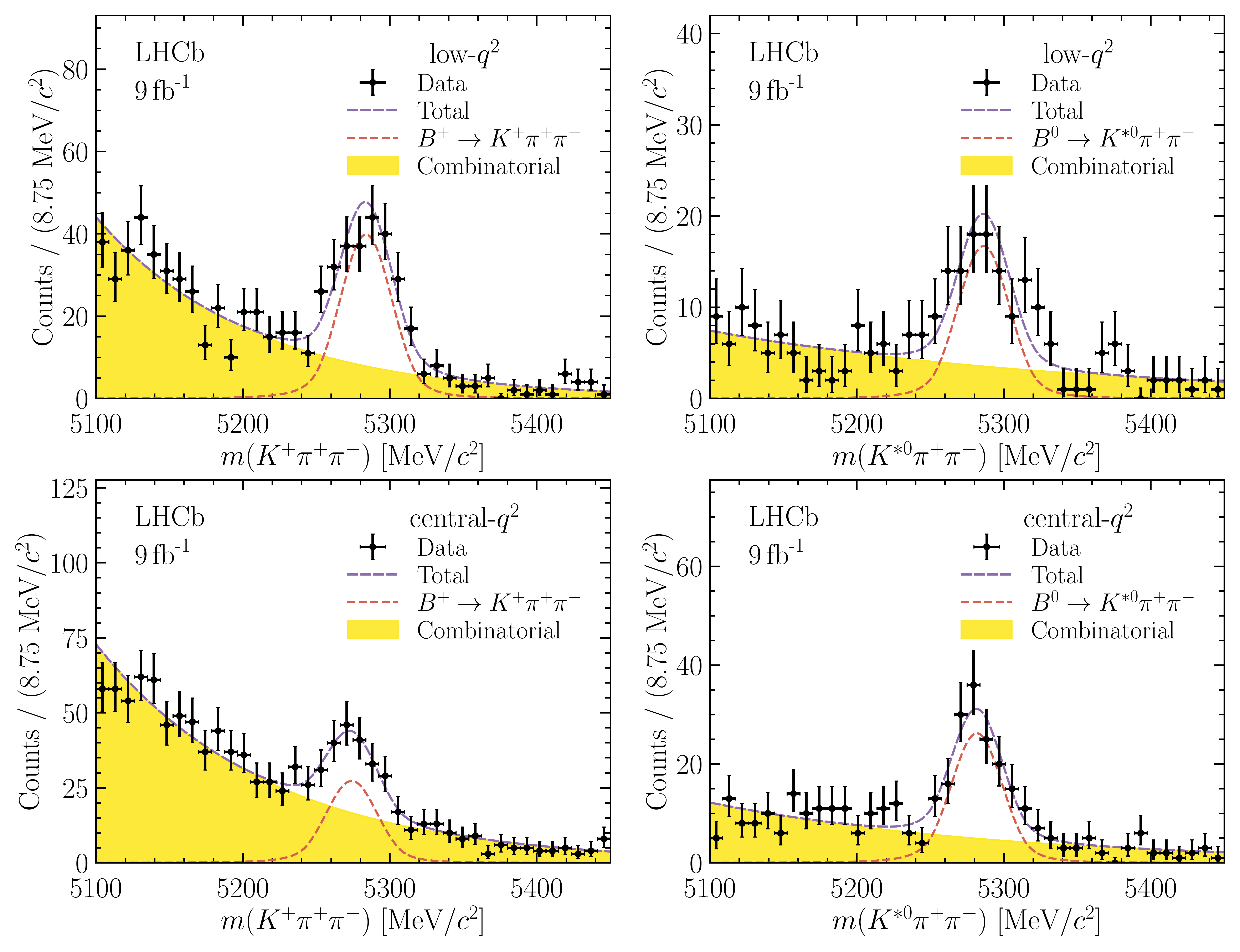}
\caption{\small Distributions of the invariant mass of candidates for which both electrons are in the control region, \ie having the stringent electron identification requirements inverted.
The pion mass hypothesis is applied to both electrons, without a bremsstrahlung correction.
The left and right columns correspond to the  \BuToKee and \BdToKstee modes, respectively.  
The upper and lower rows correspond to the low- and \cqsq regions, respectively.
Fit results are overlaid.}
\label{fig:fit_charmless}
\end{figure}

\begin{figure}[htb]
\centering
\includegraphics[width=0.98\linewidth]{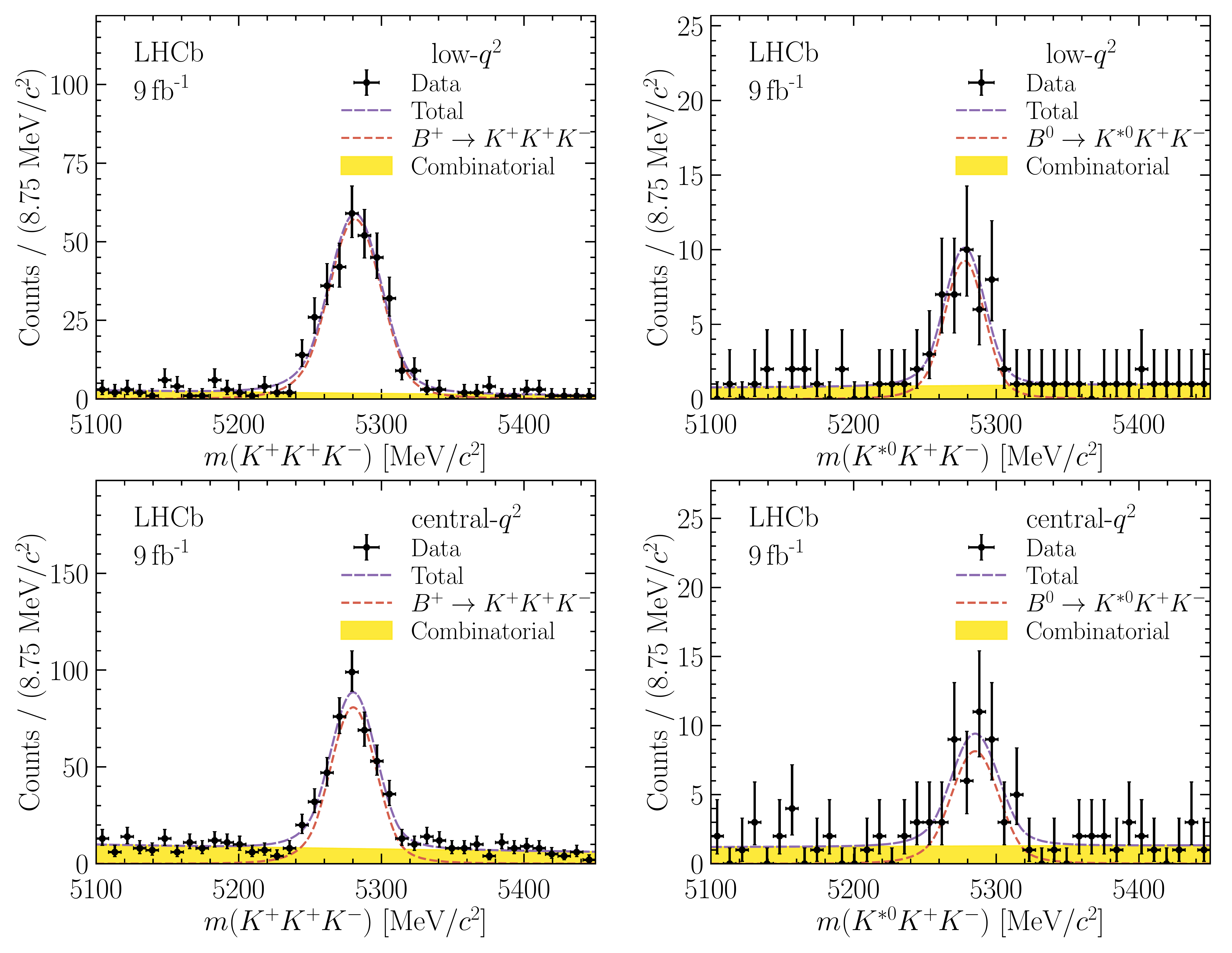}
\caption{Distributions of the invariant mass of candidates for which both electrons are in the control region, \ie having the stringent electron identification requirements inverted.
The kaon mass hypothesis is applied to both electrons, without a bremsstrahlung correction. Fit results are overlaid. (left) \BuToKee modes, (right) \BdToKstee modes. 
The upper and lower rows correspond to the low-	and \cqsq regions, respectively.}
\label{fig:fit_charmless_kk}
\end{figure}

The most straightforward backgrounds to address are the fully reconstructed misidentified decays: they are limited in number,
relatively well understood experimentally, and can be reconstructed under their own mass hypothesis leading to clear
signals in the invariant mass distribution. The background yield is estimated in this dataset by fitting to the invariant mass of $\Kp\ep\en$ ($\Kstarz \ep\en$) candidates 
where electrons are assigned the pion or kaon mass hypothesis and the bremsstrahlung correction is ignored. The fit results are shown in Fig.~\ref{fig:fit_charmless} and in Fig.~\ref{fig:fit_charmless_kk}
for the $B \to K^{(*)} \pip \pim$ and $B \to K^{(*)} \Kp \Km$ backgrounds, respectively.
The $B \to K^{(*)} \pip \pim$ peaks are parametrised by a double-sided Crystal Ball function, and non-peaking background components are modeled by an exponential function. 
Calibration samples are  used to extrapolate the misidentification rate from the amount measured in this control region, with the full analysis selection criteria applied.
The rate for misidentifying two hadrons as electrons  in the nominal dataset is found to be about 2\% of that in the control dataset. This procedure is repeated for each trigger category and data-taking
period, separately for low- and \cqsq regions. 
For the dielectron final states, it is found to be non-zero. The residual contribution is found to be higher in the \lqsq region, 
and this difference is due to contributions from low mass hadronic resonances. 
It is found that this expected contamination is compatible with zero for the dimuon final states. 

\begin{figure}[htb]
\centering
\includegraphics[width=1.0\linewidth]{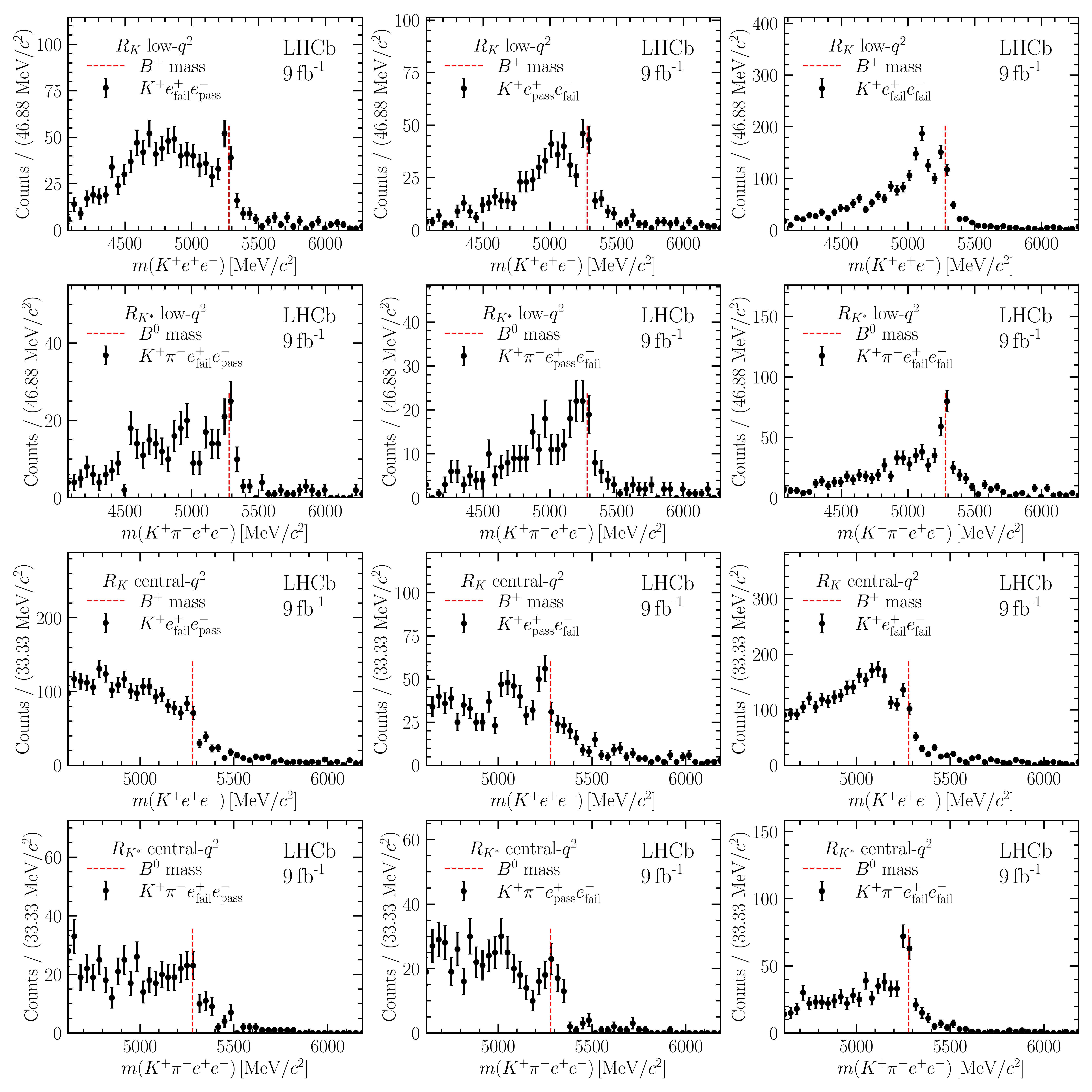}
\caption{\small Invariant mass distribution of candidates in the inverted lepton identification control region. From top to bottom: 
\RK \lqsq, \RK \cqsq, \RKst \lqsq, \RKst \cqsq.
(Left) candidates for which the lepton is in the control region and has the same charge as the kaon, 
(middle) candidates for which the lepton is in the control region and has a charge which is the opposite of that of the kaon, 
(right) candidates for which both leptons are in the control region.
}
\label{fig:misid_fail_region}
\end{figure}

In contrast to fully reconstructed backgrounds, backgrounds of the type
\mbox{\decay{\Bp}{\Kp \pim (\piz,\gamma) X}} or \mbox{\decay{\Bd}{\Kstarz \pim (\piz,\gamma) X}} do not have distinctive invariant-mass distributions, even with inverted PID criteria. 

\begin{figure}[tb]
\centering
\includegraphics[width=0.9\linewidth]{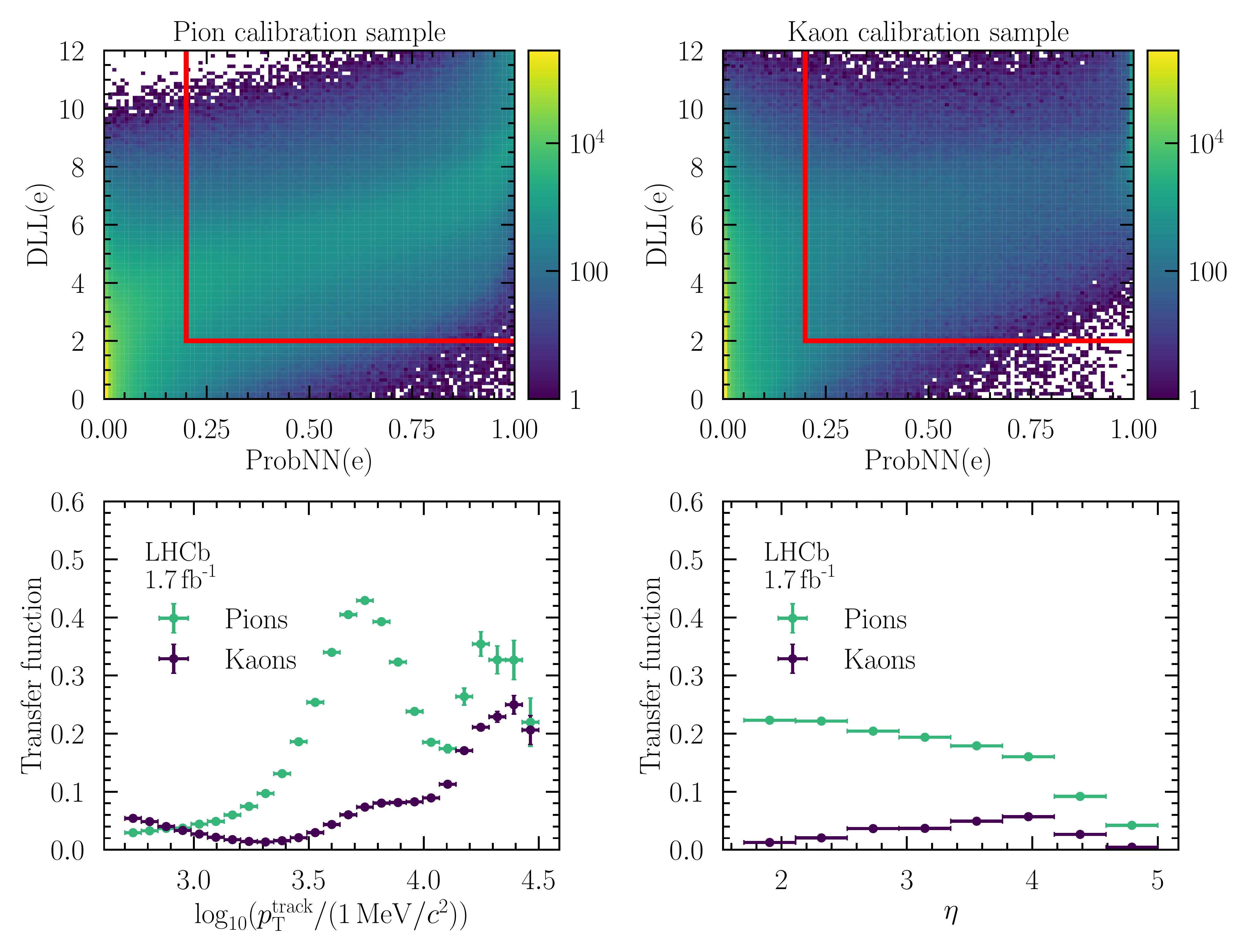}
\caption{Upper: distributions of the PID variables in the (left) pion and (right) kaon calibration samples using 2017 data. 
 The red lines separate control regions (left and below the line) from fit regions (right and above the line).
 Lower: the fraction of control region events that is expected to appear in the fit region (transfer function) as functions of track \pt and $\eta$.} 
\label{fig:transfer_maps_fail_region_2017}
\end{figure}

Figure~\ref{fig:misid_fail_region} shows the invariant mass shape in the 
control region with inverted lepton identification criteria, as defined above.  
This control region contains a combination of: fully reconstructed misidentified backgrounds, singly misidentified and/or 
partially reconstructed backgrounds, combinatorial backgrounds, and genuine signal which passes the 
preselection but fails the analysis selection criteria. Calibration samples are divided into intervals of transverse momentum and pseudorapidity
and used to extrapolate the yields and invariant mass shape of these components, given the full analysis selection criteria from the 
events in this control region. Events in the control region can contain both misidentified pions and kaons.
The probability to misidentify a kaon as an electron is significantly different from the probability to
misidentify a pion as an electron. Consequently the same multivariate criterion used to separate kaons and pions is used
to arbitrate whether a given control region event should be treated as a pion or as a kaon when extrapolating it to the signal region.
Example calibration sample maps for 2017 data and the resulting ``transfer functions'' that allow the control region events to be extrapolated to the 
fit region with nominal lepton identification criteria 
are shown in Fig.~\ref{fig:transfer_maps_fail_region_2017}.

\begin{figure}[htb]
\centering
\includegraphics[width=0.95\linewidth]{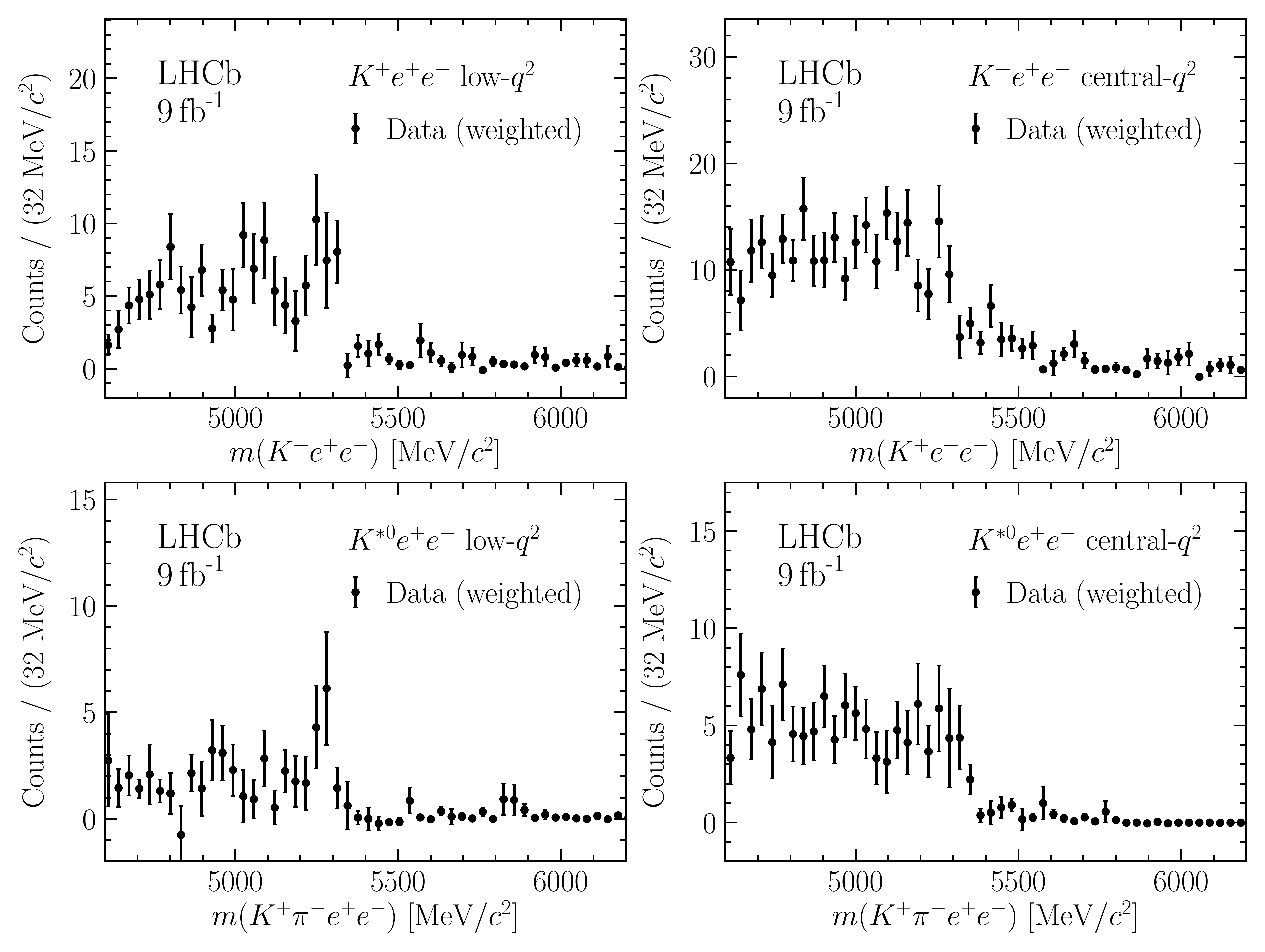}
\caption{\small Distributions of misidentified background events predicted for the $\Kp \epem$ and $\Kstarz \epem$ samples with full selection criteria applied.
}
\label{fig:misid_prediction_nominal}
\end{figure}

The residual signal that passes the preselection but fails the analysis selection criteria is subtracted 
based on PID efficiencies from calibration samples and on an initial signal yield estimate from a simplified invariant mass fit.
This procedure has a negligible effect on the final result. The final extrapolated misidentified backgrounds are shown in Fig.~\ref{fig:misid_prediction_nominal} for the two electron final states at \lqsq and \cqsq. Given that the extrapolation employs data calibration samples, the depicted shapes model the ensemble effect 
of \decay{\Bp}{\Kp \pim (\piz,\gamma) X} or \decay{\Bd}{\Kstarz \pim (\piz,\gamma) X} decays, without being susceptible to mismodeling of relative yields and kinematics.
The narrow excesses seen between 5200 and 5300\mevcc are attributed to the previously estimated, fully reconstructed misidentified backgrounds,
and are statistically compatible with those dedicated estimates. No clear structure is seen below 5200\mevcc, however the observed shape cannot be 
explained by combinatorial background events alone. 
Although the contribution from each individual \decay{\Bp}{\Kp \pim (\piz,\gamma) X} 
or \decay{\Bd}{\Kstarz \pim (\piz,\gamma) X} process is negligible, their total sum is not and needs to be accounted for in the invariant mass fit.

\clearpage

\section{Calibration of simulation and determination of efficiencies}
\label{sec:effs}
Simulated events must be calibrated to reproduce fully all aspects of the LHC production environment
and \lhcb detector performance. The calibration consists of a set of weights, the product of which is applied to the simulation to ensure both reliable 
modeling of the different components that enter the invariant mass fits, described in the
next section, and the accurate determination of detector efficiencies used to calculate
\RK and \RKst. For each data-taking year the simulation is calibrated using abundant, high-purity, 
control samples from data. As no single data control sample can calibrate all aspects of the simulated detector
performance, a multi-step sequential procedure is followed, each with its own weight, $w$, as summarized below.

\begin{enumerate}[rightmargin=10mm]
    \item The PID performance (\wpid) is calibrated as a function of track kinematics and detector
    occupancy using control samples of \mbox{\NBToKJPsll}, \DstDpi, and $\jpsi\to\mup\mun$ decays;
    \item The electron track reconstruction performance (\wtrk) is calibrated
    using control samples of \NBToKJPsee decays. As hadron and muon efficiencies are found to agree 
    well between data and simulation the calibration is only applied to electrons;
    \item The event multiplicity and \B meson kinematics (\wmco) are calibrated using control samples of \mbox{\NBToKJPsll} decays;
    \item The \lone trigger efficiency (\wlo) is calibrated using control samples of \mbox{\NBToKJPsll} decays;
    \item An analogous procedure is followed for the \hlt trigger efficiency (\whlt); 
    \item A final set of calibrations, \wmcreco, are computed using control samples of \mbox{\NBToKJPsll} decays in order to correct residual differences in the description of reconstructed \B meson properties in simulation.
    \end{enumerate}

The full chain of calibrations is applied when computing the \qsq selection efficiency to ensure reliable modeling of the migration of events between \qsq regions. 
With the exception of \wmco, which uses a dedicated prior calibration chain as input, 
each calibration step uses as input the output of the preceding step. 
Calibrations are calculated separately using $\Bz$ and $\Bp$ decays and are shown to be 
interchangeable. 
As the same sample of \NBToKJPsll decays is used to normalize 
decay rates in the (\RK,\RKst) double ratios and to compute the calibrations, 
the \Bz calibrations are applied to \RK and the \Bp calibrations are applied to \RKst to remove correlations arising from the statistical overlap between the normalization and calibration sample.

\subsection{Particle identification}
\label{sec:effs_pid}
The performance of hadron and muon PID is calculated using a weight \wpid computed as the efficiency with which the analysis
criteria correctly identify a given particle type. These efficiencies are evaluated using a three-dimensional binning in particle momentum, particle pseudorapidity and the track multiplicity of the event, with the latter acting as a proxy for the detector occupancy. 
Multiplicity bins are chosen such that they are uniformly populated;
 the momentum and pseudorapidity binning is optimized in each bin of multiplicity.
Bins are required to be sufficiently narrow to ensure that the efficiency is 
uniform within uncertainties across each bin,
while being sufficiently broad that the statistical uncertainties are approximately Gaussian.
Each dimension is therefore divided initially into equally populated bins; using an iterative procedure, adjacent bins are 
merged where their efficiencies differ by less than five standard deviations.  For a small number of bins at the corners of
the ($p$,$\eta$) phase space that remain empty, the nearest neighbor efficiency is used, while the efficiencies are rounded to 0 or 1 for those with unphysical values of efficiencies. An analogous procedure is followed to evaluate misidentification efficiencies for backgrounds. 

\begin{figure}[t]
    \centering
    \includegraphics[width=.7\textwidth]{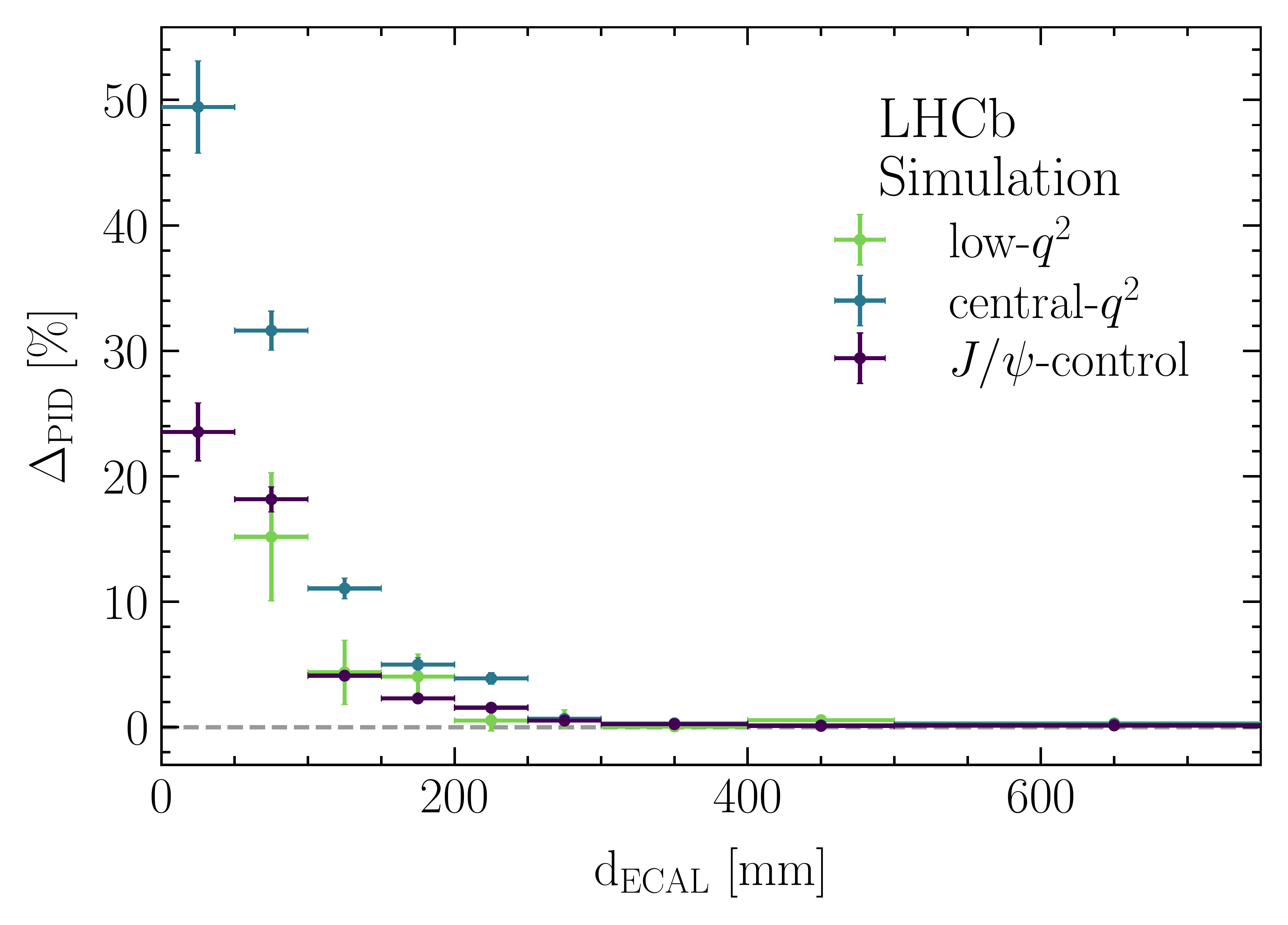}
    \caption{Factorization bias as a function of the separation between a pair of electrons in the electromagnetic calorimeter in the various $q^{2}$ regions determined using simulated $B^{+}\rightarrow K^{+}e^{+}e^{-}$  decays.}
    \label{fig:PIDFact_GenVCorr_18RKDECAL}
\end{figure}

The simulation is used to verify that the identification efficiency for a given hadron or muon is 
independent of PID  requirements applied to other hadrons and muons in the 
same event. This factorization ensures that the overall efficiency is the product of the individual \wpid.
This is not the case for electrons because their identification depends on the association of particle tracks with electromagnetic calorimeter clusters which, due to the calorimeter cell sizes, may receive contributions from more than one electron. This leads to significant correlations in their PID performance.
The probability for two electrons to leave energy deposits in the same calorimeter cell strongly
depends on the opening angle of the dilepton system and the momenta of the electrons, and is therefore found
to be significantly higher in the \cqsq  signal region than in  either the \lqsq or the \jpsi-control region. The bias \DelPID is determined using simulated \runtwo \NBuToKJPsee and \BuToKee events as the relative difference between the true PID  efficiency and that obtained under the assumption of full factorization. This is illustrated in Fig.~\ref{fig:PIDFact_GenVCorr_18RKDECAL}, where \DelPID is shown as a function of \dECAL, the distance separating two electrons at the electromagnetic calorimeter after extrapolation of their trajectories to its upstream surface. 

The non-factorization of electron efficiencies is sufficiently  different in the signal
and control regions that a dedicated treatment is required. This is most significant for those candidates having $\dECAL<100$~mm. As only a few percent of signal candidates
fall into this region they are excluded from the analysis in order to avoid having to model the effect of the overlap when computing efficiencies. 
Electron efficiencies are evaluated from truth-level information to account for
non-factorization, but must be corrected for imperfections in modeling by simulation. 
Therefore, electron PID efficiencies are evaluated in 
both data and simulation using identically selected control samples. 
These efficiencies are computed
in bins of $\pt$, $\eta$, and  $n_{\rm Tracks}$ and are further determined in separate categories depending on whether the electron has an associated bremsstrahlung photon or not. 
In each bin, \wpid is defined as the ratio of the efficiency in data to that in simulation. These weights 
 are used to correct the PID  efficiency of the dielectron system determined using simulation.

The \lone calorimeter and muon triggers employ a simplified PID 
algorithm to select events, which can lead to biases in the measured PID  
performance. The calibration samples are therefore selected requiring a TIS \lone decision.

\subsection{Track reconstruction}
The track reconstruction performance for muons is evaluated using samples of \JPsmumu decays detached
from the PV, in which one muon is fully reconstructed in the tracking system and the presence of the other
muon is inferred from activity in the muon stations and TT detector only~\cite{LHCb-DP-2013-002}.
The rate at which this second muon is also reconstructed as a track in the full detector
gives the track reconstruction efficiency. The muon efficiencies are found to  be described well by  
simulation, and no additional calibration factors are applied. Differences between
data and simulation in the reconstruction performance are assumed to be the same for hadrons and muons, 
and to cancel in the double LU ratios.

Energy losses induced by bremsstrahlung cause a lower track reconstruction efficiency for electrons than for muons, depending on momentum and pseudorapidity.
For this reason a dedicated calibration has been developed~\cite{LHCb-DP-2019-003}, which uses 
control samples of \NBuToKJPsee\
decays in which one electron is fully reconstructed in the
tracking system and the other is only reconstructed in the vertex detector. 
The rate at which the second electron is also reconstructed as a track in the full tracking system gives the track reconstruction efficiency. These efficiencies are evaluated in data and simulation
in bins of electron momentum and pseudorapidity, as well as regions in the vertex detector which
contain more or less detector material and therefore induce more or less bremsstrahlung.
In each bin, \wtrk is defined as the ratio of the efficiency in data to that in  simulation.  These are used 
to correct the efficiency of the dielectron system measured in simulation.

\subsection{Multiplicity and kinematics}

The kinematics of the \B hadron and the particle multiplicity of the underlying event are imperfectly
simulated, partly due to limitations in how well the output of \pythia reflects $pp$ collisions
at LHC energies, and partly due to limitations in the description of the LHCb detector material
and the production of low-momentum particles from secondary interactions with the detector.
The detector material is simulated with varying degrees of accuracy for different constituent
parts of the LHCb detector, so that no single occupancy proxy can perfectly calibrate the
observed event multiplicities in the detector as a whole. In common with the rest of the analysis,
the calibration is performed using the track multiplicity as a proxy, and systematic uncertainties
are assigned for residual imperfections in the modeling of other multiplicity observables.
The kinematics are calibrated in three dimensions: the momentum, transverse momentum, and 
pseudorapidity of the \B hadron.

A dedicated boosted 
decision tree from the \texttt{\hepml} library \cite{Rogozhnikov:2016bdp} is trained to align the simulation with data in the three kinematic observables and the occupancy proxy observable. The outputs of this
decision tree are \wmco weights which encode the relative statistical importance that the final
efficiency determination should assign to each simulated event.

The calibration is performed using simulated and data samples of \mbox{\NBToKJPsmm} decays selected 
by the \lone muon trigger. This is both the most abundant and the highest-purity sample available; 
since multiplicity and kinematic corrections are by construction independent of the \B hadron 
decay it is appropriate to use the muonic decay as a proxy for the electron modes. 
For this independence to hold, residual data-simulation disagreements 
caused by PID and trigger performance must be reduced to a minimum.
While data are recorded with a range of trigger configurations, only a small number of these are simulated
in order to reduce the operational burden of their production and analysis.
A separate correction chain of \wpid, \wlo, and \whlt is therefore computed as input to the 
determination of \wmco, using only data taken with the simulated trigger configurations.

\subsection{Trigger}

The \lone \tos efficiencies are calibrated as a function of muon transverse momentum
and electron transverse energy, with the electron efficiency calibrated separately for
each of the three electromagnetic calorimeter regions. 
The efficiency denominator is the number of \tis events, 
while the numerator is the number of \tis events which are also \tos on the lepton trigger of interest~\cite{LHCb-PUB-2014-039}. 
In order to minimize non-factorizable effects, the muon efficiency is computed with 
hadron or electron \tis events as the denominator, while the electron efficiency 
is computed with hadron or muon \tis events as the denominator. Alternative definitions of the
denominator are used as cross-checks and give compatible results.
Efficiencies are calculated on data and simulation, and the \wlo weights
encode the ratio of data and simulation efficiencies in each kinematic bin.
Since there are two leptons in each event, the final per-event weight has to be corrected
in order not to count twice events in which both leptons satisfy the \tos criteria, where

\begin{equation}
\label{eq:efficiencies:trigger_combination_tos}
\wlo^{\tos} = \frac{\varepsilon_{\tos}^\text{data}}{\varepsilon_{\tos}^\text{MC}} = \frac{ 1- \left( 1- \varepsilon_\text{\tos}^\text{data}(\ellp) \right) \cdot  \left( 1- \varepsilon_\text{\tos}^\text{data}(\ellm) \right)}{1- \left( 1- \varepsilon_\text{\tos}^\text{MC}(\ellp) \right) \cdot  \left( 1- \varepsilon_\text{\tos}^\text{MC}(\ellm) \right)} \,.
\end{equation}

The \lone \tis efficiencies are calibrated as a function of the \B hadron transverse
momentum and the event track multiplicity, since the \tis trigger is by definition more likely to
select events with higher activity in the detector. The efficiency denominator is the number of
lepton and hadron \tos events, to maximize the available control sample yields. The efficiency
numerator is the number of those events which are also \tis. 
As with the \tos weights, the efficiencies are calculated on data and simulation, and the \wlo weights
encode the ratio of data and simulation efficiencies in each kinematic bin.

The \hlt efficiencies are calibrated analogously to the \lone efficiencies, as a function of the
event track multiplicity. The same control samples are used, with separate \whlt calibrations for 
the \lone \tis and \tos categories. The efficiencies are calculated on data and simulation, and 
the \whlt weights are defined as the ratio of data and simulation efficiencies in each bin.

\begin{figure}[t]
\centering
\includegraphics[width=0.49\linewidth]{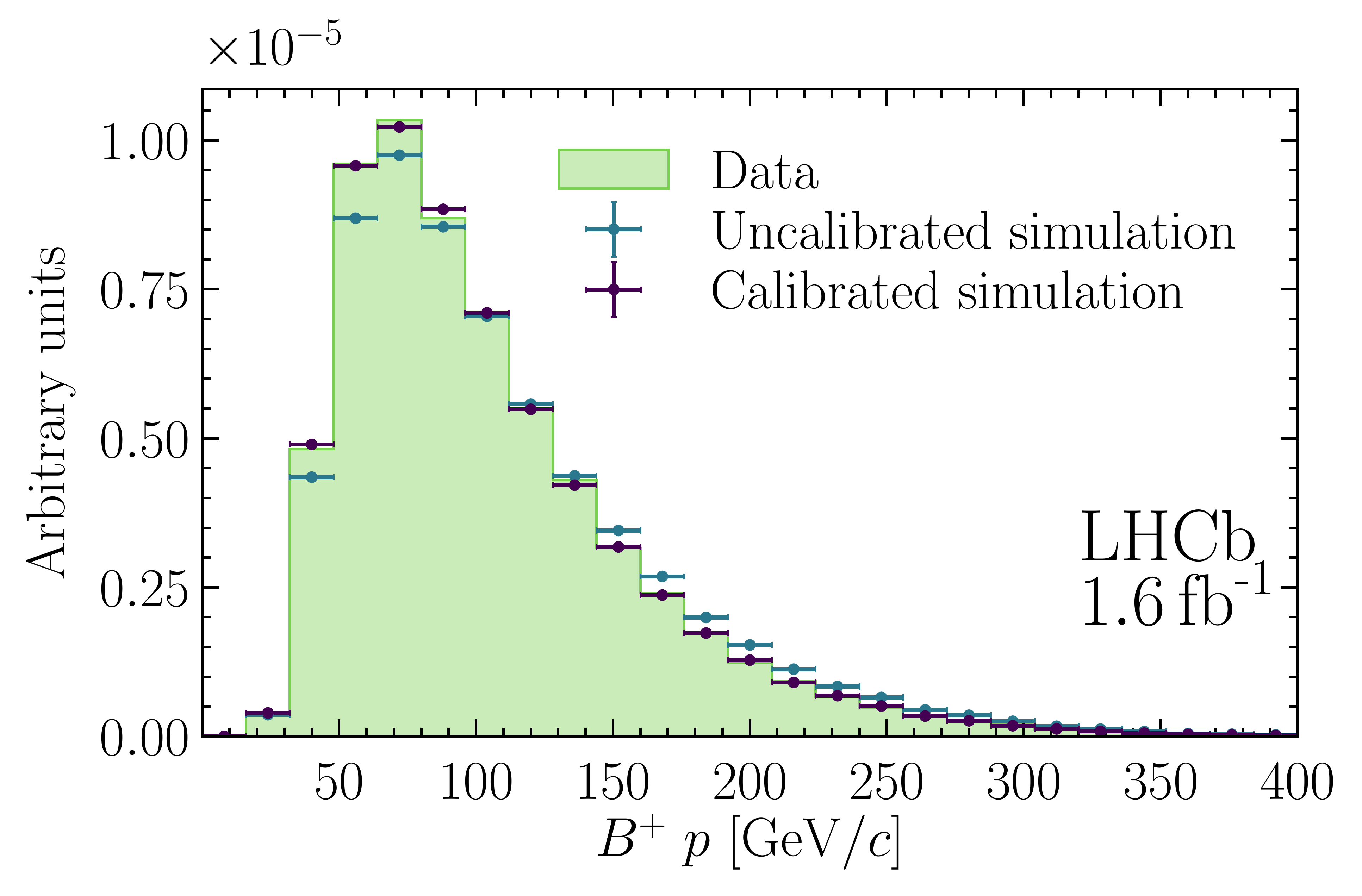}
\includegraphics[width=0.49\linewidth]{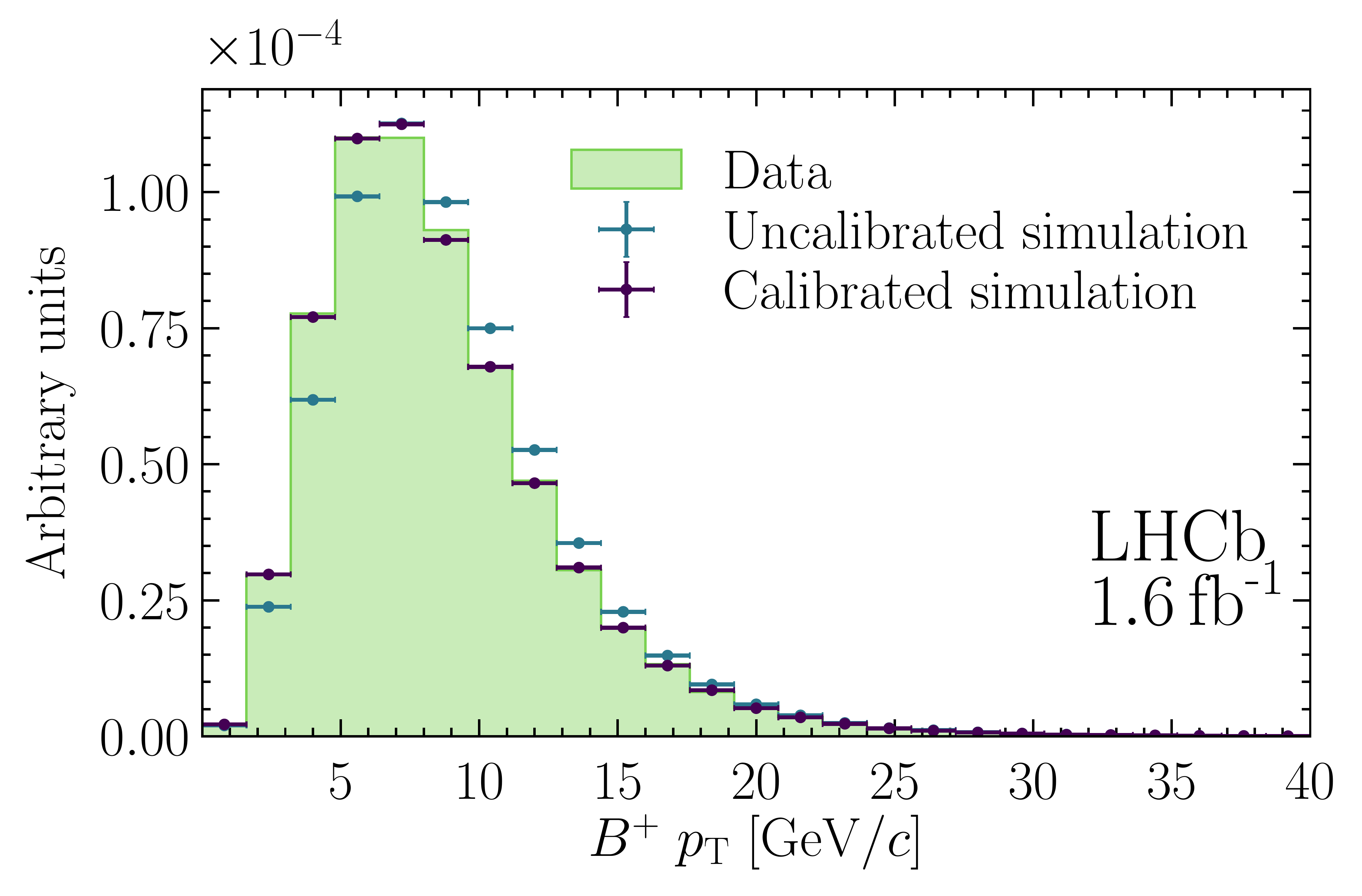}
\includegraphics[width=0.49\linewidth]{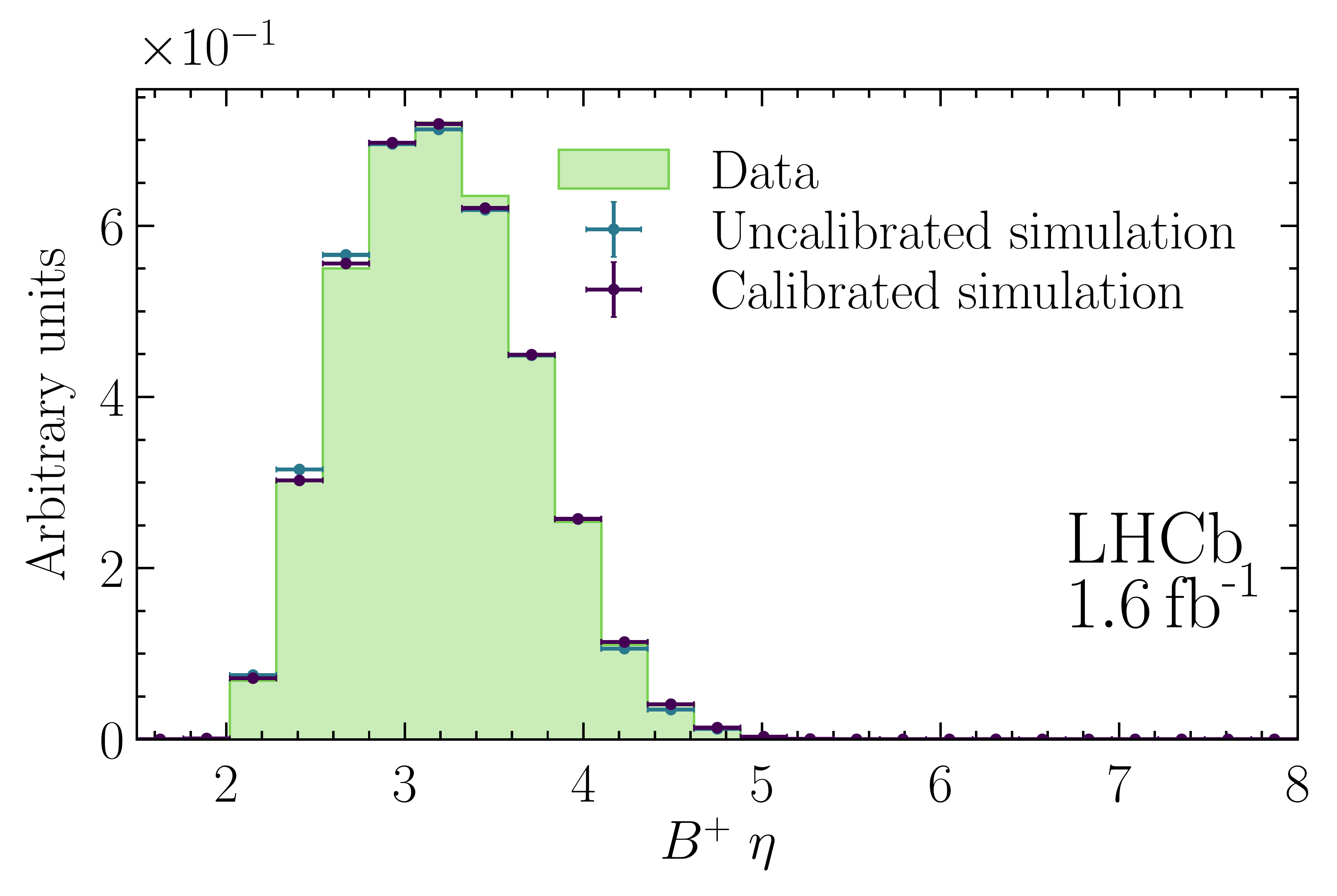}
\includegraphics[width=0.49\linewidth]{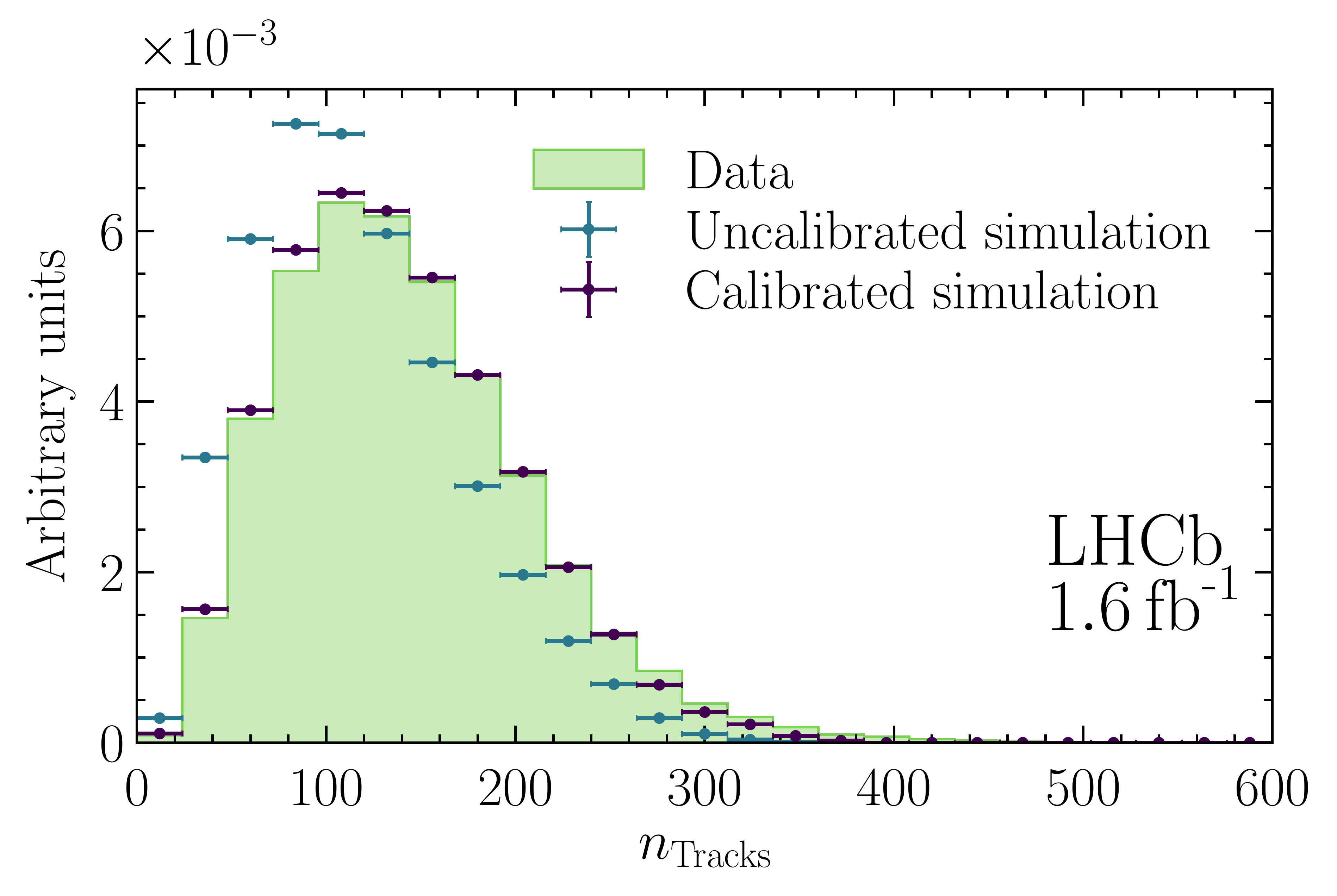}
\includegraphics[width=0.49\linewidth]{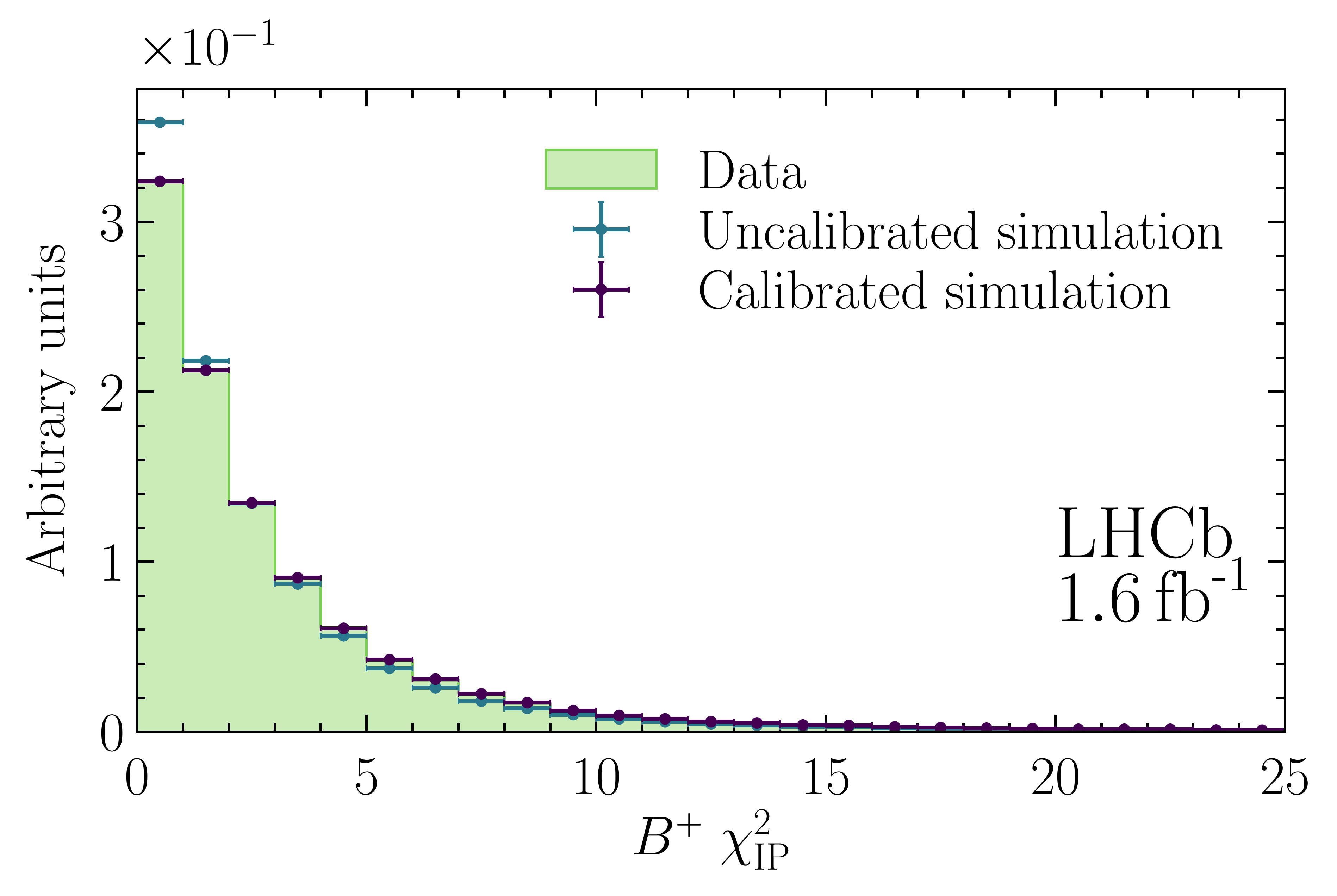}
\caption{\small Distributions of selected measured quantities for \tos \BuToKJPsee candidates in the 2016 data,   
compared with simulation before (red) and after (blue) calibrations.}
\label{fig:data_mc_ee}
\end{figure}

\subsection{Candidate reconstruction}
Residual discrepancies between data and simulation arise from differences in the performance of the reconstruction, particularly in the uncertainties assigned to track trajectories
that in turn affect derived quantities such as the vertex fit quality.
A second boosted decision tree, trained analogously to \wmco, is used to improve further the data-simulation agreement. As reconstruction differences are sensitive to the
particle species being calibrated, their calibration is performed separately for the electron
and muon final states, and separately for each \lone trigger category. The reweighting is
performed as a function of five variables: the same three kinematic quantities used for \wmco as well as
the \chisqip of the \B and \jpsi mesons, where \chisqip for a given particle is defined as the  difference in the \chisq of the PV fit with and without that particle. Examples of the final agreement between data and simulation are presented in Fig.~\ref{fig:data_mc_ee}. 

\subsection{Migration in \texorpdfstring{\boldmath{\qsq}}{q2}}
As a result of these calibrations the simulation accurately models most features of the data. However, 
the migration of electron candidates across the \qsq spectrum is sensitive to residual misalignments between data and simulation that affect the \qsq resolution and its behavior in the tails; particular attention is required when evaluating the impact of bremsstrahlung.
The \qsq of simulated candidates is therefore smeared using a function with parameters determined by fitting the dielectron mass spectra from \NBToKJPsee decays in data and simulation.
The smeared dilepton mass for each candidate in simulation is given by 
\begin{equation}
    \msmeared = \mtrue + s_{\sigma} \cdot (\mreco - \mtrue) + \Delta\mu + (1-s_{\sigma}) \cdot (\mu^{\mathrm{MC}} - \MjpsiPDG),
\end{equation}
\noindent where \mtrue is the generated dilepton mass calculated using the difference between the generated kinematics of the parent \B hadron and of the $K^{(+,*0)}$; 
\mreco is the reconstructed dilepton mass in simulation; $s_{\sigma}$ is the ratio of the widths of the reconstructed mass distributions in data and simulation; $\Delta\mu$ is the difference in the means of the reconstructed mass distributions in data and simulation; $\mu^{\mathrm{MC}}$ is the mean mass determined from a fit to simulated data.

\begin{figure}[t]
\centering
\includegraphics[width=0.98\linewidth]{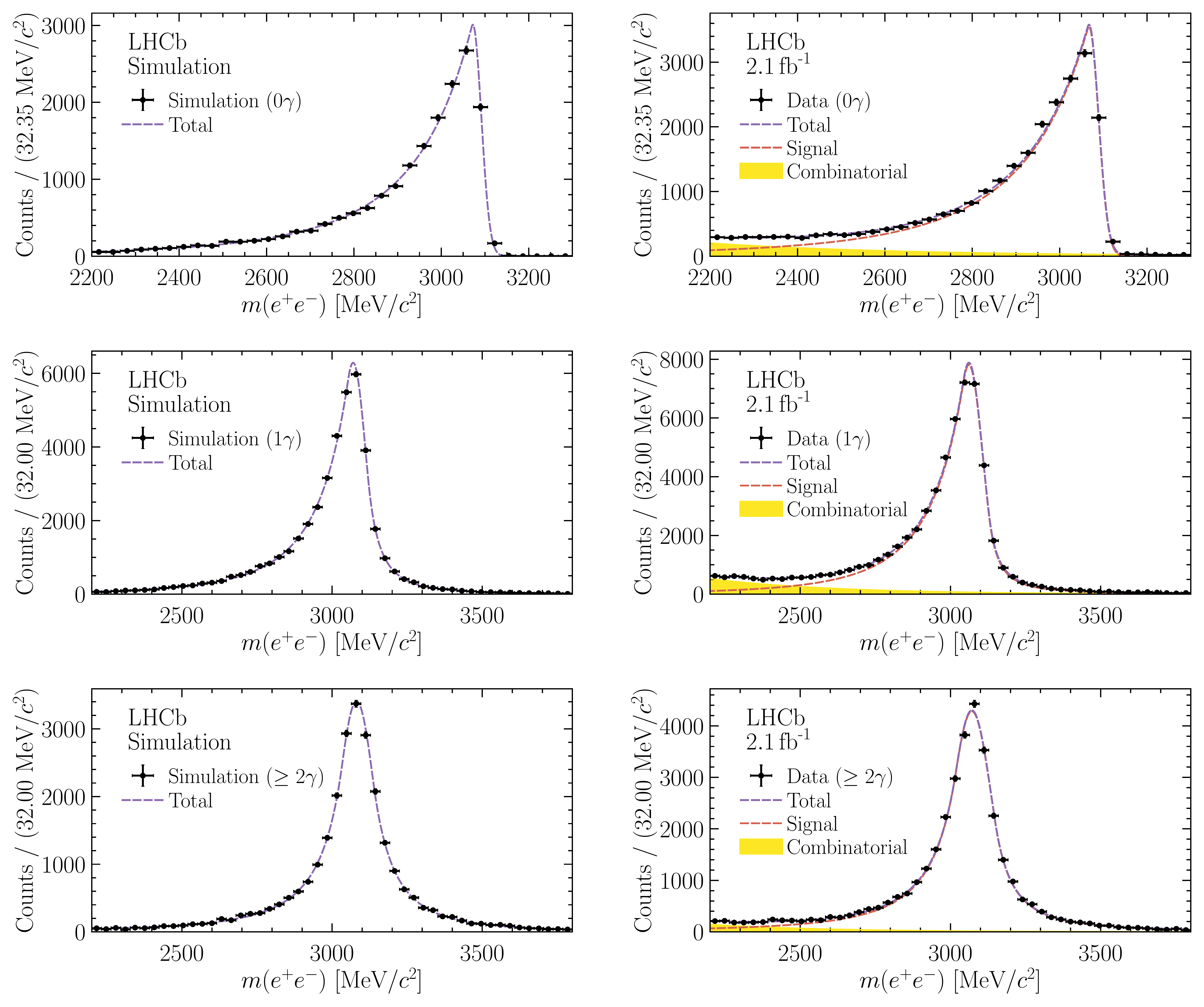}
\caption{\small Distributions of the dielectron invariant mass for \BuToKJPsee candidates in (left) simulation and (right) data, for each of the three bremsstrahlung categories (from top to bottom), overlaid with the projections of the fit model. The candidates correspond to the 2018 data and the TIS trigger category.}
\label{fig:eff_smearing_fits}
\end{figure}

Unbinned maximum-likelihood fits are performed separately for the \Bp and \Bz modes in each
trigger and bremsstrahlung category and for each data-taking year. 
The full selection is applied leading to  excellent sample
purity. A modified Crystal Ball function~\cite{Skwarnicki:1986xj} with power law tails both
above and below the mean mass value (DSCB) is used to model the dielectron spectrum, with  the remaining combinatorial background modeled using an exponential 
function.
The high quality of the fit is illustrated by comparing 2018 data and simulation in Fig.~\ref{fig:eff_smearing_fits}.
The smeared mass allows the efficiency measured 
in a given range of reconstructed \qsq to be transformed into the corresponding range of true \qsq, defined before emission of final state photon radiation, as required for the measurement of the lepton universality ratios. This correction is denoted as \wsmear.

\subsection{Determination of efficiencies}
The overall efficiency for the signal and resonant control modes is determined using fully
calibrated simulation samples for each data-taking year and trigger category. 
Efficiencies for background samples that are modeled in the
invariant mass fit are determined in the same way. To make the best use of computing resources, only 
events in which all of the decay products of a \B candidate are generated within the geometric acceptance of the \lhcb detector are processed by the detector simulation. The efficiency, \egeo, of this generator selection is evaluated for each signal mode as a function of \qsq using dedicated samples generated without \lhcb detector acceptance requirements. The overall efficiency is then given by 
\begin{equation}\label{eq:toteffs}
\etot = \egeo \times (\varepsilon_\mathrm{MVA} \times \varepsilon_\mathrm{Presel} \times \varepsilon_\mathrm{Trg} \times \varepsilon_\mathrm{PID}|\mathrm{geo}),
\end{equation}
\noindent where $\varepsilon_\mathrm{MVA}$ is the efficiency of the multivariate selection,
$\varepsilon_\mathrm{Presel}$ is the efficiency of the preselection excluding PID  criteria, $\varepsilon_\mathrm{Trg}$ is the trigger efficiency, and
$\varepsilon_\mathrm{PID}$ is the PID  efficiency. 

The strategy of applying \Bz calibrations to \Bp final states and vice versa
reduces correlations in the total efficiency determination but can not eliminate them entirely.
The most significant irreducible correlation is caused by the fact that the same simulated samples
are used to compute both the resonant mode efficiencies and the data-simulation calibrations.

Further residual correlations occur because of calibrations that are shared between the muon and electron final states, because the \tis and \tos samples used in the calibrations are not required to be mutually exclusive in order to increase the control sample sizes,
and because the resonant control modes are used to compute the trigger efficiencies
and train the algorithms that produce the \wmco and \wmcreco weights.
These correlations are evaluated using a bootstrapping procedure as follows. Each reconstructed data or simulation candidate is assigned
one hundred different Poisson-distributed weights with a mean value of 1. 
The generation of weights is performed using a common seed for each event based on a unique
event identifier. This allows 100 different correction maps to be generated and
their correlations assessed by comparing the simulation efficiencies and data sample yields
for each bootstrapped data sample. The distributions of the bootstrapped efficiencies are verified to be well described by Gaussian functions.
The relative efficiencies of nonresonant  and resonant modes in both low- and \cqsq  are found to vary between 0.7 and 0.9 for both electron and muon modes.

\section{Simultaneous invariant mass fit}
\label{sec:simfit}

The signal and resonant control mode yields in Eqs.~\ref{eq:doubleratiorx}--\ref{eq:rawrjpsik},
as well as those of the \psitwos equivalents, are obtained using simultaneous maximum-likelihood fits to the invariant mass
distributions of selected \B meson candidates. The invariant mass is calculated using the decay tree fitter algorithm to constrain the momentum vector of the \B meson to be aligned with its displacement vector. The fits to the signal modes are unbinned, whereas the fits to the more abundant resonant modes are performed to data that are binned in the invariant mass. The fits are based on the \roofit \cite{Verkerke:2003ir} and \root \cite{Brun:1997pa} frameworks, with a custom implementation of the probability density functions (PDFs)~\cite{Gligorov:2021sry} that eliminates biases in binned fits caused by sharp PDF variations within a given bin. 
Events selected in the \tis and \tos trigger categories are fit simultaneously. The structure allows the fit to be performed either for the signal mode yields; for the resonant mode yields; simultaneously for the signal and resonant mode yields; or simultaneously for \RK and \RKst 
by using the efficiencies, determined on calibrated simulation samples, and the covariance matrix, obtained by bootstrapping the efficiencies, as constraints in the invariant mass fit. 

Similarly, the fit can be executed for each of the \runone, \runtwopo, or \runtwopt data-taking periods, or for all three simultaneously. 
The configuration in which \RK and \RKst are fitted simultaneously in all trigger categories and data-taking periods is referred to as ``nominal'' and used to produce the results reported in Sec.~\ref{sec:results}.
All constraints described are implemented as Gaussian functions 
with mean and width corresponding to the central value and the uncertainty associated with the parameter 
being constrained. Systematic uncertainties and their correlations are instead accounted for including a multiplicative factor to the \RK and \RKst values in each fit projection category, which is constrained using a Gaussian function with mean of unity and a width representing the relative uncertainty of the relevant source. Multidimensional Gaussian constraints are implemented for correlated parameters.
\begin{table}[!t]
	\centering
	\caption{Invariant mass ranges used in the fits.
 The fit type indicates where the dilepton invariant mass is constrained to the known \jpsi (\psitwos) mass.}
	\label{tab:fit_mass_ranges}
	\renewcommand\arraystretch{1.3}
	\begin{tabular}{l|clc}
		\textbf{Lepton} & \textbf{\boldmath{\qsq} region} & \textbf{Fit type} & \textbf{Range (\mevcc)} \\
		\hline
        \multirow{4}{*}{Electron} & low, central & unconstrained & 4600--6200 \\
         & \multirow{2}{*}{\jpsi} & unconstrained & 4600--6200 \\
         & & constrained & 4900--6200 \\
         & \psitwos & constrained & 5100--5750 \\
        \hline
        \multirow{4}{*}{Muon}  & low, central & unconstrained & 5150--5850 \\
         & \multirow{2}{*}{\jpsi} & unconstrained & 5100--6100 \\
         & & constrained & 5100--6100 \\
         & \psitwos & constrained & 5100--5750 \\
	\end{tabular}
\end{table}
\begin{table}[t]
	\centering
	\caption{Observed yields of the six signal and control modes and their statistical uncertainties.}
	\label{tab:fit_result_nominal_yields}
	\renewcommand\arraystretch{1.3}
	\begin{tabular}{r|cc}
	 \textbf{LU observable} & \textbf{Muon (\boldmath{$\times 10^3)$}} & \textbf{Electron (\boldmath{$\times 10^3)$}} \\
	 \hline
low-\qsq \RK       &  \phantom{$\left(4\right.$}$1.25 \pm 0.04$\phantom{$\left.2\right)\times 10^3$} &
                   \phantom{$\left(\right.$}$0.305 \pm 0.024$\phantom{$\left.\right)\times 10^3$}   \\
low-\qsq \RKst     &  \phantom{$\left(4\right.$}$1.001 \pm 0.034$\phantom{$\left.2\right)\times 10^3$}  &
                  \phantom{$\left(\right.$}$0.247 \pm 0.022$\phantom{$\left.\right)\times 10^3$}   \\
central-\qsq \RK   &  \phantom{$\left(4\right.$}$4.69 \pm 0.08$\phantom{$\left.2\right)\times 10^3$}     &
                \phantom{$\left(9\right.$}$1.19 \pm 0.05$\phantom{$\left.5\right)\times 10^3$}   \\
central-\qsq \RKst &  \phantom{$\left(4\right.$}$1.74 \pm 0.05$\phantom{$\left.2\right)\times 10^3$}    &
                \phantom{$\left(\right.$}$0.443 \pm 0.028$\phantom{$\left.\right)\times 10^3$}  \\
\jpsi \RK          &  $\left(2.964 \pm 0.002\right) \times 10^{3}$   & $\left(7.189 \pm 0.015\right) \times 10^{2}$ \\
\jpsi \RKst        &  $\left(9.733 \pm 0.010\right) \times 10^{2}$   & $\left(2.517 \pm 0.009\right) \times 10^{2}$ \\
	\end{tabular}
\end{table}

The invariant mass resolution of the resonant control modes can be improved by constraining the
dilepton invariant mass to be equal to that of the \jpsi or \psitwos resonance, and this improvement
is particularly large for electrons because of their poorer intrinsic resolution. The unconstrained
dilepton invariant mass is used in the nominal fits in order to match the modeling of the nonresonant mode 
and reduce systematic uncertainties in the double ratio, whereas the constrained mass is used 
for cross-checks and systematic studies. The fit ranges used in the analysis are given in
Table~\ref{tab:fit_mass_ranges}; where studies of specific systematic uncertainties use different fit ranges, these are noted in Section~\ref{sec:systematics}.

\begin{figure}[tb]
\centering
\includegraphics[width=0.98\textwidth]{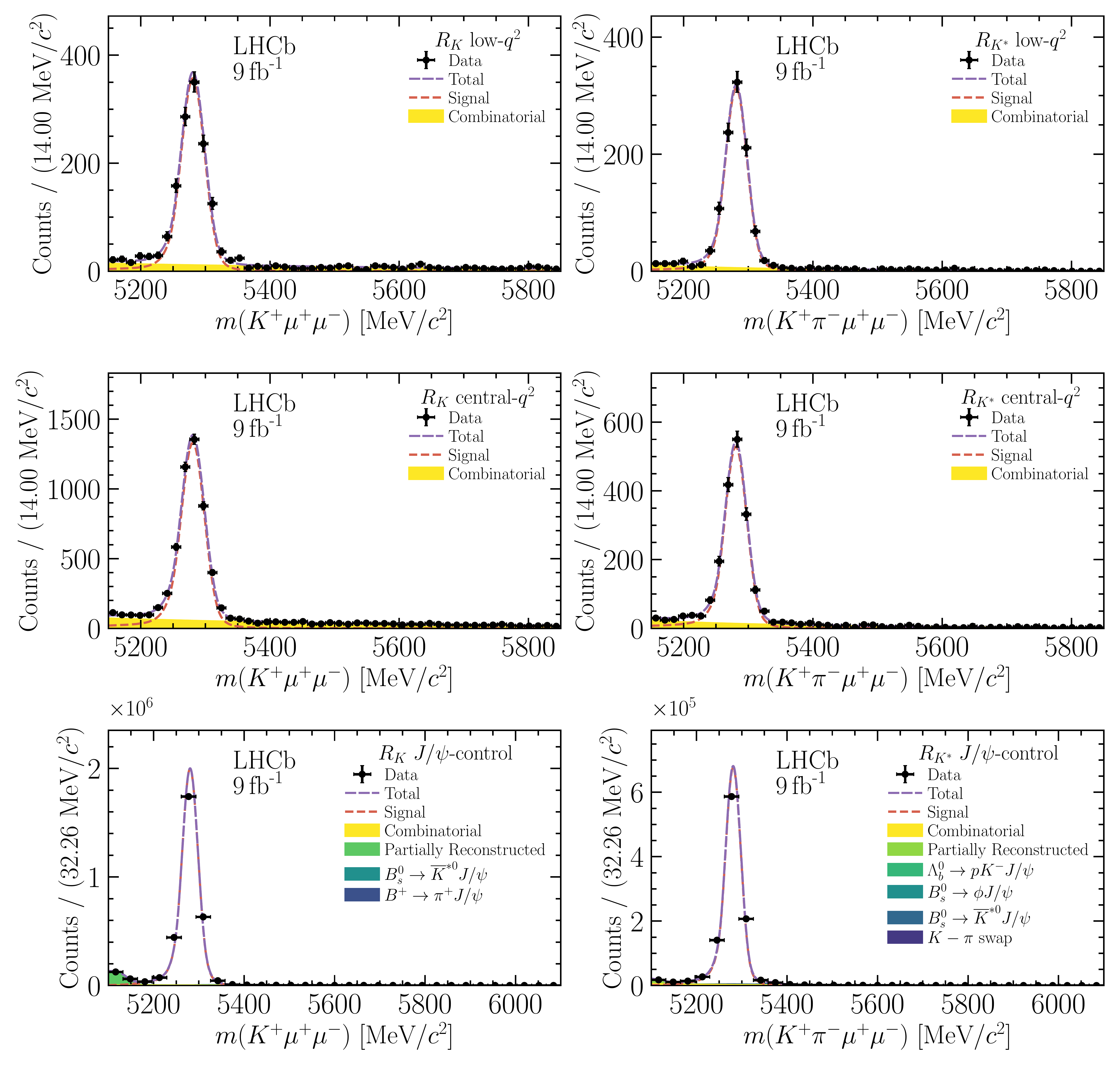}
\caption{\small Distributions of (left) $m(K^{+}\mu^{+}\mu^{-})$  and (right) $m(K^{+}\pi^{-}\mu^{+}\mu^{-})$ of \lqsq, \cqsq, and \jpsi-control regions (from top to bottom), overlaid with the projections of the fit model. Each of the fit components are discussed in Section~\ref{sec:simfit:fit_components}.}
\label{fig:fit_result_nominal_mm}
\end{figure}

\begin{figure}[tb]
\centering
\includegraphics[width=0.98\textwidth]{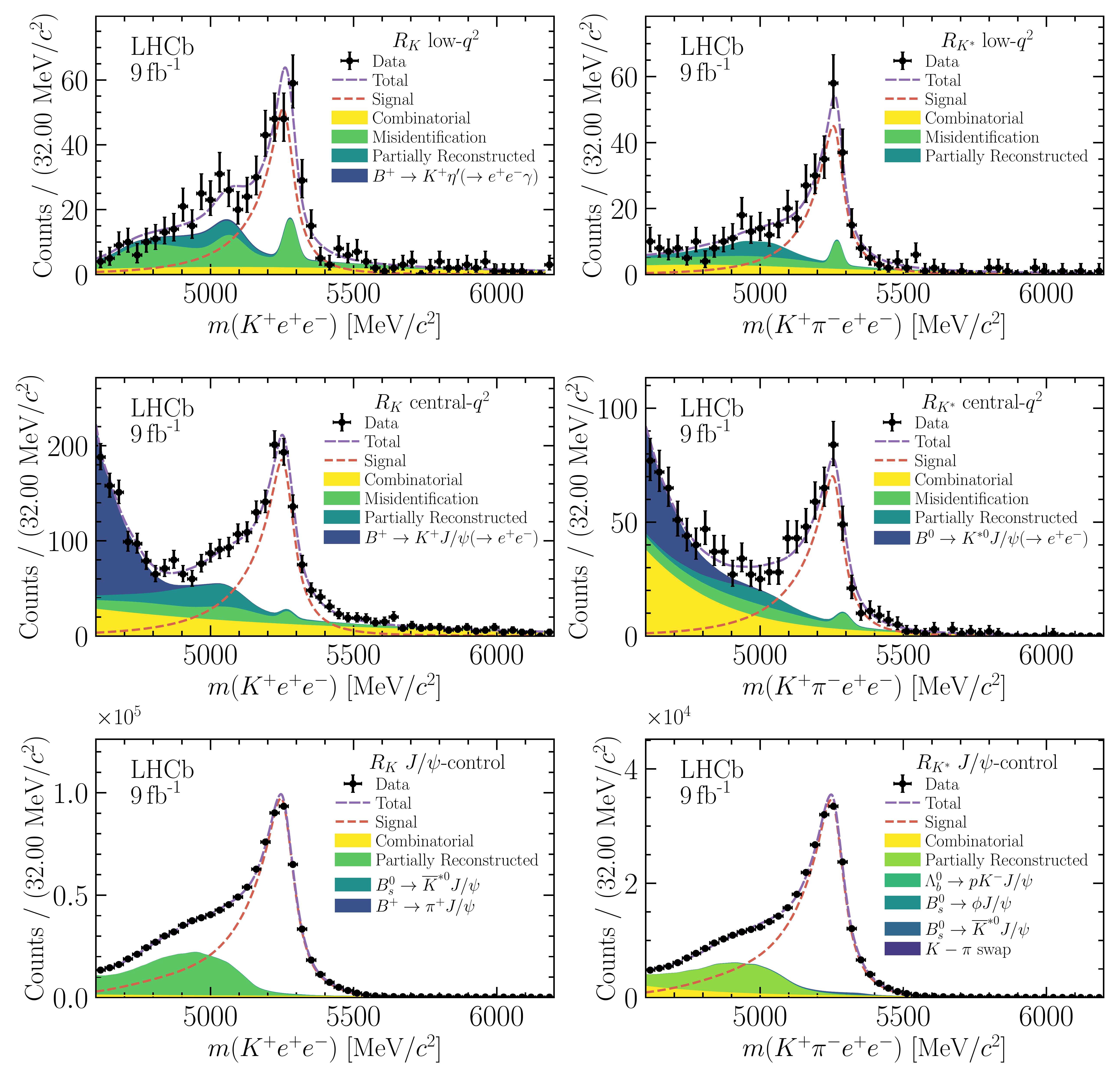}
\caption{\small Distributions of (left) $m(K^{+}e^{+}e^{-})$  and (right) $m(K^{+}\pi^{-}e^{+}e^{-})$ in the (top to bottom) \lqsq, \cqsq, and \jpsi-control regions, overlaid with the projections of the fit model. Each of the fit components are discussed in Sec.~\ref{sec:simfit:fit_components}.
}
\label{fig:fit_result_nominal_ee}
\end{figure}

Large ensembles of pseudodata generated with the component yields observed in data are used
to validate that the fit is unbiased and gives accurate uncertainties. The uncertainties on the
\RK and \RKst double ratios are found to be asymmetric, which is accounted for when reporting the results in Sec.~\ref{sec:results}.
The result of the nominal simultaneous fit to the signal and \jpsi modes is shown
in Fig.~\ref{fig:fit_result_nominal_mm} for the muon and Fig.~\ref{fig:fit_result_nominal_ee} for the electron final states. The observed yields of the six signal and control modes, as well as their statistical uncertainties, are reported in Table~\ref{tab:fit_result_nominal_yields}.

\FloatBarrier

\FloatBarrier

\subsection{Fit components}
\label{sec:simfit:fit_components}
\subsubsection{Signal and control modes}
For fits where the dilepton invariant mass is unconstrained, the signal and resonant control mode PDFs are obtained by fitting analytic 
functions to fully calibrated simulated samples in each data-taking period and trigger category. The 
best-fit values of the function parameters are subsequently fixed in nominal fits to data and varied in pseudoexperiments to estimate the associated systematic uncertainties, which are found to be negligible. 
For the final states with electrons, individual PDFs are obtained by fitting simulated samples separated according 
to their bremsstrahlung category; these are subsequently added in proportion to the abundance
of bremsstrahlung categories observed in fully calibrated simulation samples to obtain an overall 
PDF for use in data fits. 
The relative abundance of the bremsstrahlung categories with zero, one, and two or more reconstructed bremsstrahlung photons corresponds to $25\%$, $50\%$, and $25\%$, respectively. A systematic uncertainty is assigned to account for the finite knowledge of these fractions. 
Fits constraining the \jpsi\ mass 
are only used for cross-checks and systematic uncertainties and have a 
better mass resolution which does not depend significantly on the dielectron 
bremsstrahlung category. For this reason the data fit PDFs are obtained in a simplified way in this case by fitting 
analytical functions to uncalibrated simulation samples without any separation for the
bremsstrahlung category. 

The analytical functions used to define the signal and resonant control mode PDFs are listed in
Table~\ref{tab:fit_mass_functions}. The overall PDF normalizations vary freely 
for each data-taking period and trigger category. The different treatment of the electron signal and \jpsi 
PDFs obtained without a constraint on the dielectron invariant mass is motivated by a combination
of two effects. First, the \jpsi-\qsq range is significantly narrower above
the mean \jpsi meson mass than below it, which deforms the right-hand tail of the PDF. Second,
the veto on cascade \decay{B}{\psitwos(\to\jpsi X)K^{(*)}} decays deforms the left-hand tail of the PDF. 
An acceptable fit quality can therefore only be obtained by adding either one or two Gaussian
functions, depending on the bremsstrahlung category, to the PDF. The normalization of these
Gaussian functions relative to the principal DSCB component is a free parameter of the fit to data and simulation. 

Residual differences between data and simulation are parametrized
through a shift in the mean value of the signal PDF and a scale
factor applied to the width of this PDF. These parameters are independent for muons and electrons,
and independent for each data-taking period and trigger category, 
but shared between the signal and control modes.
Prior to calibrating the simulation the scale factors are typically between 1.1 and 1.15, while
the mean value of the PDF is shifted by $\order(10\mevcc)$.
After the simulation is calibrated, the scale factors are found to be compatible with unity
while the mean value shifts are reduced to $\order(1 \mevcc)$. The scale factors and mean values
are left as free parameters within the nominal fit to account for systematic uncertainties
caused by residual imperfections in the calibration of simulation.

\begin{table}[t]
	\centering
	\caption{Analytical functions used to describe the signal and resonant control modes. The fit type refers to whether the dilepton invariant mass is constrained to that of the \jpsi (\psitwos) resonance or not. Category refers to the bremsstrahlung category in the case of electron modes. Hypatia refers to a two-sided version of a generalized Crystal Ball distribution introduced in Ref.~\cite{Santos:2013gra}.}
	\label{tab:fit_mass_functions}
	\renewcommand\arraystretch{1.3}
	\begin{tabular}{l|cccl}
		\textbf{Lepton} & \textbf{\qsq Region} & \textbf{Fit type}
		                       & \textbf{Category} & \textbf{Function} \\
		\hline
        \multirow{6}{*}{Electron} & low, central & unconstrained & \phantom{$\ge$2}all &  DSCB \\
         & \multirow{4}{*}{\jpsi\phantom{$(2)$}} & \multirow{3}{*}{unconstrained} & \phantom{$\ge$all}0 & DSCB + Gaussian \\
         &  &  & \phantom{$\ge$all}1 & DSCB + two Gaussians \\
         &  &  & $\phantom{al}\ge$2 & DSCB + two Gaussians \\
         &  & \phantom{un}constrained & \phantom{$\ge$2}all &  DSCB \\
         & \psitwos & \phantom{un}constrained & \phantom{$\ge$2}all & DSCB \\
        \hline
        \multirow{4}{*}{Muon} & low, central & unconstrained & & DSCB + two Gaussians \\
         & \multirow{2}{*}{\jpsi\phantom{$(2)$}} & unconstrained & & DSCB + two Gaussians\\
         &  & \phantom{un}constrained & & Hypatia + Gaussian\\
         & \psitwos & \phantom{un}constrained & & Hypatia + Gaussian \\
	\end{tabular}
\end{table}

\subsubsection{Combinatorial background}
\label{sec:simfit_combo}
The combinatorial background is described by a single exponential function for the resonant modes and the nonresonant muon modes.
For the nonresonant electron modes, the multivariate selections and the \mcorr criteria are found to induce a deviation from an exponential shape by introducing a sculpting of the invariant mass spectrum within the fit ranges considered. 
Same-sign lepton data are exploited to calibrate the modeling of the combinatorial shape in the low- and  \cqsq bins for each data-taking period and trigger category. 
The sculpting is described by a factor $1/\bigl(1+\exp(s(m-m_{0})\bigr)$ that multiplies the exponential function and where the parameters ($s,m_{0}$) are obtained from fits to same-sign data and fixed in fits to the nonresonant electron signal modes. 
Systematic uncertainties associated with the procedure are evaluated by varying $(s,m_{0})$ according to the uncertainty determined in same-sign data fits. 
The slope of the exponential function is left as a free parameter in the fit. 
The PDF normalization is allowed to vary independently for each data-taking period and trigger category in all cases.

\subsubsection{\texorpdfstring{\boldmath{\jpsi}}{J/psi} leakage in the central-\texorpdfstring{\boldmath{\qsq}}{q2} region}
A significant fraction of \jpsi decays which leak into the central \qsq region also fall within
the invariant mass fit range for the electron mode. The energy loss which causes their invariant
dielectron mass to fall within the central \qsq region also causes their invariant \B meson mass
to be shifted to values much lower than the signal. The extended fit range in signal electron modes,
down to 4600~\mevcc, allows the interplay between this background component, the combinatorial
background, and specific physics backgrounds to be well modeled. The \jpsi leakage PDFs are
described using unbinned templates derived from fully calibrated simulation samples. 
The normalization of the PDF is also obtained from fully calibrated simulation samples
for each data-taking period and trigger category. It is constrained in fits to data,
with a $20\%$ uncertainty which reflects not only the measured uncertainties on simulation
but also accounts for any residual disagreement between the data and simulation.

\subsubsection{Specific backgrounds at low- and central-\texorpdfstring{\boldmath{\qsq}}{q2}}
No significant specific backgrounds are present in the low- and \cqsq muon modes. 

For the electron case, 
the remaining specific backgrounds are \decay{B^{+,0}}{(\Kp\pim\pi^{+,0})\epem} in the case of 
the \Bz mode, \decay{B^{+,0}}{ (\Kp\pi^{0,-}) \epem} in the case of the \Bp mode, 
and misidentified
backgrounds for both modes.
At low-\qsq, there is an additional small contribution to the \Bp\ mode from \decay{B^{+}}{K^{+}\eta^{\prime}(\decay{}{\epem\gamma})} decays which is included in the fit model with a shape determined from simulation generated accounting for the $\eta^{\prime}(\decay{}{\epem\gamma})$ dynamics~\cite{BESIII:2015zpz} and constrained to its expectation~\cite{CLEO:1999kfw}.

The \Bz mode backgrounds that are not affected by misidentification are described using unbinned templates 
obtained from fully calibrated simulation of \decay{\Bp}{\Kp\pim\pip\ep\en} decays. 
Their normalization is allowed to vary freely for each data-taking period and trigger category in all cases. 

The \Bp mode backgrounds that are not affected by misidentification are also described using unbinned 
templates obtained from fully calibrated simulation samples. As the \Bp mode backgrounds include the 
\decay{\Bz}{K^{*0} \epem} signal, it is desirable to constrain their normalization using the \Bz mode in the simultaneous fit. This both improves sensitivity and, 
more importantly, enforces that the two measurements of LU are coherent: one is 
measured at the best-fit point of the other. To enable this, individual components of the   
\decay{B^{+,0}}{ (\Kp\pi^{0,-}) \epem} background are considered separately, with their  
normalizations constrained relative to that of the \decay{\Bz}{K^{*0}\epem} signal as explained below.  
The specific contributions identified are: 
\begin{enumerate}[rightmargin=10mm]
\item \decay{\Bz}{(\Kp\pim) \epem} where the \Kp\pim invariant mass is within 100\mevcc of the
\Kstarz invariant mass: these correspond directly to the \Bz mode signal and their normalization is constrained from its normalization corrected by the relative efficiencies obtained from fully calibrated
simulation.
\item \decay{\Bz}{(\Kp\pim) \epem} where the \Kp\pim invariant mass is more than 100\mevcc from the
\Kstarz invariant mass, and the \Kp\pim pair originates from the \Kstarz resonance: these correspond
to the tail of the Breit-Wigner distribution describing the \Kstarz resonance. Their normalization is 
also constrained from the \Bz mode signal normalization corrected by the relative efficiencies obtained 
from fully calibrated simulation samples and by the relative efficiency of the 100\mevcc mass 
window applied to the \Kstarz Breit-Wigner distribution.
\item \decay{\Bz}{(\Kp\pim) \epem} where the \Kp\pim invariant mass is less than 1200\mevcc and the
\Kp\pim pair does not originate from the \Kstarz resonance: these correspond to the non-resonant
(S-wave) \decay{\Bz}{(\Kp\pim) \epem} counterpart of the \Bz mode signal. Their relative decay rate
has been directly measured in Ref.~\cite{LHCb-PAPER-2016-012} for muonic modes and that measurement, together with
relative efficiencies obtained from fully calibrated simulation samples, is used to constrain 
the normalization of this background component.
\item \decay{\Bz}{(\Kp\pim)\epem} where the \Kp\pim invariant mass is greater than 1200\mevcc
and the \Kp\pim pair does not originate from the \Kstarz resonance: these include S-wave counterparts
of the signal, for which the decay rate is estimated in two different ways: extrapolating linearly the known branching ratios for $m(\Kp\pim)$ below 1200\mevcc up to 2400\mevcc into four regions of $m(\Kp\pim)$, and using the full amplitude 
model of \Kp\pim\jpsi decays developed in Ref.~\cite{LHCb-PAPER-2014-014}. These estimates are found to 
be compatible, and the linear extrapolation, together with relative efficiencies obtained from fully
calibrated simulation samples, is used to constrain the normalization of this background component. 
A 50\% relative systematic uncertainty is assigned to this extrapolation.
\item \decay{\Bp}{(\Kp\piz) \epem}: this is the isospin partner of the neutral signal decay, with analogous
resonant and non-resonant $K\pi$ components. The normalization of its components is constrained to that 
of the analogous $(\Kp\pim)$ components, corrected by relative efficiencies obtained from fully calibrated
simulation, and scaled by an isospin extrapolation factor which accounts for differences in the \Bp and \Bz 
lifetimes as well as \decay{\Kstarz}{\Kp\pim} and \decay{\Kstarp}{\Kp\piz} relative decay rates.
A relative systematic uncertainty of 10\% is assigned to the isospin extrapolation factor.
\end{enumerate}

\begin{figure}[tb]
    \centering
    \includegraphics[width=0.95\linewidth]{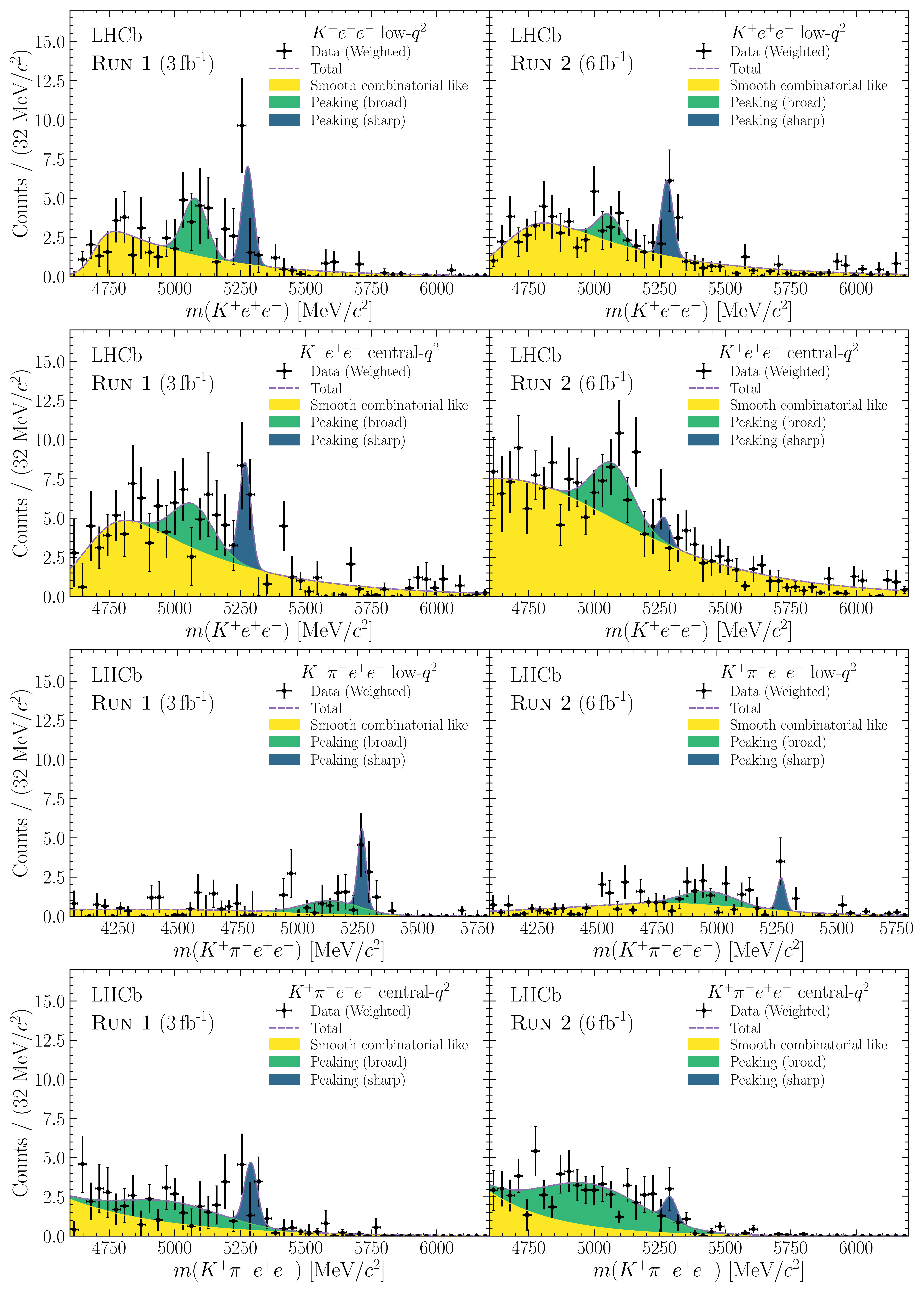}
    \caption{Template shapes for misidentified backgrounds obtained from data. The shapes for \runone are given on the left, the shapes for \runtwo are given on the right. 
From top to bottom, the shapes for \RK in \lqsq, \RK in \cqsq, \RKst in \lqsq and \RKst in \cqsq regions are given.}
    \label{fig:templatePDFs}
\end{figure}

Misidentified backgrounds are described using shapes constructed from the histograms in Fig.~\ref{fig:misid_prediction_nominal} in Sec.~\ref{sec:misidentifiedbkgds}. 
The baseline approach models these backgrounds with a Gaussian component for the fully reconstructed doubly misidentified
backgrounds that peak between 5200 and 5300\mevcc, a second empirical Gaussian component to describe
the non-combinatorial backgrounds below 5200\mevcc, and an exponential component sculpted in the same way described in Sec.~\ref{sec:simfit_combo}. 
Due to similar and compatible misidentification rates and data-taking conditions between \runtwopo and \runtwopt, the determination of the misidentified background component model is obtained combining the predicted background events in these periods.
The resulting nominal misidentified background components are shown in Fig.~\ref{fig:templatePDFs}. 
An alternative approach based on kernel density estimates is used to assign a systematic uncertainty to the choice of model. 

The yields of these backgrounds in the nominal fit are constrained using the sum of the 
weighted entries in the histograms and the associated uncertainty.
The lepton identification requirements used to define the control region dataset, as well as
the threshold used to assign an event as pion- or kaon-like, are varied to compute systematic uncertainties.

\subsubsection{Specific backgrounds in \texorpdfstring{\boldmath{\Bp}}{B+} resonant modes}
Specific backgrounds in \Bp resonant modes are listed in Table~\ref{tab:BkgforfitsRK}.
The \mbox{\BuToPiJPsll} background is modeled using a DSCB function for which the parameters are obtained from uncalibrated simulation samples. The
\mbox{\BsToKstJPsll} background is described using an unbinned template obtained from simulation samples. 

Backgrounds from \BToXJPsmm and \BToXPsimm decays are modeled using unbinned templates obtained 
from fully calibrated simulation samples. Inclusive simulated samples of \Bs, \Bp, and \Bz decays
are used to construct the templates.
The relative normalization of these background samples is
fixed to the known relative production fractions and decay rates, while their
overall normalization freely varies in the fit.
An analogous procedure is followed in the case of \mbox{\BToXJPsee} and \mbox{\BToXPsiee}, with adjustments
for the significantly wider mass range used in the unconstrained \jpsi electron fits.

The \psitwos electron mode fits require two additional backgrounds to be modeled: 
the leakage of \BuToKJPs decays into the \psitwos~\qsq range, and partially reconstructed 
\decay{\Bu}{\Kp\psitwos(\to\jpsi(\decay{}{\epem}) X)}
decays. 
Both are modeled using unbinned templates obtained from fully 
calibrated simulation samples, and their normalizations are allowed to vary freely in the fit.

\subsubsection{Specific backgrounds in \texorpdfstring{\boldmath{\Bz}}{B0} resonant modes}
Specific backgrounds in \Bz resonant modes are listed in Table~\ref{tab:BkgforfitsRkst}.
Backgrounds from \Lb processes are corrected for the known inaccuracies in the \pythia  modeling 
of \Lb kinematics using the same correction factors as in Ref.~\cite{LHCb-PAPER-2019-040}. Corrections
are applied as a two-dimensional function of the \Lb transverse momentum and pseudorapidity, separately
for samples simulated at center-of-mass energies of 7, 8, and 13\tev. 
The \BToXJPsmm and \BToXPsimm backgrounds are modeled analogously to the procedure followed
in the \Bp resonant modes. Similarly, the \BdToKstJPs leakage in the \psitwos electron fits
and the background from partially reconstructed \decay{\Bd}{\Kstarz\psitwos(\to\jpsi(\epem) X)}
 decays are modeled 
analogously to the procedure followed in the \Bp resonant modes.

The \LbTopKJPsll background is described using unbinned templates obtained from fully
calibrated simulation samples. In addition to the correction of the \Lb kinematics, the \LbTopKJPsll simulated samples are also corrected for the amplitude 
structure of the \Lb decay measured in Ref.~\cite{LHCb-PAPER-2015-029}. The relative normalization
is constrained from its known decay rate~\cite{LHCb-PAPER-2015-032}, the measured \Lb production fraction~\cite{LHCb-PAPER-2018-050,LHCb-PAPER-2011-018}, 
and the selection efficiency measured using fully 
calibrated simulation samples. The same strategy is used for both muon and electron modes.

The \BsToKstJPsll background is described using the same PDF as the \jpsi control mode,
shifted to account for the known difference in \Bs and \Bz masses. 
The normalization of this background is allowed to vary freely in the fit, shared between the electron and muon mode.

The \BsToPhiJPsll background and backgrounds in which the kaon and pion from a genuine 
$\Bz\to \Kstarz \ellp\ellm$ decay are swapped are modeled using unbinned templates obtained from fully
calibrated simulation samples. 
The normalization of these backgrounds is constrained to their expectation. 

\subsection{Impact of correlations between data samples}
The invariant mass fit is used to extract directly the \RK and \RKst double ratios
by including the efficiencies obtained from the fully calibrated simulated samples. Statistical
uncertainties and their correlations between trigger categories
and the different final states are obtained from bootstrapping. The statistical uncertainties are 
uncorrelated between data-taking years.
The covariance matrix of 
systematic uncertainties associated with the efficiency determination, described in Sec.~\ref{sec:systematics}, is computed and added
to this covariance matrix of bootstrapping uncertainties in order to obtain the full constraints
on the efficiencies used as inputs to the fit. An analogous approach
is used to measure the \RJPsK and \RJPsKst resonant mode ratios, or the
\RPsiK and \RPsiKst resonant mode double ratios.

The data samples  selected in the different decay modes and \qsq ranges must be fully disjoint to obtain accurate uncertainties from the simultaneous fit. This is verified from data using the unique
event identifier assigned to each candidate. The signal mode samples are found to be fully disjoint.
The resonant mode samples in which the dielectron mass is constrained are found to contain a 
percent-level overlap, while the resonant mode samples in which the dielectron mass is not constrained
are found to have an overlap at the level of up to ten percent. The impact of this overlap on the 
reported uncertainties is evaluated and found to be negligible.
\section{Cross-checks}
\label{sec:crosschecks}
\subsection{Resonant mode decay rates}
\begin{figure}[t]
\centering
\includegraphics[width=0.88\linewidth]{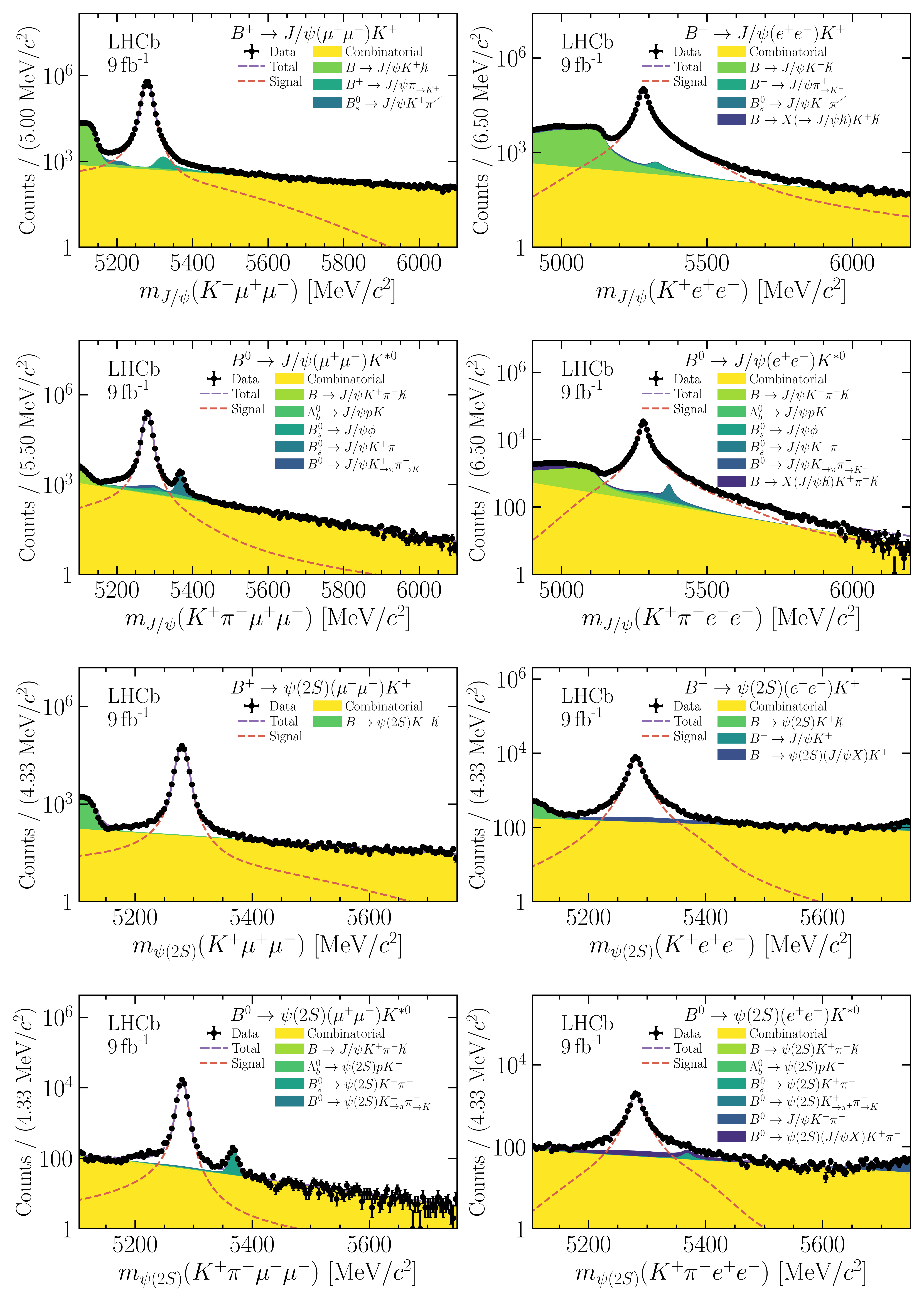}
\caption{\small Invariant mass fit to the resonant control modes, from top to bottom: \jpsi mode in \BuToKll, \jpsi mode in \BdToKstll, \psitwos mode in \BuToKll, \psitwos in \BdToKstll. 
The muon (electron) modes are given on the left (right).}
\label{fig:fit_result_cross_check}
\end{figure}

In common with previous \lhcb analyses of LU, measurements of the relative decay rates of the resonant modes, \RJPsKorKst and \RPsiKorKst, are used to validate the analysis procedure. 
The stability of \RJPsKorKst, measured as a function of
different kinematic and geometric properties of the decays, both validates
the analysis procedure and allows to quantify residual inaccuracies in the analysis chain 
and to assign systematic uncertainties. The compatibility of the \Bp and \Bz simulation
calibrations is demonstrated by performing all cross-checks using both calibration chains.
In addition, the cross-checks are repeated using the \tosinc trigger category, for which \tos is the primary trigger category and no requirements are imposed on the \tis classification of events.

The single ratios \RJPsK and \RJPsKst are sensitive to residual imperfections 
in the simulation of electron and muon mode efficiencies, as well as those  in the modeling
of the resonant modes in the invariant mass fit. These ratios are expected to be equal to unity
in the SM and have been determined precisely in previous measurements.  
Corrections can arise at the per mille level from the wider \qsq range in the electron mode, which could affect the decay rate due to subleading contributions from the FCNC process. 
 Agreement of the \RJPsK and \RJPsKst ratio with 
 predictions of the SM, compatibility between data-taking periods, trigger categories, and when computed with the \Bp or the \Bz simulation correction
 was a prerequisite to evaluating the \NRKorKst observables.
The ratios \RPsiK and \RPsiKst are also used to validate that residual
imperfections in the computation of efficiencies indeed cancel in the double muon-electron ratio.

\begin{figure}[!ht]
\centering
\includegraphics[width=1.0\linewidth]{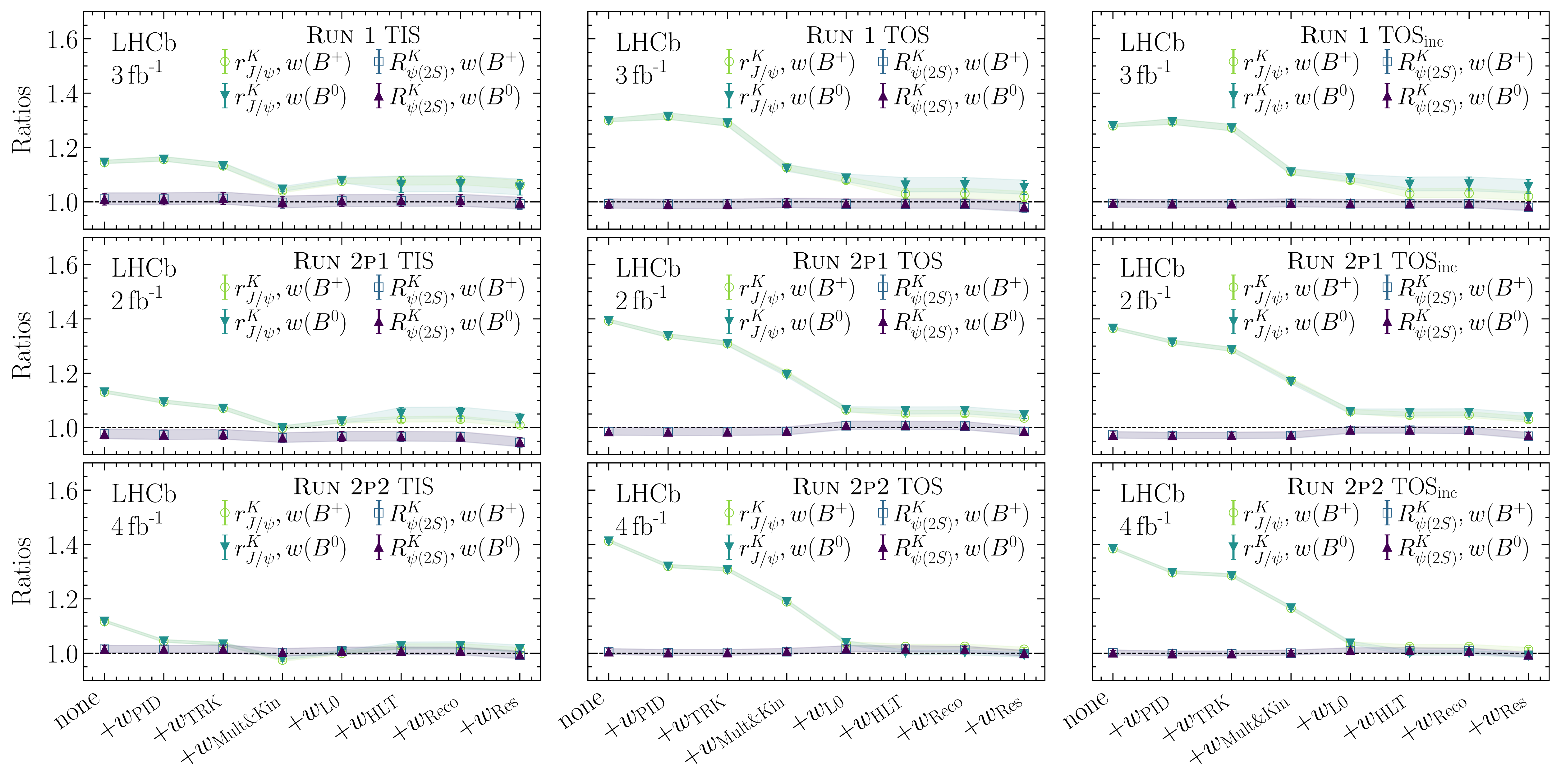}
\caption{\small Evolution of the \RJPsK single and \RPsiK 
double ratios with each step of the simulation calibration procedure as labeled on the $x$-axis.
The data-taking period and trigger category are indicated in the legend of each plot.}
\label{fig:cross_check_evolution_rk}
\centering
\includegraphics[width=1.0\linewidth]{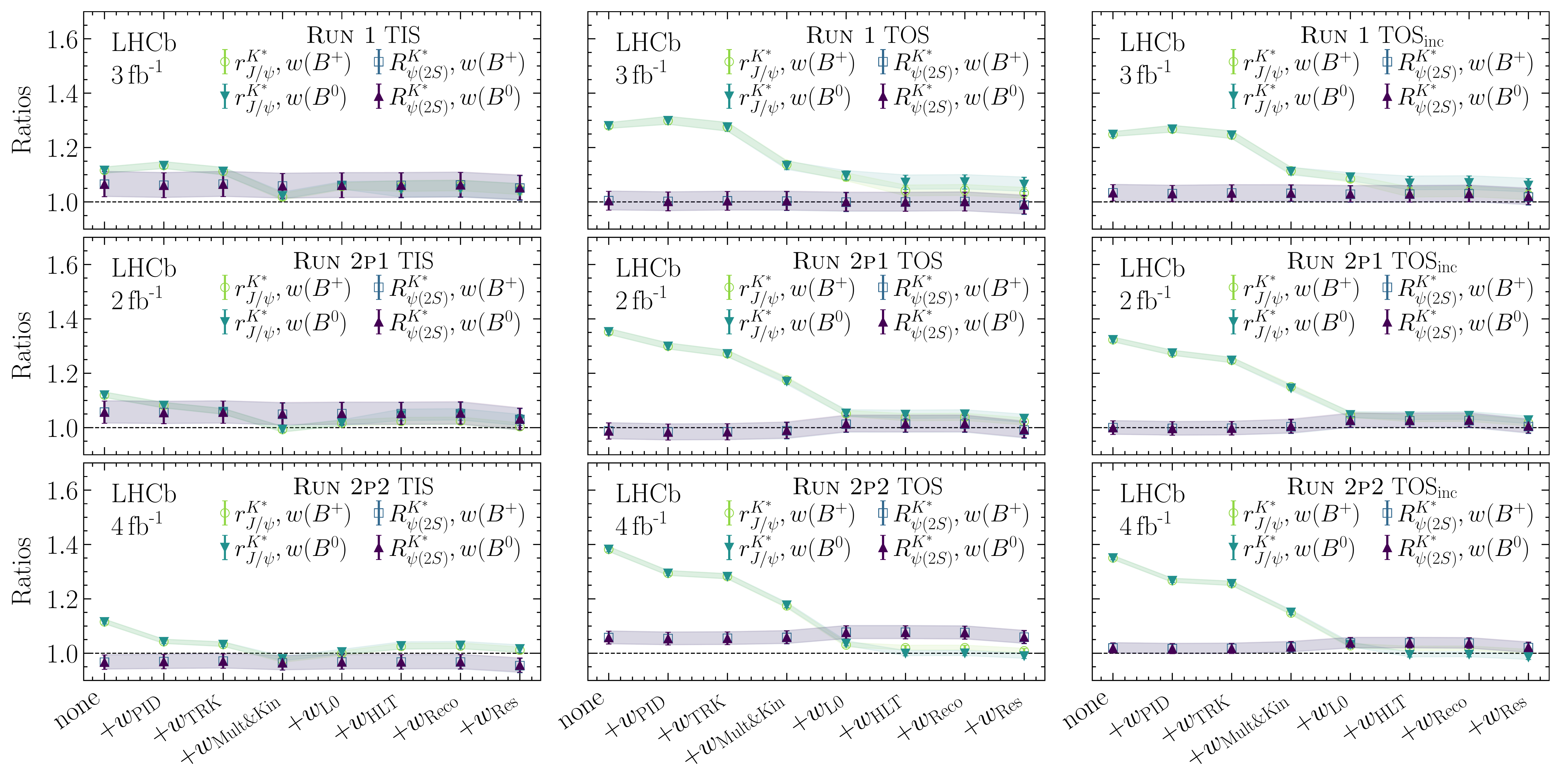}
\caption{\small Evolution of the \RJPsKst single and \RPsiKst 
double ratio with each step of the simulation calibration procedure 
as labeled on the $x$-axis. 
The data-taking period and trigger category are indicated in the legend of each plot.}
\label{fig:cross_check_evolution_rkst}
\end{figure}

All invariant mass fits are performed constraining the invariant mass of the dilepton system to 
the \jpsi or \psitwos mass, as appropriate, where $m_{J/\psi}$ and $m_{\psi(2S)}$ labels denote the application of constraints on the dilepton system. The fits to data are illustrated in Fig.~\ref{fig:fit_result_cross_check}. The cross-check results are presented 
in Table~\ref{tab:cross_check_ratios} and shown in Fig.~\ref{fig:cross_check_evolution_rk} and Fig.~\ref{fig:cross_check_evolution_rkst} for the \Bp and \Bd, respectively, where only systematic uncertainties associated to the simulation sample and calibration sample statistics are included. Figures~\ref{fig:cross_check_evolution_rk} and~\ref{fig:cross_check_evolution_rkst} also show the incremental effect of corrections to simulation in the determination of \RJPsKorKst.
The uncertainties are dominated by the bootstrapping uncertainty on the simulation calibrations. As expected, both the single and double ratios are compatible with unity in all cases.
The single ratio is incompatible with unity for the uncalibrated simulation, and its compatibility improves
gradually as each calibration is applied.
In contrast, the double ratio is compatible with unity from the outset
and is practically unaffected by the calibrations applied to simulation.

\begin{table}[!ht]
\centering
\caption{Values of the \RJPsK and \RJPsKst single ratios, as well as \RPsiK and \RPsiKst double
ratios, calculated in different data-taking
periods, trigger categories, and using the $w(\Bp)$ or $w(\Bz)$ calibration chains. The three
uncertainties are, from left to right: statistical from the invariant mass fits, statistical from 
the finite simulated sample sizes, and the bootstrapping uncertainty on the simulation calibrations.}
\label{tab:cross_check_ratios}

\resizebox{0.95\textwidth}{!}{
\begin{tabular}{lcc}
Sample &  \RJPsK &  \RJPsKst \\
\hline
 \runonetable \tistable  $w(\Bp)$   &  $1.063 \pm 0.005 \pm 0.003 \pm 0.015$ & $1.046 \pm 0.010 \pm 0.004 \pm 0.016$ \\
 \runonetable \tistable  $w(\Bz)$   &  $1.054 \pm 0.005 \pm 0.003 \pm 0.028$ & $1.038 \pm 0.010 \pm 0.004 \pm 0.027$ \\
 \runonetable \tostable  $w(\Bp)$   &  $1.020 \pm 0.004 \pm 0.003 \pm 0.017$ & $1.033 \pm 0.008 \pm 0.004 \pm 0.018$ \\
 \runonetable \tostable  $w(\Bz)$   &  $1.053 \pm 0.004 \pm 0.003 \pm 0.025$ & $1.065 \pm 0.008 \pm 0.004 \pm 0.025$ \\
 \runonetable \tosinctable  $w(\Bp)$   &  $1.021 \pm 0.004 \pm 0.002 \pm 0.016$ & $1.026 \pm 0.007 \pm 0.003 \pm 0.017$ \\
 \runonetable \tosinctable  $w(\Bz)$   &  $1.056 \pm 0.004 \pm 0.002 \pm 0.025$ & $1.061 \pm 0.007 \pm 0.003 \pm 0.024$ \\
\hline
 \runtwopo \tistable  $w(\Bp)$ & $1.010 \pm 0.004 \pm 0.003 \pm 0.009$ & $1.003 \pm 0.008 \pm 0.004 \pm 0.010$ \\
 \runtwopo \tistable  $w(\Bz)$ & $1.033 \pm 0.004 \pm 0.003 \pm 0.019$ & $1.028 \pm 0.008 \pm 0.004 \pm 0.018$ \\
 \runtwopo \tostable  $w(\Bp)$ & $1.035 \pm 0.004 \pm 0.003 \pm 0.010$ & $1.022 \pm 0.007 \pm 0.005 \pm 0.010$ \\
 \runtwopo \tostable  $w(\Bz)$ & $1.046 \pm 0.004 \pm 0.003 \pm 0.012$ & $1.033 \pm 0.007 \pm 0.005 \pm 0.012$ \\
 \runtwopo \tosinctable  $w(\Bp)$ & $1.030 \pm 0.003 \pm 0.002 \pm 0.010$ & $1.017 \pm 0.006 \pm 0.004 \pm 0.010$ \\
 \runtwopo \tosinctable  $w(\Bz)$ & $1.039 \pm 0.003 \pm 0.002 \pm 0.012$ & $1.028 \pm 0.006 \pm 0.004 \pm 0.012$ \\
\hline
 \runtwopt \tistable  $w(\Bp)$ & $1.012 \pm 0.003 \pm 0.003 \pm 0.007$ & $1.011 \pm 0.006 \pm 0.005 \pm 0.007$ \\
 \runtwopt \tistable  $w(\Bz)$ & $1.016 \pm 0.003 \pm 0.003 \pm 0.012$ & $1.016 \pm 0.006 \pm 0.005 \pm 0.011$ \\
 \runtwopt \tostable  $w(\Bp)$ & $1.014 \pm 0.003 \pm 0.003 \pm 0.006$ & $1.009 \pm 0.005 \pm 0.006 \pm 0.004$ \\
 \runtwopt \tostable  $w(\Bz)$ & $0.993 \pm 0.003 \pm 0.003 \pm 0.007$ & $0.990 \pm 0.005 \pm 0.006 \pm 0.006$ \\
 \runtwopt \tosinctable  $w(\Bp)$ & $1.014 \pm 0.002 \pm 0.003 \pm 0.006$ & $1.006 \pm 0.004 \pm 0.005 \pm 0.005$ \\
 \runtwopt \tosinctable  $w(\Bz)$ & $0.991 \pm 0.002 \pm 0.003 \pm 0.007$ & $0.985 \pm 0.004 \pm 0.005 \pm 0.007$ \\
\hline
\phantom{-}\\
Sample  &  \RPsiK &  \RPsiKst \\
\hline
 \runonetable   \tistable  $w(\Bp)$   & $0.993 \pm 0.021 \pm 0.005 \pm 0.001$  & $1.051 \pm 0.044 \pm 0.009 \pm 0.002$ \\
 \runonetable   \tistable  $w(\Bz)$   & $0.996 \pm 0.021 \pm 0.005 \pm 0.001$  & $1.053 \pm 0.044 \pm 0.009 \pm 0.002$ \\
 \runonetable  \tostable  $w(\Bp)$    & $0.979 \pm 0.016 \pm 0.004 \pm 0.002$  & $0.988 \pm 0.033 \pm 0.007 \pm 0.002$ \\
 \runonetable  \tostable  $w(\Bz)$    & $0.982 \pm 0.016 \pm 0.004 \pm 0.003$  & $0.990 \pm 0.033 \pm 0.007 \pm 0.004$ \\
 \runonetable  \tosinctable  $w(\Bp)$      & $0.980 \pm 0.014 \pm 0.003 \pm 0.001$  & $1.018 \pm 0.029 \pm 0.006 \pm 0.003$ \\
 \runonetable  \tosinctable  $w(\Bz)$      & $0.983 \pm 0.014 \pm 0.003 \pm 0.002$  & $1.020 \pm 0.029 \pm 0.006 \pm 0.003$ \\
\hline  
 \runtwopo   \tistable  $w(\Bp)$  & $0.945 \pm 0.017 \pm 0.004 \pm 0.001$  & $1.030 \pm 0.039 \pm 0.008 \pm 0.002$ \\
 \runtwopo   \tistable  $w(\Bz)$  & $0.947 \pm 0.017 \pm 0.004 \pm 0.001$  & $1.032 \pm 0.039 \pm 0.008 \pm 0.002$ \\
 \runtwopo  \tostable  $w(\Bp)$  & $0.986 \pm 0.014 \pm 0.003 \pm 0.003$  & $0.991 \pm 0.029 \pm 0.006 \pm 0.004$ \\
 \runtwopo  \tostable  $w(\Bz)$  & $0.987 \pm 0.014 \pm 0.003 \pm 0.003$  & $0.993 \pm 0.029 \pm 0.006 \pm 0.005$ \\
 \runtwopo  \tosinctable  $w(\Bp)$  & $0.969 \pm 0.012 \pm 0.003 \pm 0.002$  & $1.004 \pm 0.025 \pm 0.006 \pm 0.003$ \\
 \runtwopo  \tosinctable  $w(\Bz)$  & $0.970 \pm 0.012 \pm 0.003 \pm 0.002$  & $1.006 \pm 0.025 \pm 0.006 \pm 0.004$ \\
\hline  
 \runtwopt  \tistable  $w(\Bp)$  & $0.992 \pm 0.013 \pm 0.004 \pm 0.001$  & $0.954 \pm 0.025 \pm 0.006 \pm 0.001$ \\
 \runtwopt  \tistable  $w(\Bz)$  & $0.994 \pm 0.013 \pm 0.004 \pm 0.001$  & $0.956 \pm 0.025 \pm 0.006 \pm 0.001$ \\
 \runtwopt \tostable  $w(\Bp)$  & $0.999 \pm 0.010 \pm 0.003 \pm 0.002$  & $1.059 \pm 0.023 \pm 0.006 \pm 0.002$ \\
 \runtwopt \tostable  $w(\Bz)$  & $1.000 \pm 0.010 \pm 0.003 \pm 0.002$  & $1.060 \pm 0.023 \pm 0.006 \pm 0.002$ \\
 \runtwopt \tosinctable  $w(\Bp)$  & $0.993 \pm 0.009 \pm 0.003 \pm 0.001$  & $1.020 \pm 0.018 \pm 0.005 \pm 0.002$ \\
 \runtwopt \tosinctable  $w(\Bz)$  & $0.994 \pm 0.009 \pm 0.003 \pm 0.001$  & $1.022 \pm 0.018 \pm 0.005 \pm 0.002$ \\
\end{tabular}
}
\end{table}
\FloatBarrier

\begin{figure}[!htb]
\centering
\includegraphics[width=0.9\linewidth]{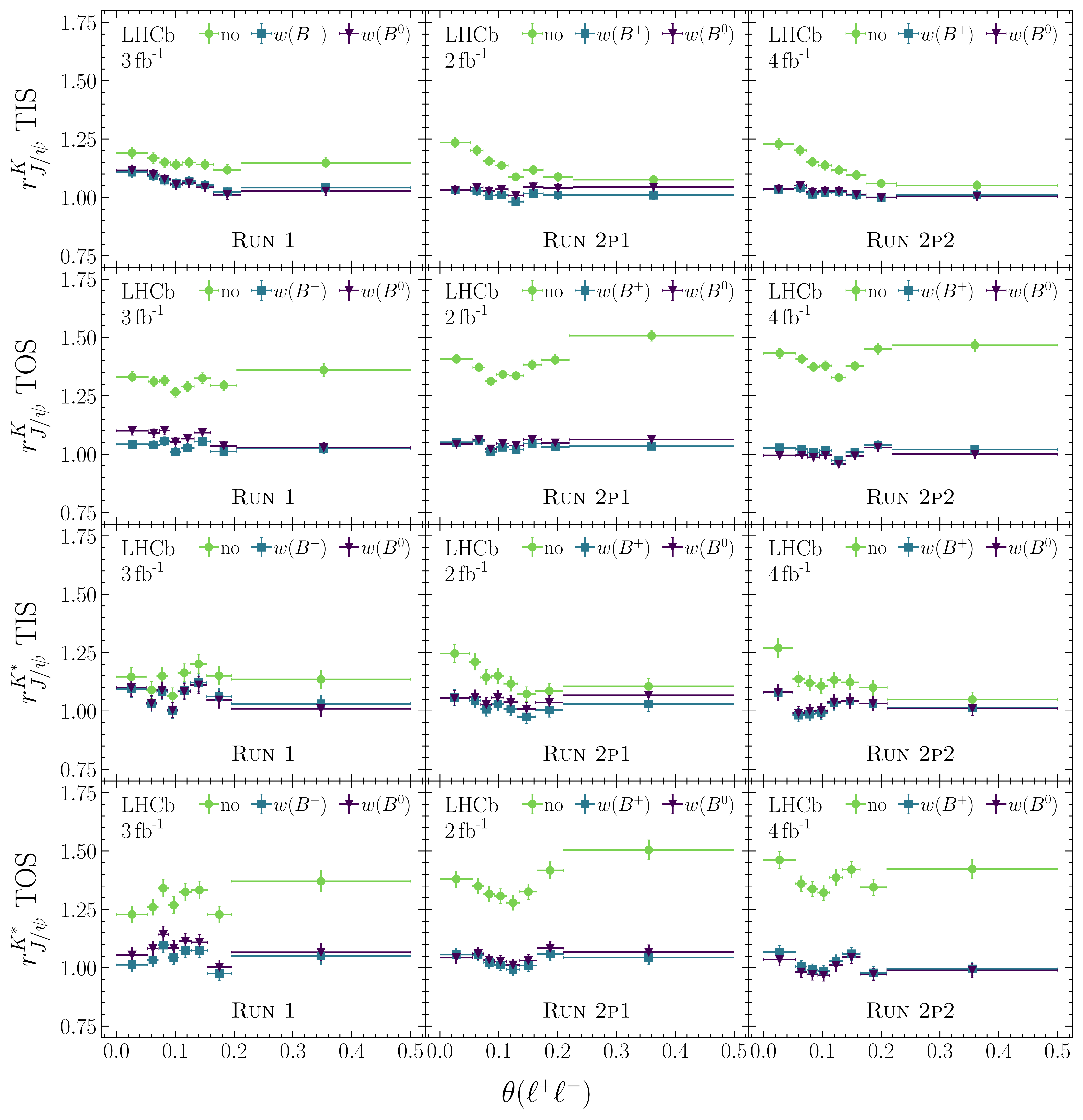}
\caption{\small Values of the \RJPsK and \RJPsKst single ratios as a function of the 
dilepton opening angle $\theta(\ellell)$. From top to bottom: \RJPsK TIS, \RJPsK TOS,
\RJPsKst TIS, and \RJPsKst TOS. From left to right: the \runone, \runtwopo and \runtwopt
data-taking periods. The ratios are shown without simulation calibrations, 
with \Bp calibrations, and with \Bz calibrations.\label{fig:cross_check_flatness}}
\end{figure}

\FloatBarrier
The stability of the single ratios \RJPsK and \RJPsKst is tested by repeating the single-ratio
cross-check as a function of 44 quantities related to the kinematics, geometry, or vertex
quality of the decay, as well as to the event occupancy. For each quantity, the data are divided into eight intervals, each with comparable statistical precision on the single ratios. If the simulation is perfectly calibrated, 
the dependence on each quantity will be compatible with a straight line with slope zero and intercept one.
Residual imperfections do not necessarily indicate a bias in the LU
observables as long as the underlying distribution of the quantity in question is similar
between the low-, central-, and $\jpsi$-$\qsq$ regions. Figure~\ref{fig:cross_check_flatness} shows the
stability of \RJPsK and \RJPsKst as a function of the dilepton opening angle $\theta(\ellell)$, one of the
quantities whose distribution is most different between the low-, central-, and \jpsi-\qsq 
regions. The potential for small residual bias to be reflected on the LU observables is evaluated and discussed in Section~\ref{sec:systematics}.

\begin{figure}[!ht]
\centering
\includegraphics[width=0.49\linewidth]{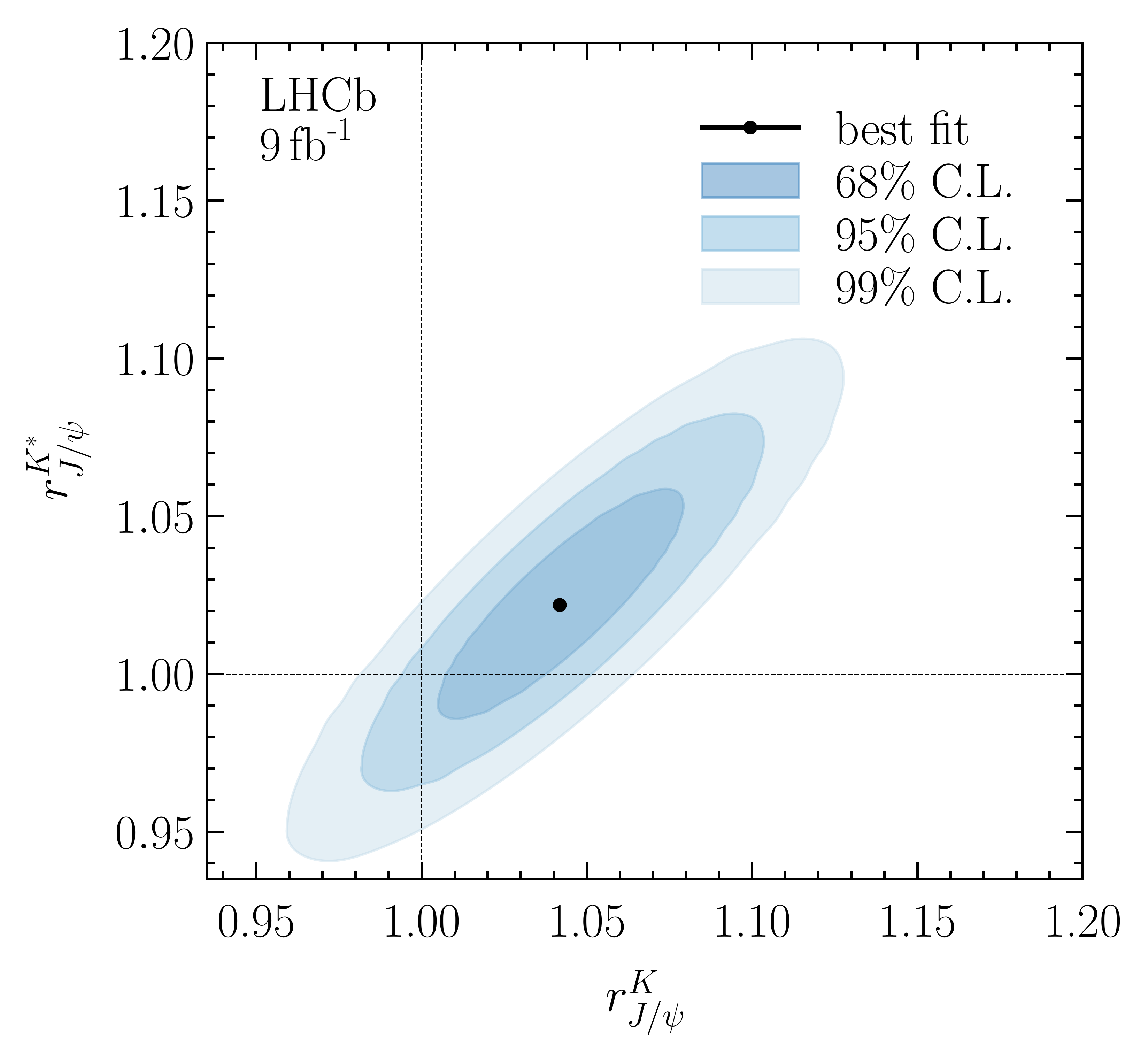}
\includegraphics[width=0.49\linewidth]{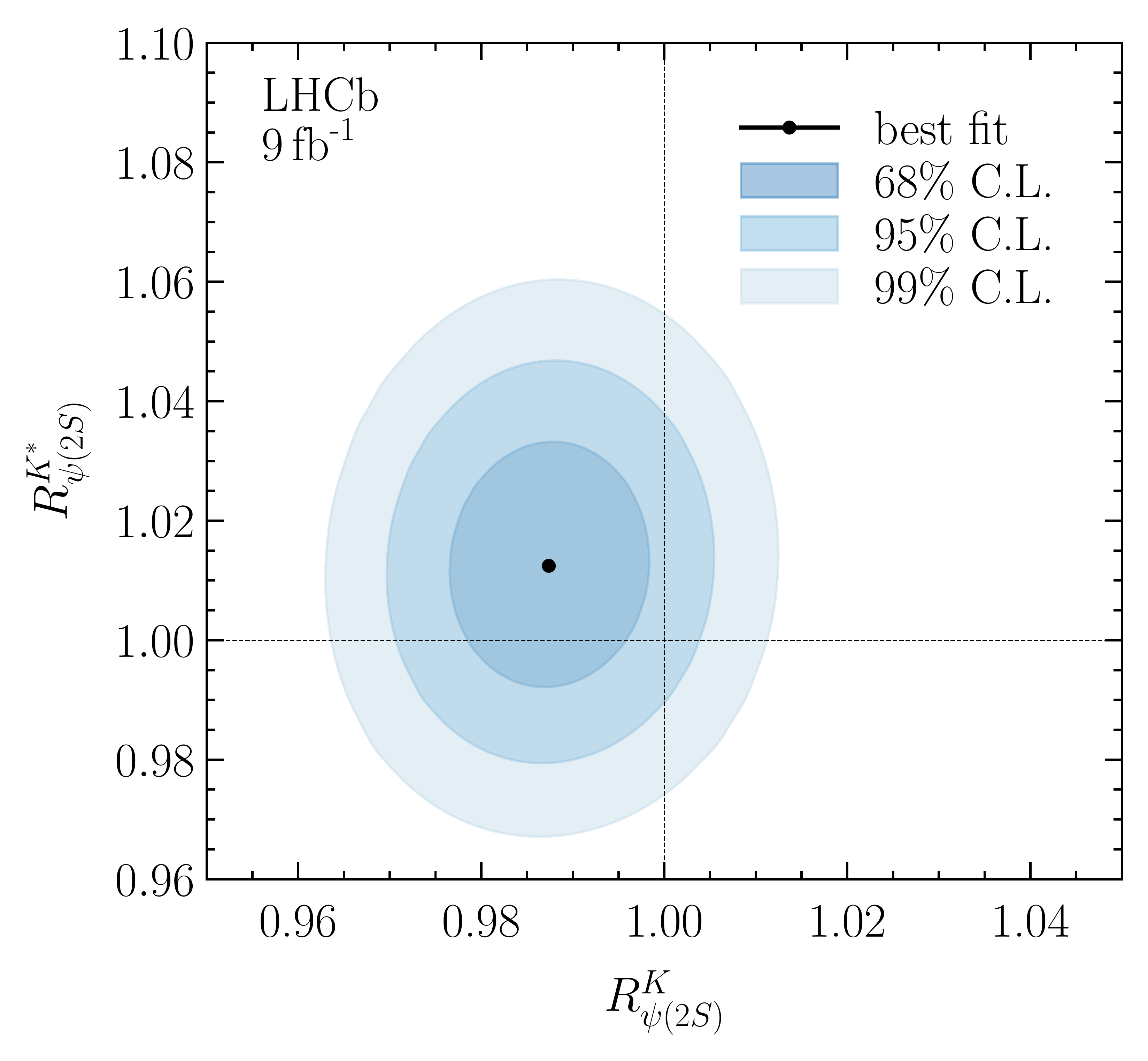}
\caption{\small Two dimensional likelihood scans of (left) \RJPsK \vs \RJPsKst and (right) \RPsiK \vs \RPsiKst. The contours show the 68\%, 95\% and 99\% C.L.\ regions and the solid markers show the best fit values.}
\label{fig:rJPsi_rPsi2S_2D_LL_Scan}
\end{figure}

Finally the \RJPsK and \RJPsKst single ratios, as well as \RPsiK and \RPsiKst double
ratios, are computed including all relevant systematic uncertainties described in
Sec.~\ref{sec:systematics}. The two-dimensional likelihood scans are shown in Fig.~\ref{fig:rJPsi_rPsi2S_2D_LL_Scan}.
For likelihood scans of \RPsiK, \RPsiKst, no systematic uncertainties on the fit model are included.
Both the single and double ratios agree with the Standard Model predictions at better
than two standard deviations.

\FloatBarrier
\subsection{Stability of results with respect to PID criteria}

\begin{figure}[tbp]
\centering
\includegraphics[width=1.0\linewidth]{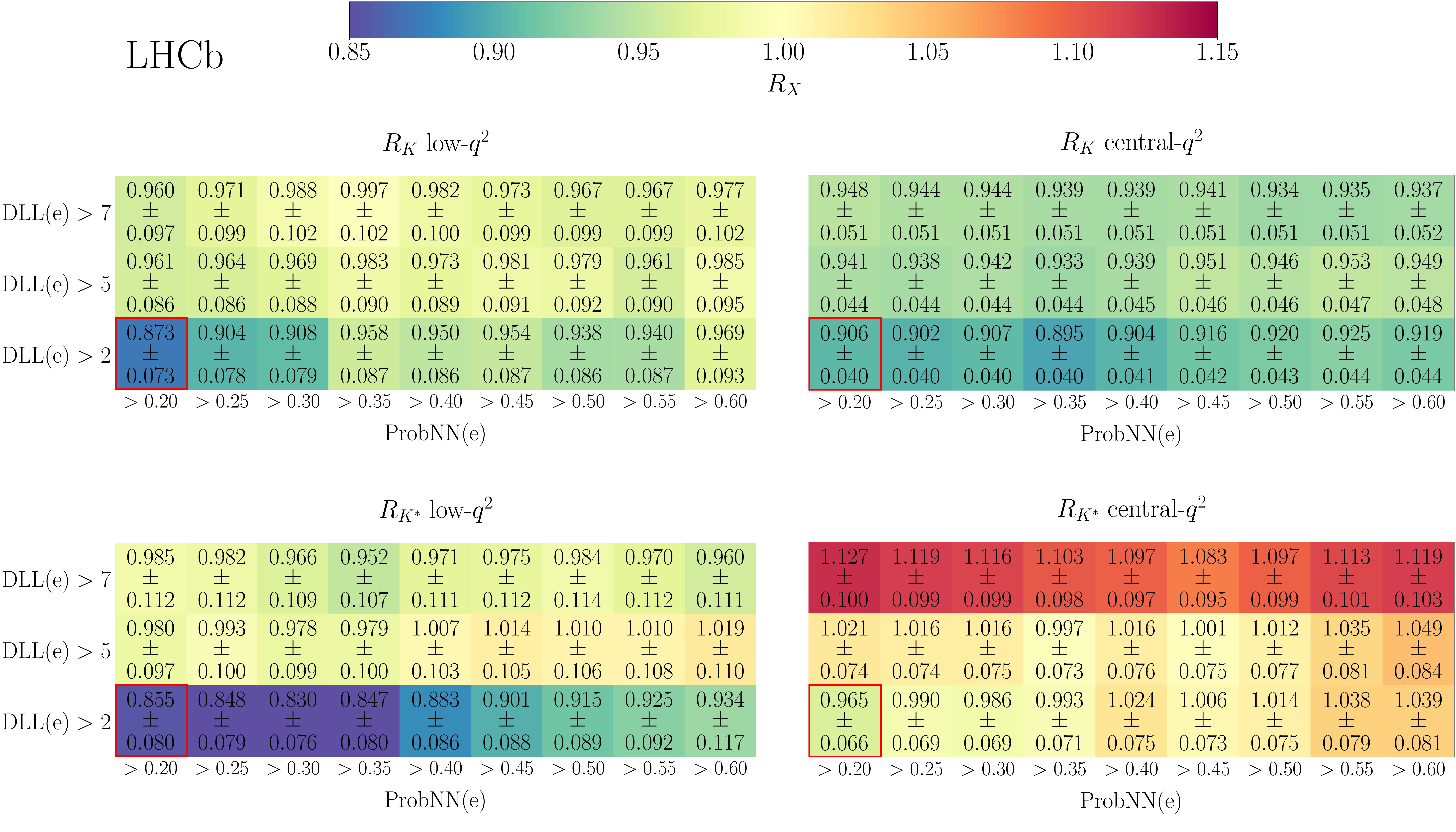}
\caption{\small Results of the nominal fit without modeling of misidentified backgrounds as a function of the PID criteria used. The bins are, from top left to bottom right: \RK low-\qsq, \RK \cqsq, \RKst \lqsq and  \RKst \cqsq. The nominal set of criteria is highlighted in red. The quoted uncertainties are statistical only.}
\label{fig:pid_scan}
\end{figure}

To check the modeling of misidentified backgrounds, the nominal fit is performed without including these backgrounds.
The fit is then repeated, progressively tightening the PID criteria. The results are shown in Fig.~\ref{fig:pid_scan}.
The clear trends observed when loosening PID  criteria 
demonstrate the importance of including these backgrounds in the nominal fit. Past a certain
point, however, the fit results plateau in all four LU observables. When all uncertainties
are taken into account, the fit values in this plateau region are fully compatible with the nominal 
fit result discussed in Sec.~\ref{sec:results}, where the misidentified backgrounds are explicitly modeled.

\begin{figure}[tbp]
\centering
\includegraphics[width=0.75\linewidth]{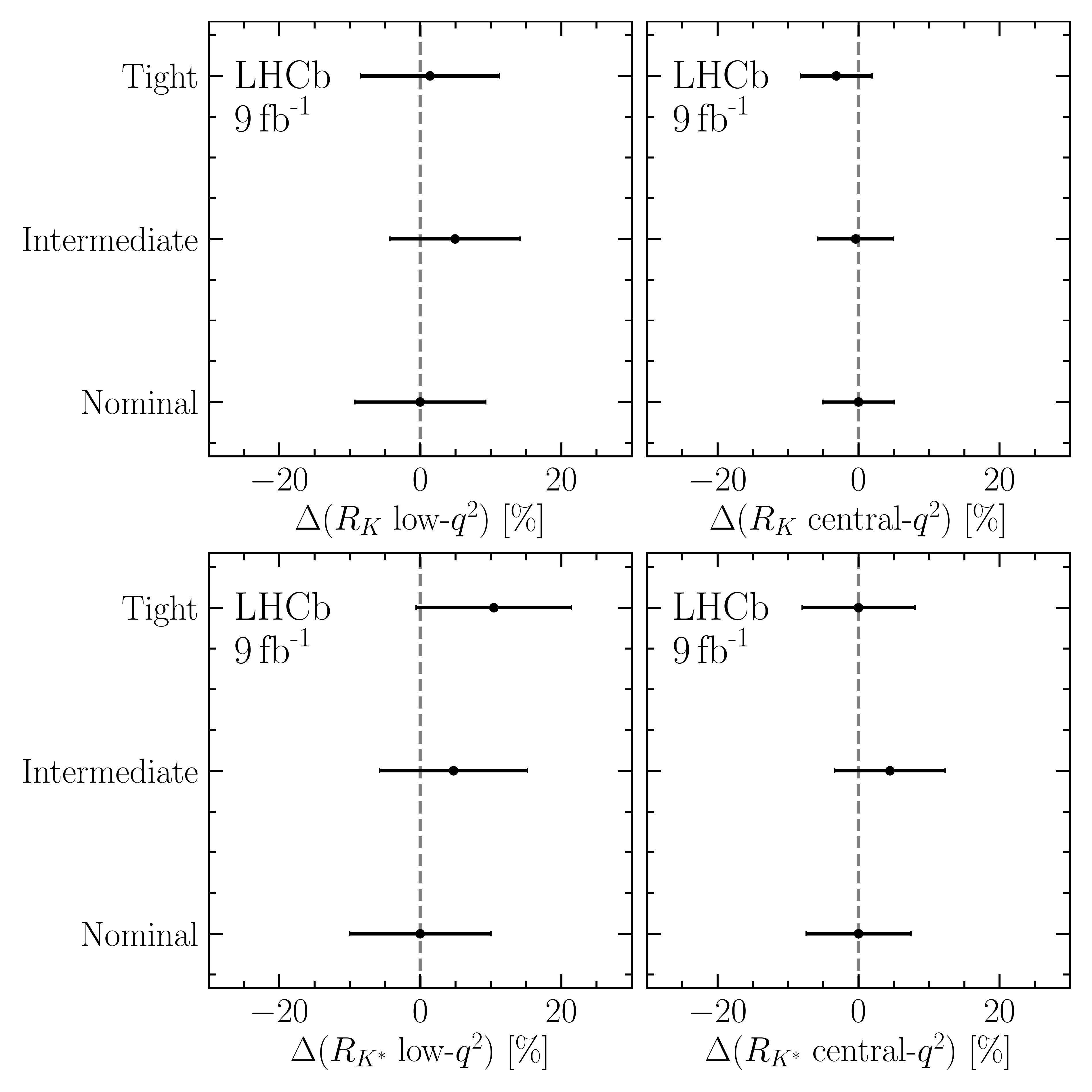}
\caption{\small Shifts of the central value results from varying the PID  criteria on electrons while modeling misidentified backgrounds in the fit. Particle selection criteria are varied to an intermediate working point reducing the expected contamination by a factor two, and to a tighter working point reducing the contamination by more than 75\%. The bins are, from top left to right: \RK low-\qsq, \RK \cqsq, \RKst \lqsq and  \RKst \cqsq. The quoted relative uncertainties are statistical only.}
\label{fig:pid_scan_variation_includemodel}
\end{figure}

The same procedure is repeated at two working points while including the modeling of the backgrounds in the fit model. 
The results are shown in Fig.~\ref{fig:pid_scan_variation_includemodel}. Here the ``intermediate'' working point is 
$\textrm{DLL(e)}>3$ and $\textrm{ProbNN(e)}>0.4$, while the ``tight'' working point is $\textrm{DLL(e)}>5$ and $\textrm{ProbNN(e)}>0.5$,
for comparison to Fig.~\ref{fig:pid_scan}. The overall expected contamination from misidentified backgrounds at the intermediate working point
is half of the contamination at the nominal working point, while the contamination at the tight working point is
expected to be nearly negligible. No trends are observed, giving confidence in the extrapolation and modeling of misidentified backgrounds in the fit.

\subsection{Study of \texorpdfstring{\boldmath{\BdToKstee}}{B0 -> K*0 e+e-} at very low-\texorpdfstring{\qsq}{q2}}
As an additional test of the portability of efficiencies from the \jpsi region to other \qsq regions,
the branching fraction of the \BdToKstee mode normalized to the \Bd resonant mode is measured~\cite{Lemettais:2826428}
in the very \lqsq region of $[0.0001,0.1]\gevgevcccc$. Since there are practically no relevant hadronic backgrounds in
this \qsq region, this cross-check also further tests our understanding of misidentified backgrounds in the nominal analysis.
The selection criteria and efficiency determination are the same as for the rest of the analysis. 
The measured branching fraction is determined to be equal to $(1.57\pm 0.12)\cdot 10^{-7}$, where the uncertainty includes only the statistical component. The result agrees perfectly with the SM prediction which has been evaluated multiplying the known world best average branching ratio of \BdToKstG~\cite{PDG2022} to the ratio of decay rates of \BdToKstee in the very \lqsq and \BdToKstG. The latter has been evaluated using the \texttt{flavio} package~\cite{Straub:2018kue}. Moreover no significant trends in the efficiency corrected yields are 
observed when varying the PID requirements.
\section{Systematic uncertainties}
\label{sec:systematics}
Systematic uncertainties can be divided into two broad categories: those associated with the determination
of the signal and control mode efficiencies that directly enter
Eq.~\ref{eq:doubleratiorx}
and those associated with the simultaneous invariant mass fit. 
Bootstrapping uncertainties associated with the nominal calibration procedure are
described in Sec.~\ref{sec:effs} and are
considered separately from effects discussed further in this section.

Systematic uncertainties associated with efficiencies are generally determined by varying 
assumptions made in the calibration of simulated samples and measuring the corresponding 
shifts in \RK and \RKst. Systematic uncertainties associated with the fit model are generally 
determined by generating large ensembles of pseudoexperiments,
varying assumptions made in the fit procedure, and measuring the corresponding shifts in \RK and \RKst
between the nominal and varied fit configuration.
In both cases correlations are inferred from observing the coherence of the measured shifts.
All systematic uncertainties are assumed to follow Gaussian distributions and are  evaluated separately
for each LU observable, data-taking period and trigger category.
The correlations between observables, data-taking periods, and trigger categories are
also evaluated for each source of systematic uncertainty. The final outcome is a
$24\times 24$ covariance matrix that can be used as an additional constraint in the 
simultaneous fit to calculate the likelihood for each observable including both statistical 
and systematic uncertainties.

\subsection{Systematic uncertainties on efficiencies}
\subsubsection*{Truth-level information}
The nominal analysis procedure associates reconstructed candidates in simulated events
to the underlying ``truth-level'' information which describes which generated particles
left hits in the LHCb detector. This association is used to filter out misreconstructed
candidates and ensures that each PDF constructed from simulated events represents only the
decay mode of interest for that PDF. The association criteria are varied and the efficiencies
recomputed. 
\subsubsection*{Multiple candidates}
The results for \RJPsK and \RJPsKst are recomputed keeping all
candidates for each event, rather than selecting a single candidate at random. All
deviations are found to be compatible with zero and no systematic 
uncertainty is therefore assigned.
\subsubsection*{Form factors used in simulation}
The simulated samples used in this analysis are generated for \Bp and 
\Bz decays according to the form factor model given in  Ref.~\cite{Ball:2004rg}. These form factors affect both the signal efficiencies and  the migration of 
events between \qsq regions. 
The associated systematic uncertainty is evaluated by deriving differential decay rates across \qsq and the decay angles defined as in Ref.~\cite{LHCb:2013zuf}, where only the efficiency dependence on the decay angle describing the lepton system is considered. The nominal form factors of Ref.~\cite{Ball:2004rg} are then compared with those of  Refs.~\cite{Bharucha:2015bzk} and \cite{Gubernari:2018wyi}, for the \Bd and \Bp decays respectively, by multiplying the resulting differential decay rate with the relevant efficiency distribution.
The theory uncertainty on the differential decay rate for the form factors taken from Refs.~\cite{Bharucha:2015bzk,Gubernari:2018wyi}, is also propagated to the efficiency ratios, but the resulting deviation is found to be generally smaller than that due to the difference in central values between the two models. 

\subsubsection*{Particle identification efficiencies}
Systematic uncertainties associated with PID  efficiencies arise from two
sources: residual non-factorization of the electron efficiencies, and the binning scheme used
to compute the PID  efficiencies on data and simulation. The first effect is quantified by comparing efficiencies obtained from truth-level
information with efficiencies obtained using the nominal calibration procedure on
simulated signal samples. The binning scheme systematic is evaluated separately for muons,
hadrons, and electrons. In the case of muons and hadrons a kernel density estimator
is used to provide an unbinned efficiency parametrization in momentum and pseudorapidity,
while the number of track multiplicity bins is varied. This is possible because
the muon and hadron calibrations are derived from high-purity background-subtracted
samples of charm hadron and charmonia decays, and the weights used to subtract background
also allow a per-event efficiency to be determined.

Dielectron calibration samples 
have a lower purity because of bremsstrahlung, which also introduces correlations between the reconstructed dielectron
mass and the properties of its constituent electrons, including their probability to
pass a given PID criterion. Their efficiencies therefore have to be calculated
using a fit-and-count approach in the defined binning scheme. For this reason no per-event
background-subtracted efficiency can be determined and consequently no unbinned parametrization 
is possible. The systematic uncertainty is therefore derived by interpolating the binned efficiency 
maps and measuring the difference in efficiencies between this interpolated parametrization and 
the binned maps. The \pt, pseudorapidity, and track
multiplicity binning schemes are also varied. The factorization- and the binning scheme-effects are
assumed to be uncorrelated when determining the overall systematic uncertainty.

\subsubsection*{Kinematic and multiplicity calibration}
The \wmco weights are re-evaluated in two ways: first using events in the TIS trigger category
rather than the nominal approach of using events in the \lone muon TOS category,
and second using as multiplicity proxy the number of tracks reconstructed in the vertex detector, rather than in  the whole tracking system.

\subsubsection*{Trigger efficiencies}
Systematic uncertainties on the \lone efficiencies are associated with the binning scheme, 
the use of muon mode TIS efficiencies as a proxy for the electron mode, and the factorization 
of electron TOS efficiencies. The binning systematic uncertainty is evaluated by measuring
the difference in efficiencies between an interpolated parametrization and the binned maps.
The TIS efficiencies are computed for the electron mode, and compared to the proxy efficiencies
obtained from the muon mode. The factorization systematic uncertainty is evaluated by directly measuring the
\lone dielectron TOS efficiencies instead of multiplying the nominal per-electron efficiencies.

Systematic uncertainties on the \hlt efficiencies are associated with the binning scheme and
the decision to parametrize the efficiency as a function of track multiplicity. The potential systematic uncertainty on the  \hlt efficiencies are estimated by parametrizing in terms of the \B hadron transverse momentum instead of track multiplicity and varying the binning scheme.

\subsubsection*{Stability of \texorpdfstring{\boldmath{\RJPsK}}{rKJ/psi} and \texorpdfstring{\boldmath{\RJPsKst}}{rK*J/psi}}
The fact that the single ratios \RJPsK and \RJPsKst are not perfectly flat when evaluated as
a function of the properties of the corresponding \jpsi control mode decay implies the presence
of residual imperfections in the calibration of simulated samples. The corresponding systematic
uncertainties on \RK and \RKst are quantified with a flatness parameter, $d_{f}$, defined as
\begin{equation}
    d_{f} = \left( \frac{\sum^{8}_{i}\varepsilon_{\text{Rare},\mu}^{i}\cdot\mathcal{Y}^{i}_{\mu}}{\sum^{8}_{i}\varepsilon_{\text{Rare},\mu}^{i}}
            \cdot 
            \frac{\sum^{8}_{i}\varepsilon_{\jpsi,\mu}^{i}}{\sum^{8}_{i}N^{i}_{\mu}}~\bigg/~
            \frac{\sum^{8}_{i}\varepsilon_{\text{Rare},e}^{i}\cdot\mathcal{Y}^{i}_{e}}{\sum^{8}_{i}\varepsilon_{\text{Rare},e}^{i}}
            \cdot
            \frac{\sum^{8}_{i}\varepsilon_{\jpsi,e}^{i}}{\sum^{8}_{i}N^{i}_{e}} \right) - 1,
\label{equation:cross_check:flatness}
\end{equation}
where $\varepsilon^i_{\text{Rare},\ell}$ and $\epsilon^i_{\jpsi,\ell}$ refer to the signal and control mode efficiencies in
bin $i$ respectively; $N^{i}_{\ell}$ denotes the control mode yield measured in bin $i$; and $\mathcal{Y}^{i}_{\ell}$ is the efficiency corrected control mode yield in bin $i$,
\begin{equation}
\mathcal{Y}^{i}_{\ell} = \frac{N^{i}_{\ell}}{\varepsilon_{\jpsi,\ell}^{i}}.
\end{equation}
The $d_{f}$ parameter can be considered a proxy for the double ratios \RK and \RKst in which the signal
mode yields are replaced by the \jpsi mode yields. The $d_{f}$ parameter is evaluated for each
of the 44 quantities used to describe the \jpsi control mode decay. The vast majority of the 
quantities considered result in $d_{f}$ values of a few per mille that are compatible with zero. 
The two quantities that show the greatest deviations 
from zero are the dilepton opening angle and the impact parameter $\chisq$ of the dilepton system. The dilepton opening angle $d_{f}$ values are larger
than those computed from the impact parameter $\chisq$ of the dilepton system in all cases, 
and are consequently used to define the resulting systematic uncertainty.

\subsection{Systematic uncertainties on the invariant mass fit}
\subsubsection*{\boldmath{\jpsi} mode fit model}
The fidelity of the \jpsi mode fit model is limited by the knowledge of the numerous
partially reconstructed backgrounds that contribute to the region below the nominal \B mass.
This is a particular problem for the \jpsi mode because of partially reconstructed 
backgrounds, such as those with a $\psitwos\to\jpsi\pi\pi$ decay chain, which have no analogue in the signal mode.
Although partially reconstructed backgrounds have missing energy and should therefore be located 
only below the nominal \B mass, poorly reconstructed candidates or candidates with wrongly
associated bremsstrahlung photons cause a long tail towards higher \B masses. Since the statistical
sensitivity of the fit, as seen from Table~\ref{tab:cross_check_ratios}, is at the few per mille level,
even contributions with mismodeling below the percent level can lead to a significant systematic
uncertainty. It is particularly important to evaluate this systematic uncertainty with care because the fit is fully correlated between the low- and central-\qsq measurements of \RK and \RKst. 
The results of four fits to the invariant mass of the \jpsi\ mode are compared: 
\begin{enumerate}
    \item The nominal fits used for the measurement of the \RK and \RKst double ratios, without any constraint on the dilepton invariant mass;
    \item Fits without a constraint on the dilepton invariant mass in which the partially reconstructed
    backgrounds in the electron mode are minimized by requiring that the \B candidate invariant mass is greater than 5200\mevcc when the dilepton invariant mass is constrained to the \jpsi mass;
    \item The fits with a constraint on the dilepton invariant mass used for all the results given in Sec.~\ref{sec:crosschecks};
    \item The same as 3.\ but extending the lower fit range of the electron mode to 4650\mevcc, in order to test the sculpting of the partially reconstructed background PDFs induced by the 6\gevgevcccc
    lower limit on the dielectron \qsq;
\end{enumerate}
These fits are grouped into two categories in order to assign a systematic uncertainty. Differences
between 1.\ and 3.\ probe uncertainties related to imperfect signal modeling, to the choice of fit range, and to residual partially reconstructed backgrounds which peak under the signal when the dilepton mass is constrained but not otherwise. Differences between 2.\ and 4.\ probe uncertainties
due to the imperfect composition of the partially reconstructed background cocktails. 
These two differences are added together in quadrature to obtain a total systematic uncertainty for the modeling and fitting of the \jpsi mode. These differences are also taken to accommodate uncertainties
associated with the finite simulated samples used to derive background PDFs, since changes in the 
background templates between constrained and unconstrained fits are far bigger than any statistical
variation.

\subsubsection*{Fixed fit parameters}
In the fit, parameters that are fixed, rather than constrained, are varied within their uncertainty
in pseudoexperiments, and a corresponding systematic uncertainty calculated. The fraction of electron
signals in each bremsstrahlung category is studied as a function of data-taking periods, trigger categories,
and the transverse momentum and pseudorapidity of the \B meson. The fraction of events with a single
    bremsstrahlung photon is found to be $(50 \pm 1)\%$ in all cases, with differences in the rate
of bremsstrahlung photon emission or in their detection efficiency causing a migration of events
from the zero photon category to the two-or-more photon category and vice versa. A systematic
uncertainty is assigned by varying the fraction of events in the zero and two-or-more bremsstrahlung
categories by $\pm 1\%$ in pseudoexperiments and observing the resulting change on the \RK and \RKst
double ratios.

\subsubsection*{Specific backgrounds}
The shape of the \BuToKPiPiee decay modeled in the \BdToKstee mass fits depends on the amplitude
model assumed for the $\Kp\pip\pim$ system. The simulated events used in this analysis are generated
with a phase-space distribution of $\Kp\pip\pim$ masses. It has been checked that weighting the $m(\Kp\pip\pim)$, $m(\Kp \pim)$ and  $m(\pip\pim)$ distributions to match those obtained from efficiency corrected and background-subtracted 
\BuToKPiPiJPsmm data does not impact the modeling of the background after all selections are applied.

A second study, to evaluate the effect on angular structures, is performed generating dedicated samples
of \BuToKPiPiee including the $K^+_1$ and $K^{*+}_2$ resonances, and once again the $\Kp\pim\ep\en$ 
mass distribution is found to be compatible. Pseudoexperiments are used to confirm that the 
systematic uncertainty associated with residual differences in the mass distribution are negligibly small.
Further residual differences in mass shapes between the 
\BuToKPiPiee and $\Bd \to \Kp \pim \piz \epem$ backgrounds, which could be caused by isospin-breaking effects, are 
also negligible due to the detector resolution. 

The relative normalizations of the different physics processes which contribute to the 
$B^{+,0}\to (K\pi)^{+,0} \ep\en$ background in the \BuToKee invariant mass are varied within the
uncertainties given in Sec.~\ref{sec:simfit:fit_components}. Pseudoexperiment studies are used to determine 
the systematic uncertainties on \RK and \RKst associated with these variations.

Systematic uncertainties are calculated for the invariant mass shapes and expected yields of misidentified
backgrounds. One group of systematic uncertainties concerns the PID weights used to extrapolate 
from the control region to the nominal fit region. The binning of the calibration histograms used to compute
these weights is varied, the weights are parameterised in particle momentum instead of transverse momentum,
and an additional correction for the detector occupancy is applied. A second type of systematic uncertainties concerns
the definition of inverted PID criteria which define the control region. Four different
variations are evaluated and the biggest observed difference taken as a systematic uncertainty. In
addition, the threshold used to define a control region event as pion- or kaon-like is varied from the nominal
approach of a very pure sample of pion-like events to an alternative choice of a very pure sample of kaon-like events.
Finally, the invariant mass shape of the misidentified backgrounds is evaluated using an alternative model 
based on unbinned templates. 

\subsection{Overall systematic uncertainties}
The individual sources of systematic uncertainty on \RK and \RKst are reported 
in Table~\ref{tab:systs_all}. The dominant source of systematic uncertainty comes from the treatment of misidentified backgrounds in the fit model.
Nevertheless the systematic uncertainties remain significantly smaller than the statistical uncertainties. Moreover, the dominant sources of
systematic uncertainty arise from the finite size of control samples and are therefore themselves statistical in nature. They will consequently
decrease in future analyses based on larger data samples.

\begin{table}[t]
	\centering
	\caption{Sources of systematic uncertainties on the \RK and \RKst measurements in the low- and \cqsq regions. All values are given in percent and relative to the measured central value. These values are indicative and are computed as weighted averages of systematic variations determined in each data-taking period and trigger category. The different sources of uncertainties are determined using a best linear unbiased estimator accounting for correlations between different data taking periods and trigger categories. The bottom row with the total systematic is estimated by combining the error matrices for each source in quadrature and performing a best linear unbiased estimation.}
	\label{tab:systs_all}
	\renewcommand\arraystretch{1.3}
        \resizebox{\linewidth}{!}{
	\begin{tabular}{l|cccc}
		\textbf{Source} & \lqsq \RK & \cqsq \RK & \lqsq \RKst & \cqsq \RKst \\
		\hline
		Form factors                     & 0.09 & 0.08 & 0.83 & 0.76 \\ 
		\qsq smearing                    & 0.30 & 0.19 & 0.28 & 0.31 \\ 
		Particle identification          & 0.17 & 0.22 & 0.10 & 0.12 \\   
		Kinematics and multiplicity      & 0.35 & 0.26 & 0.57 & 0.52 \\ 
		Trigger                          & 0.27 & 0.16 & 0.26 & 0.13 \\ 
		Stability of \RJPsK and \RJPsKst & 0.78 & 0.38 & 1.79 & 0.47 \\ 
		\jpsi fit model                  & 0.35 & 0.35 & 0.40 & 0.40 \\ 
		Fixed fit parameters             & 0.14 & 0.07 & 0.25 & 0.16 \\ 
		Combinatorial shape              & 0.99 & 0.16 & 1.39 & 0.38 \\  
		Specific backgrounds             & 0.24 & 0.20 & 1.24 & 0.51 \\ 
        Misidentified backgrounds        & 2.50 & 2.22 & 1.87 & 2.29 \\
		Modeling of \mHOP                        & 0.25 & 0.24 & 0.33 & 0.33 \\ 
		\hline  
		Total                            & 2.86 & 2.33 & 3.73 & 2.52 \\
	\end{tabular}
 }
\end{table}

\section{Results}
\label{sec:results}
\begin{figure}[t]
\centering
\includegraphics[width=0.98\textwidth]{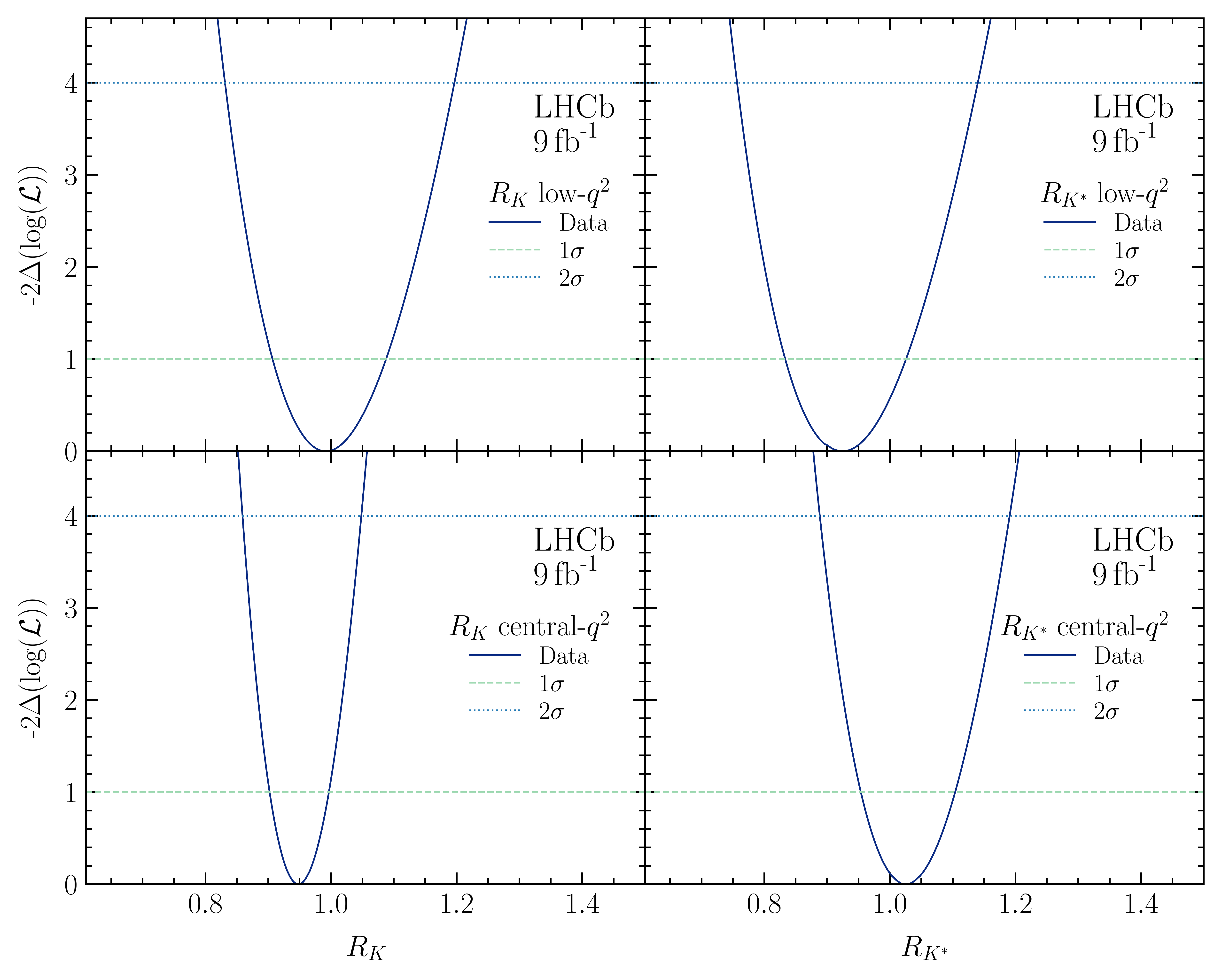}
\caption{Likelihood scans for the LU observables (left) \RK and (right) \RKst, in the (top) \lqsq and (bottom) \cqsq regions.}
\label{fig:results_likelihood_scans}
\end{figure}
\begin{figure}[t]
    \centering
    \includegraphics[width=0.6\linewidth]{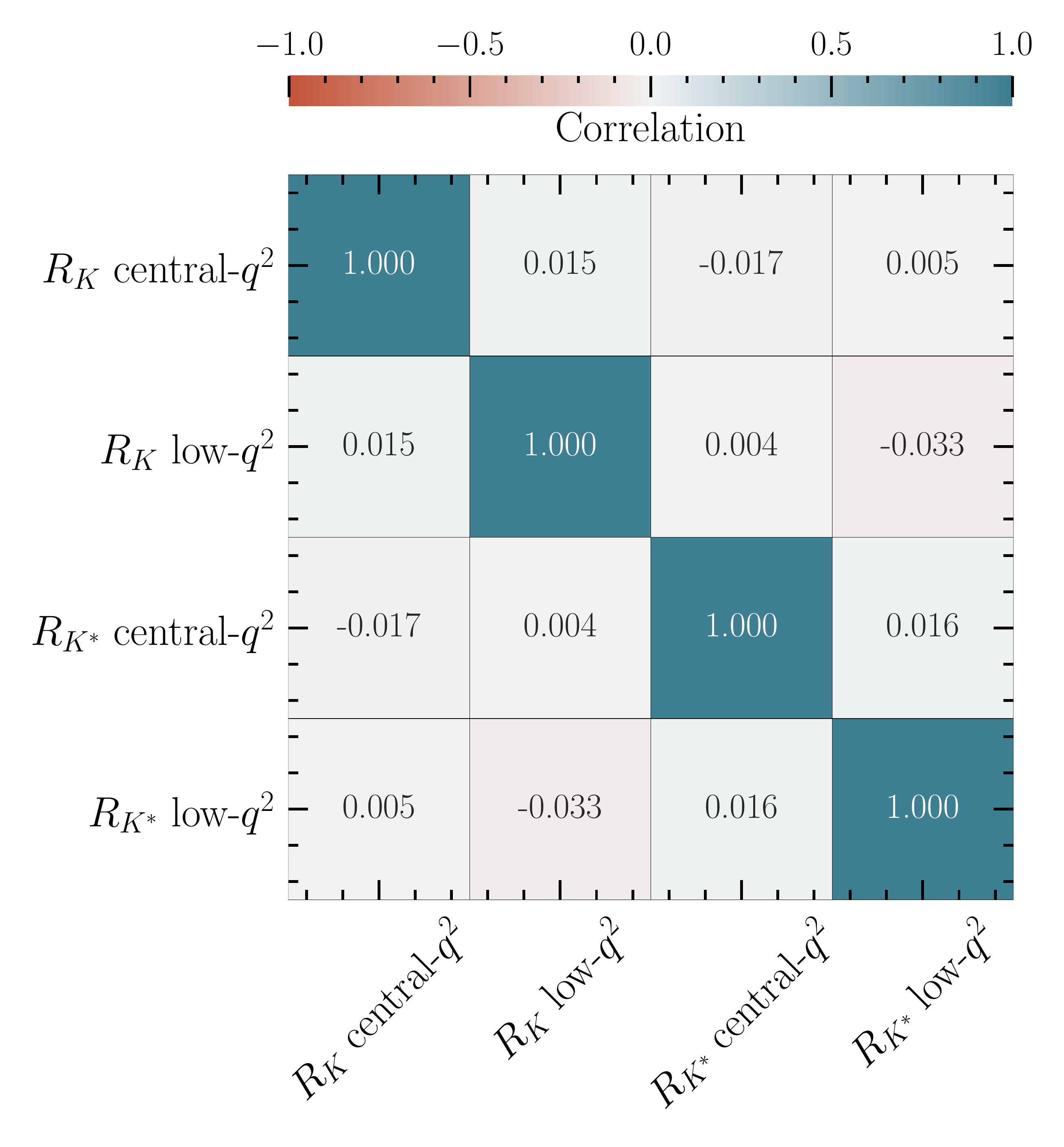}
    \caption{Correlation factors between the \RK and \RKst results in the low- and \cqsq regions.}
    \label{fig:correlation_results}
\end{figure}

The best fit point together with the statistical and systematic uncertainties
for the four LU observables are reported in Table~\ref{tab:results_cov}. The results for each
running period, given in App.~\ref{app:split_results}, are consistent with each other and with the overall result.
Each of the four relative branching fraction  measurements is the most precise to date.
The uncertainties on the lepton universality
observables are not Gaussian due to the finite sample sizes used in this analysis.
Likelihood scans for each of the double-ratio LU observables are presented in Fig.~\ref{fig:results_likelihood_scans}. The likelihood scans are used to derive the asymmetric uncertainties
reported in Table~\ref{tab:results_cov}. The correlation matrix reported by the fit to data including all uncertainties is shown in Fig.~\ref{fig:correlation_results}.
 In order to separate statistical and systematic uncertainties the
likelihood scans are performed twice, once with and once without the systematic uncertainties included in the fit
covariance matrix. The uncertainties of these are then subtracted in quadrature to obtain the contribution of
systematic uncertainties to the overall uncertainty.

\begin{table}[t]
	\centering
	\caption{Measured values of the \RK and \RKst observables in the low- and \cqsq regions, with the associated statistical and systematic uncertainties presented separately.}
	\label{tab:results_cov}
	\renewcommand\arraystretch{1.3}
	\begin{tabular}{cccc}
	 low-\qsq \RK & central-\qsq \RK & low-\qsq \RKst & central-\qsq \RKst \\ \hline 
	   $0.994\,^{+0.090}_{-0.082}\,^{+0.029}_{-0.027}$ 
	 & $0.949\,^{+0.042}_{-0.041}\,^{+0.022}_{-0.022}$
	 & $0.927\,^{+0.093}_{-0.087}\,^{+0.036}_{-0.035}$
	 & $1.027\,^{+0.072}_{-0.068}\,^{+0.027}_{-0.026}$ \\ 
     \hline
	\end{tabular}
\end{table}

\begin{figure}[tbp]
\centering
\includegraphics[width=0.48\linewidth]{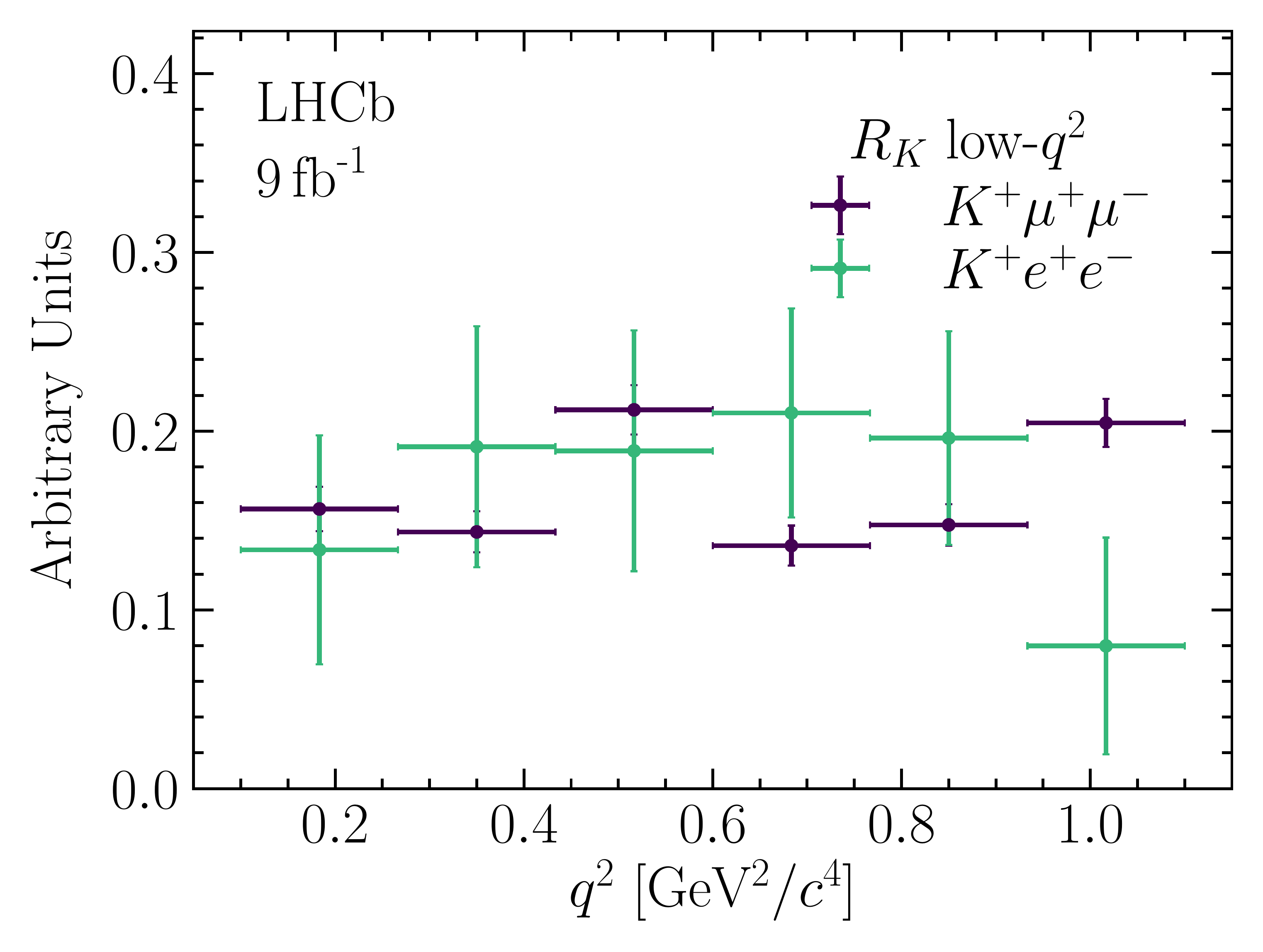}
\includegraphics[width=0.48\linewidth]{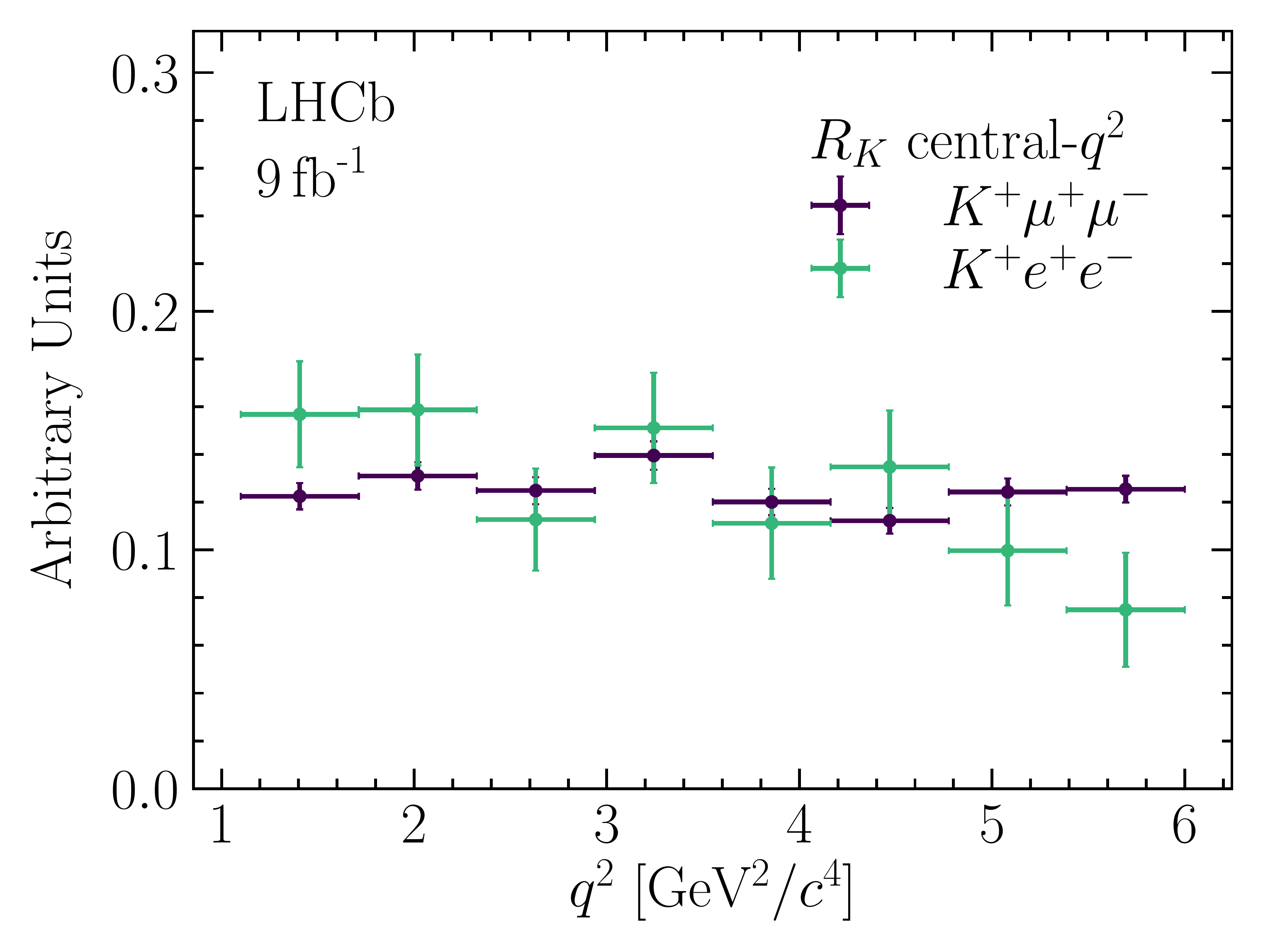}
\includegraphics[width=0.48\linewidth]{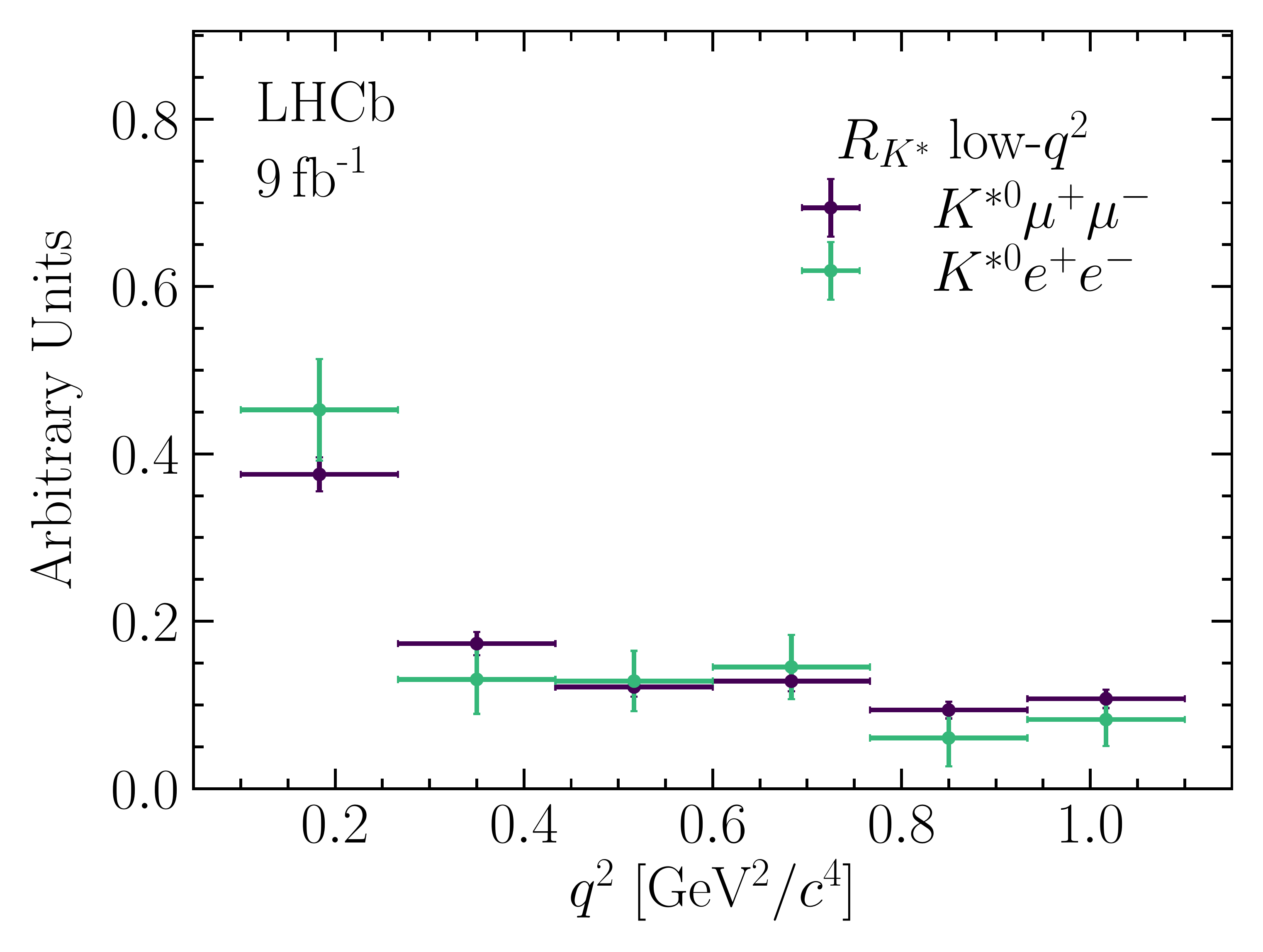}
\includegraphics[width=0.48\linewidth]{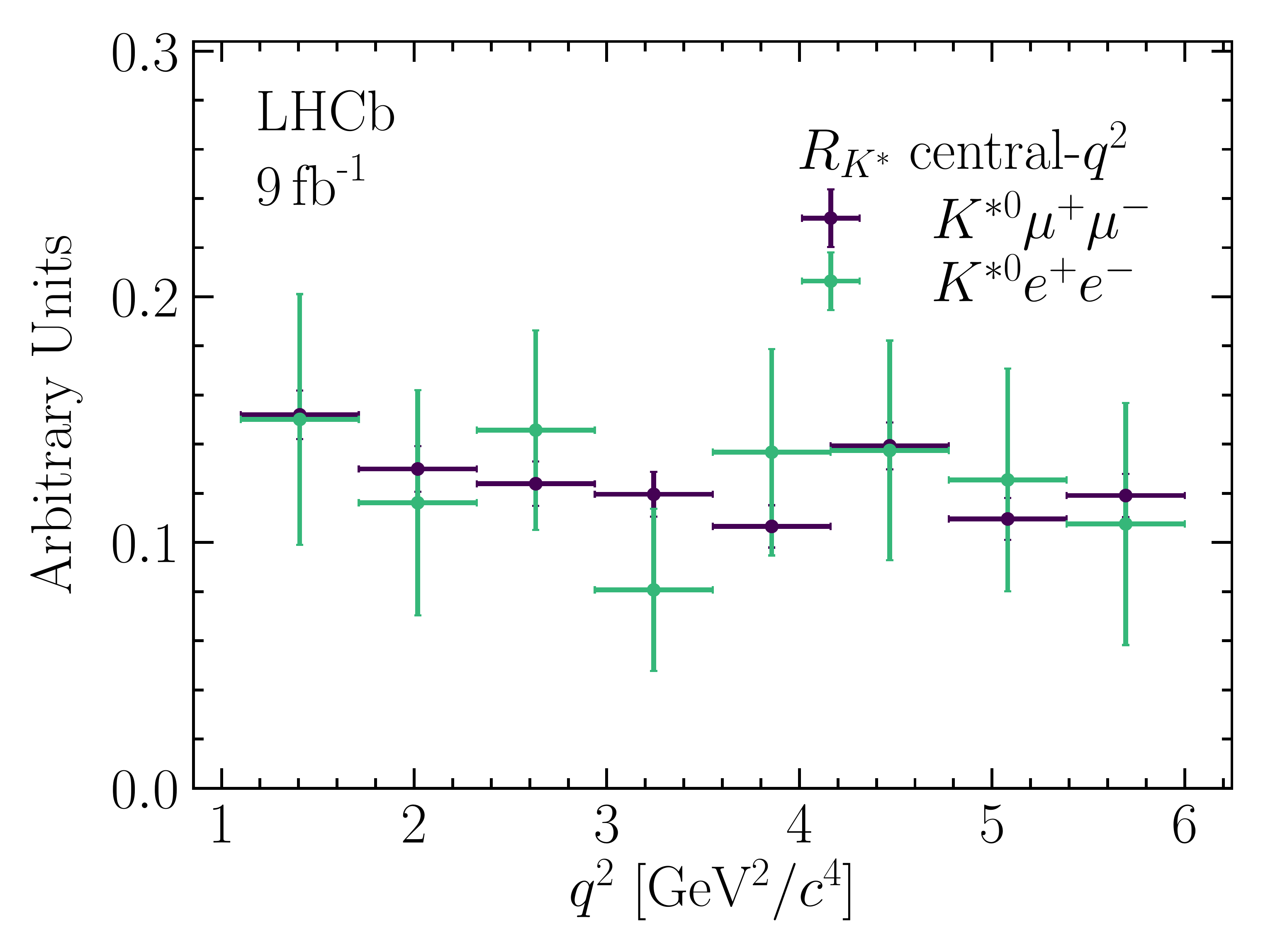}
\includegraphics[width=0.48\linewidth]{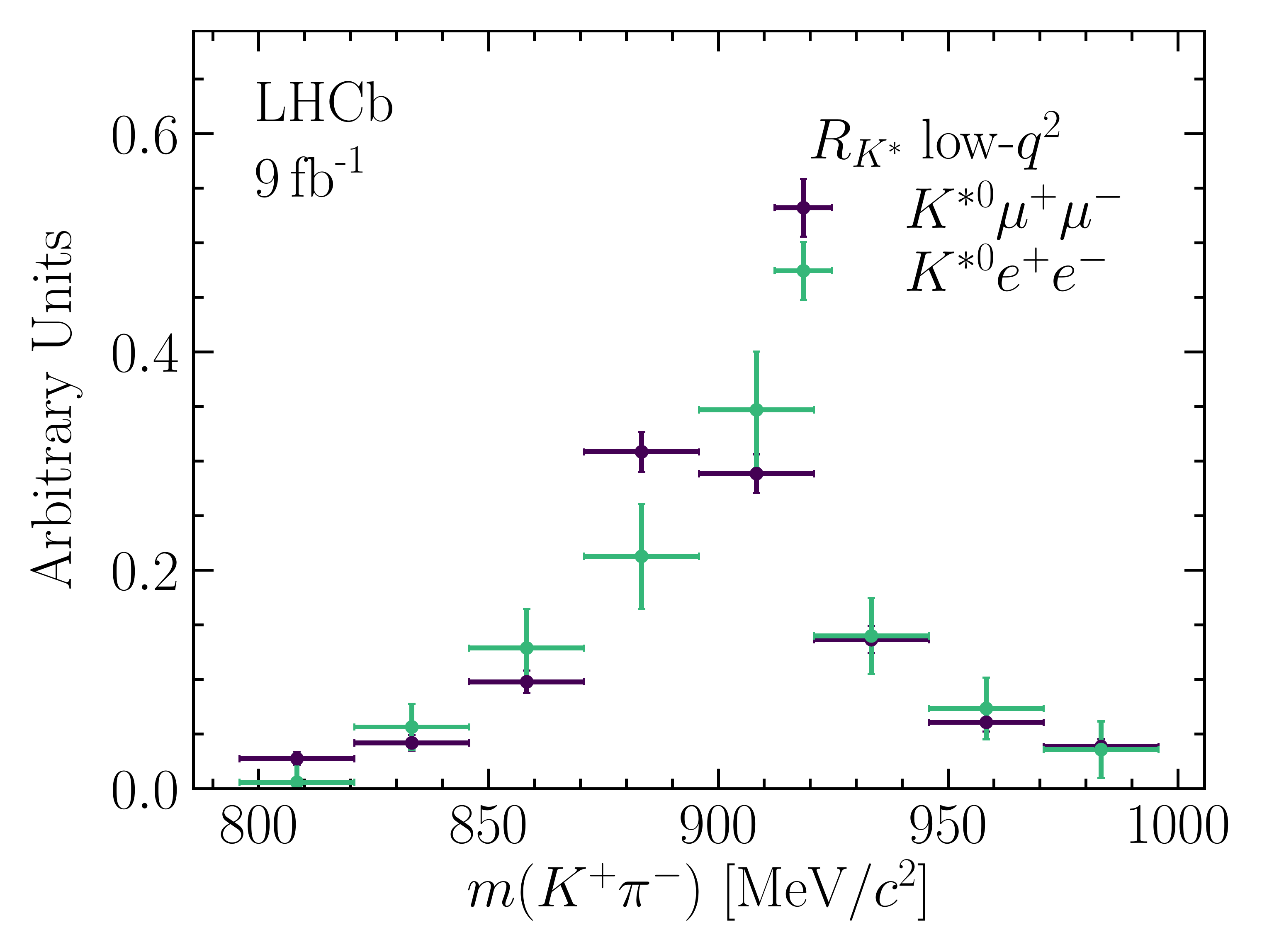}
\includegraphics[width=0.48\linewidth]{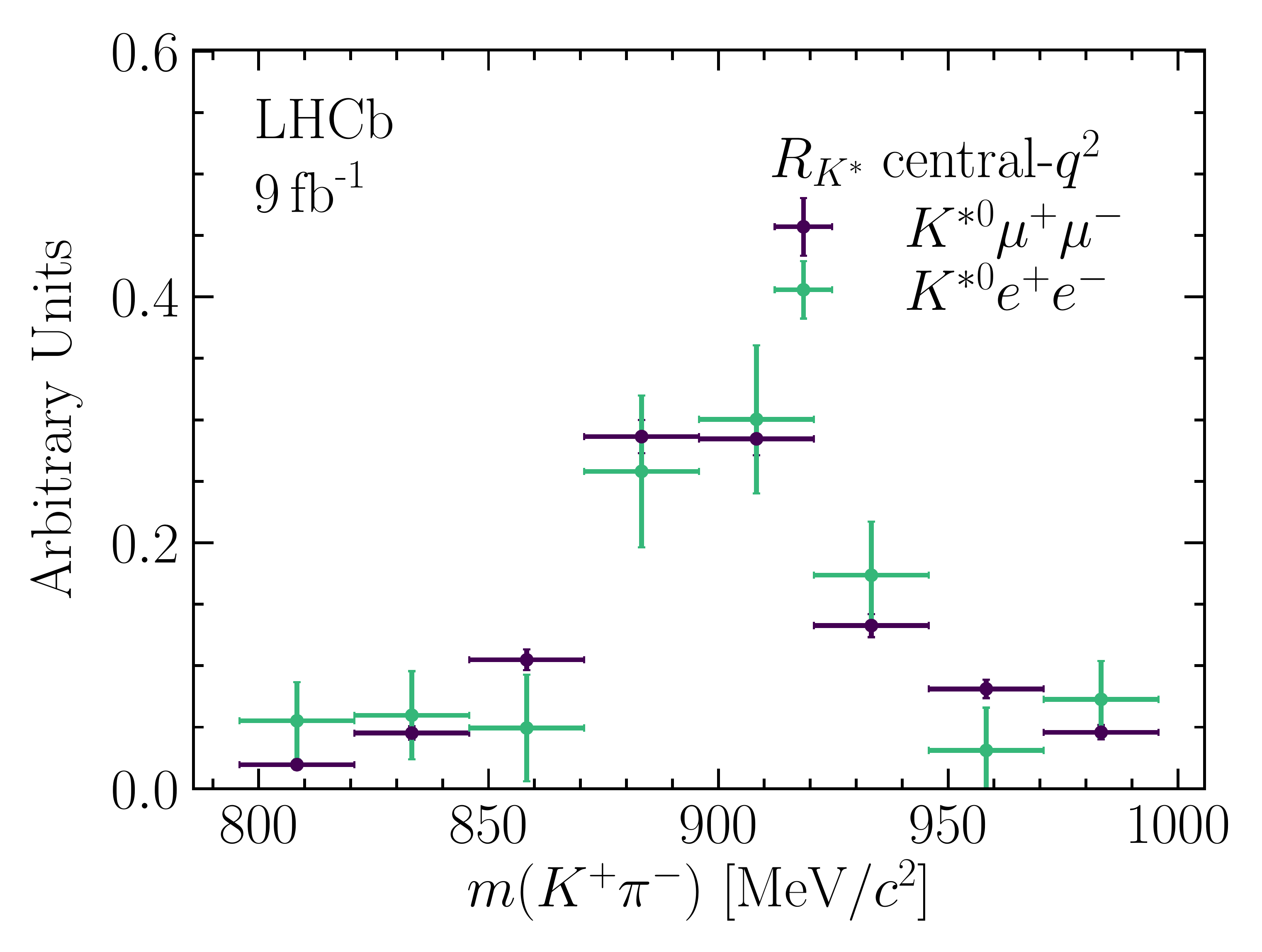}
\caption{\small Background-subtracted distributions of quantities describing the $\Bu \to \Kp \ellell$ and $\Bd \to \Kstarz \ellell$ decays. 
The \lqsq region is plotted on the left, the \cqsq region on the right. 
The top and middle rows show the distributions of \qsq for the \Bp and  \Bz signals, respectively. The bottom row shows the distribution of the \Kstarz mass for the \Bd signals.}
\label{fig:results_splotted_plots}
\end{figure}

The sPlot~\cite{Pivk:2004ty} technique is used to obtain background-subtracted distributions of quantities
describing the \Bd and \Bu decays in the four \qsq regions considered. The simulation is used to verify that this technique
allows the quantities in question to be determined accurately, despite the fact that bremsstrahlung causes significant correlations between \qsq and the mass of the \B meson candidate for both the electron signal and for the backgrounds.
Figure~\ref{fig:results_splotted_plots} shows the resulting distributions.

The LU results are used to calculate the differential branching fractions of \BuToKee and \BdToKstee decays, averaged over the central \qsq region. This is done by combining the \RK and \RKst measurements at central \qsq with the known  \BuToKmm~\cite{LHCb-PAPER-2014-006} and \BdToKstmm~\cite{LHCb-PAPER-2016-012} branching fractions; in the latter case, only the $\Kstar(892)^0$ P-wave state is considered. Similar results
are not obtained in the \lqsq region since the muonic branching fractions are not available in the same \qsq range as used in this analysis.
All systematic uncertainties on the LU ratios are assumed to be correlated with the systematic uncertainties on the branching fractions, with the exception of the effect induced by the normalization channel; this is taken to be uncorrelated. The correlations between the statistical uncertainties of the LU observables and the branching fractions are evaluated based on the overlap between \BuToKmm and \BdToKstmm data sets used in either measurements. It is found that 61\% of the \BuToKmm Run 1 sample was used in the corresponding branching-fraction measurement, whereas for \BdToKstmm this overlap is 69\%. Combined with the Run 2 yields, this leads to a correlation of 0.13 between the statistical uncertainties of the \BuToKmm branching fraction measurement and this LU measurement. We similarly find a correlation of 0.14 between the \BdToKstmm statistical uncertainties. The electron mode branching fractions, averaged over the \cqsq region, are found to be
\begin{align*}
    \frac{\mathrm d\mathcal B(\BuToKee)}{\mathrm d\qsq} &= (25.5^{+1.3}_{-1.2}\pm1.1)\times10^{-9}\,\gev^{-2}\\ 
    \frac{\mathrm d\mathcal B(\BdToKstee)}{\mathrm d\qsq}&= (33.3^{+2.7}_{-2.6}\pm2.2)\times10^{-9}\,\gev^{-2}.
\end{align*}
\section{Conclusion}
\label{sec:conclusion}
We present the first simultaneous test of LU in \BuToKll and \BdToKstll decays 
using all $pp$ collision data collected with the \lhcb detector between 2011 and 2018, corresponding to 
an integrated luminosity of 9\invfb. The ratios of the branching fractions of muon and electron modes 
are measured in both channels and in two ranges of the square of the dilepton invariant mass.
Each of these four measurements is either the first (\RK low-\qsq) or the most precise (\RKst low-\qsq, \RK central-\qsq and \RKst central-\qsq) such measurement to date. 
The measured values are

\begin{align*}
    \lqsq \begin{cases}
        \RK   &= 0.994~^{+0.090}_{-0.082} \stat 
                    \; ^{+0.029}_{-0.027} \syst,  \\
        \RKst &= 0.927~^{+0.093}_{-0.087} \stat \;
                       ^{+0.036}_{-0.035} \syst,\\
   \end{cases} \\
   \cqsq
   \begin{cases}
        \RK  &= 0.949~^{+0.042}_{-0.041} \stat\;
                      ^{+0.022}_{-0.022} \syst, \\
        \RKst &=1.027~^{+0.072}_{-0.068} \stat \;
                      ^{+0.027}_{-0.026}\syst,
   \end{cases} 
\end{align*}

\noindent where the first uncertainty in each row is statistical and the second systematic.

The central values of the SM prediction, as calculated by the \flavio software package~\cite{Straub:2018kue}, are given in Table~\ref{tab:smprediction}. 
An additional uncertainty of 1\% is assigned to take into account 
uncertainties in the modeling of QED effects 
in \BuToKll and \BdToKstll decays, following Ref.~\cite{Bordone:2016gaq}. This uncertainty is assumed to be uncorrelated between the
LU observables and dominates the covariance matrix of the SM predictions.
\begin{table}[t]
  \caption{SM predictions and uncertainties from the \flavio  software package~\cite{Straub:2018kue}. The dominant QED uncertainty from Ref.~\cite{Bordone:2016gaq} is quoted separately.\label{tab:smprediction}}
  \centering
  \begin{tabular}{l|cccc}\hline
   & \RK \lqsq & \RK  \cqsq & \RKst \lqsq & \RKst \cqsq \\\hline
SM prediction & 0.9936 & 1.0007 & 0.9832 & 0.9964\\
SM uncertainty      & 0.0003 & 0.0003 & 0.0014 & 0.0006\\
QED uncertainty~\cite{Bordone:2016gaq} & 0.01 & 0.01 & 0.01 & 0.01\\
\hline\end{tabular}

\end{table}

\begin{figure}
    \centering
    \includegraphics[width=1.0\linewidth]{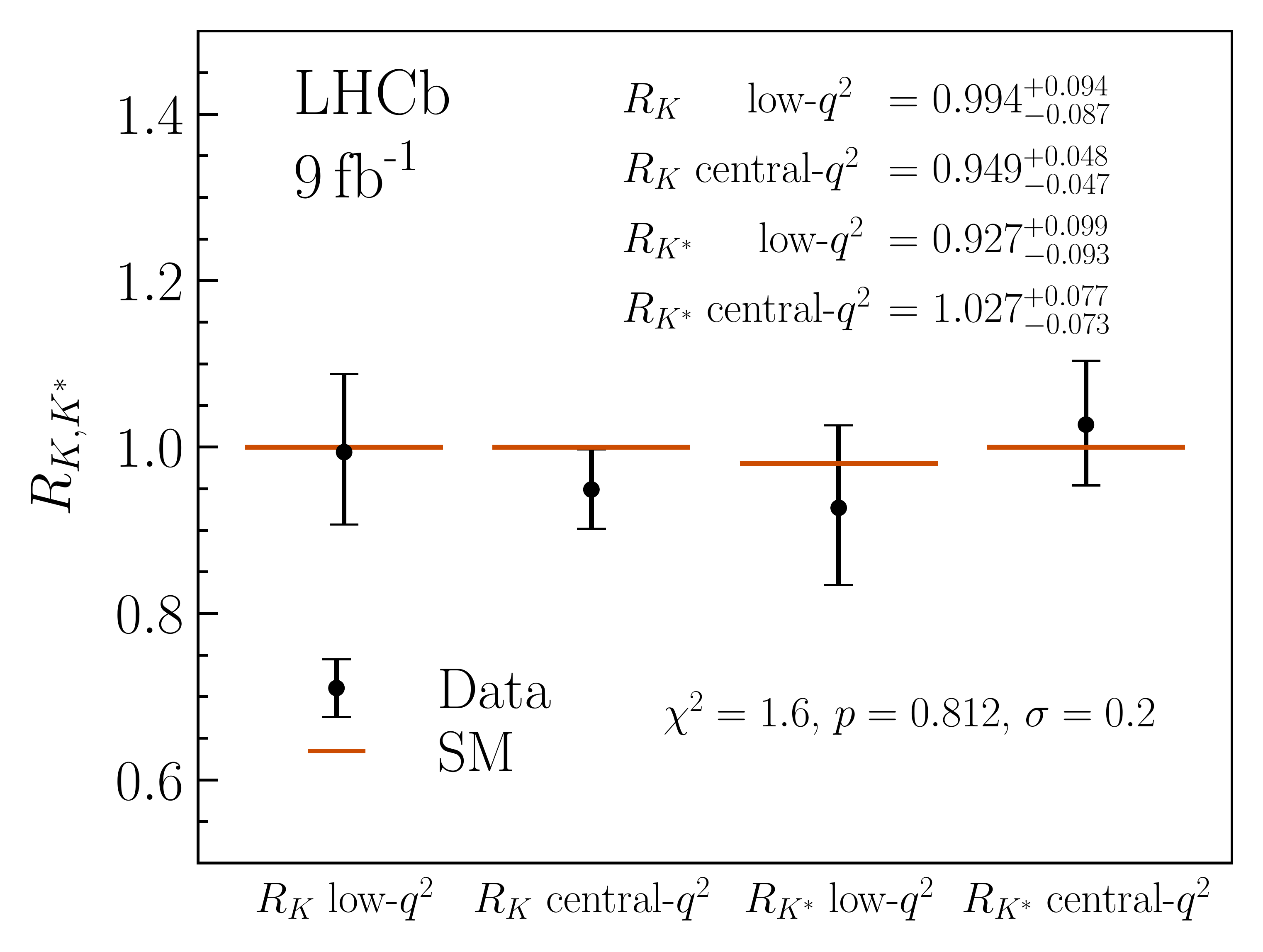}
    \caption{Measured values of LU observables in \BuToKll and \BdToKstll decays and their overall compatibility with the SM.}
    \label{fig:overall_sm_compat}
\end{figure}

Each of these four measured relative decay rates is compatible with SM  predictions~\cite{Bordone:2016gaq,Isidori:2020acz,Bobeth:2007dw,Straub:2018kue,vanDyk:2021sup,Capdevila:2016ivx,Capdevila:2017ert,Serra:2016ivr,Altmannshofer:2017fio,Jager:2014rwa,Ghosh:2014awa},
with the maximum difference between measurement and prediction being around one standard deviation. 
The results are interpreted collectively as a null test of the SM and their combined
compatibility with the SM is evaluated using a $\chi^2$ test. In this test the
distance of each measurement from the SM point is evaluated using the likelihood
obtained from the data fit. The overall compatibility is shown in Fig.~\ref{fig:overall_sm_compat} and 
agrees with the SM prediction at $0.2$ standard deviations.

The results presented here
differ from previous \lhcb measurements of \RK \cite{LHCb-PAPER-2021-004} and \RKst \cite{LHCb-PAPER-2017-013},  which they supersede.
 The measured values for \RKst (low- and \cqsq) and \RK (\cqsq) move upwards from the previous results and closer to the SM predictions. 
 Although these shifts can be attributed in part to statistical effects it is understood that the change in  \RK is primarily due to systematic effects.  
 In the case of \RK, the data sample is the same as in Ref.~\cite{LHCb-PAPER-2021-004}, but subject to a revised analysis. For \RK (\cqsq) the statistical component of the difference is evaluated using pseudoexperiments and found to follow a Gaussian distribution of width 0.033 in the absolute value of \RK.
 In the case of \RKst, the data correspond to more than a factor of five increase in the number of \bbbar pairs produced relative to Ref.~\cite{LHCb-PAPER-2017-013} and hence there is a much larger statistical component of the difference.
 For \RK (\cqsq) the expected systematic shifts caused by the improved treatment of misidentified hadronic backgrounds in the electron mode are also evaluated using pseudoexperiments.  
The biggest shift (0.064 with respect to Ref.~\cite{LHCb-PAPER-2021-004}) is found to be due to the more stringent PID criteria applied here, 
which reduce the contribution from misidentified background processes that had previously not been accounted for appropriately. 
In addition, the residual misidentified backgrounds are explicitly modeled in the fit, resulting in a further shift (0.038) compared to the previous analysis. 
These shifts
add linearly.
 The systematic shift due to misidentified backgrounds to electrons, and the uncertainties assigned to the results presented here, are greater than the systematic uncertainties in the earlier publication of \RK.
 The assigned systematic uncertainties on the new measurements presented in this paper are smaller than in previous papers, except for \RK (\cqsq) where the new result has a smaller overall relative uncertainty despite an increase in the systematic uncertainty from that of Ref.~\cite{LHCb-PAPER-2021-004}.  In all cases, the  statistical uncertainties remain significantly larger than the systematic uncertainties and therefore additional data will continue to challenge the Standard Model.

\clearpage

\section*{Acknowledgements}
%
%
\noindent We express our gratitude to our colleagues in the CERN
accelerator departments for the excellent performance of the LHC. We
thank the technical and administrative staff at the LHCb
institutes.
We acknowledge support from CERN and from the national agencies:
CAPES, CNPq, FAPERJ and FINEP (Brazil); 
MOST and NSFC (China); 
CNRS/IN2P3 (France); 
BMBF, DFG and MPG (Germany); 
INFN (Italy); 
NWO (Netherlands); 
MNiSW and NCN (Poland); 
MEN/IFA (Romania); 
MICINN (Spain); 
SNSF and SER (Switzerland); 
NASU (Ukraine); 
STFC (United Kingdom); 
DOE NP and NSF (USA).
We acknowledge the computing resources that are provided by CERN, IN2P3
(France), KIT and DESY (Germany), INFN (Italy), SURF (Netherlands),
PIC (Spain), GridPP (United Kingdom), 
CSCS (Switzerland), IFIN-HH (Romania), CBPF (Brazil),
Polish WLCG  (Poland) and NERSC (USA).
We are indebted to the communities behind the multiple open-source
software packages on which we depend.
Individual groups or members have received support from
ARC and ARDC (Australia);
Minciencias (Colombia);
AvH Foundation (Germany);
EPLANET, Marie Sk\l{}odowska-Curie Actions and ERC (European Union);
A*MIDEX, ANR, IPhU and Labex P2IO, and R\'{e}gion Auvergne-Rh\^{o}ne-Alpes (France);
Key Research Program of Frontier Sciences of CAS, CAS PIFI, CAS CCEPP, 
Fundamental Research Funds for the Central Universities, 
and Sci. \& Tech. Program of Guangzhou (China);
GVA, XuntaGal, GENCAT and Prog.~Atracci\'on Talento, CM (Spain);
SRC (Sweden);
the Leverhulme Trust, the Royal Society
 and UKRI (United Kingdom).

\addcontentsline{toc}{section}{References}
\bibliographystyle{LHCb}
\bibliography{main,standard,LHCb-PAPER,LHCb-CONF,LHCb-DP,LHCb-TDR}

\newpage
\centerline
{\large\bf LHCb collaboration}
\begin
{flushleft}
\small
R.~Aaij$^{32}$\lhcborcid{0000-0003-0533-1952},
A.S.W.~Abdelmotteleb$^{50}$\lhcborcid{0000-0001-7905-0542},
C.~Abellan~Beteta$^{44}$,
F.~Abudin{\'e}n$^{50}$\lhcborcid{0000-0002-6737-3528},
T.~Ackernley$^{54}$\lhcborcid{0000-0002-5951-3498},
B.~Adeva$^{40}$\lhcborcid{0000-0001-9756-3712},
M.~Adinolfi$^{48}$\lhcborcid{0000-0002-1326-1264},
P.~Adlarson$^{77}$\lhcborcid{0000-0001-6280-3851},
H.~Afsharnia$^{9}$,
C.~Agapopoulou$^{13}$\lhcborcid{0000-0002-2368-0147}
% ...

{\footnotesize \it

% ...
$^{9}$Universit{\'e} Clermont Auvergne, CNRS/IN2P3, LPC, Clermont-Ferrand, France\\
% ...
$^{13}$LPNHE, Sorbonne Universit{\'e}, Paris Diderot Sorbonne Paris Cit{\'e}, CNRS/IN2P3, Paris, France\\
% ...
$^{32}$Nikhef National Institute for Subatomic Physics, Amsterdam, Netherlands\\
% ...
$^{40}$Instituto Galego de F{\'\i}sica de Altas Enerx{\'\i}as (IGFAE), Universidade de Santiago de Compostela, Santiago de Compostela, Spain\\
% ...
$^{44}$Physik-Institut, Universit{\"a}t Z{\"u}rich, Z{\"u}rich, Switzerland\\
% ...
$^{48}$H.H. Wills Physics Laboratory, University of Bristol, Bristol, United Kingdom\\
% ...
$^{50}$Department of Physics, University of Warwick, Coventry, United Kingdom\\
% ...
$^{53}$School of Physics and Astronomy, University of Glasgow, Glasgow, United Kingdom\\
$^{54}$Oliver Lodge Laboratory, University of Liverpool, Liverpool, United Kingdom\\
% ...
$^{77}$Department of Physics and Astronomy, Uppsala University, Uppsala, Sweden, associated to $^{53}$\\
% ...
}
\end{flushleft}

\newpage
\begin{appendices}
\section{Results split by data-taking period}
\label{app:split_results}
\begin{figure}[!h]
\includegraphics[width=0.9\linewidth]{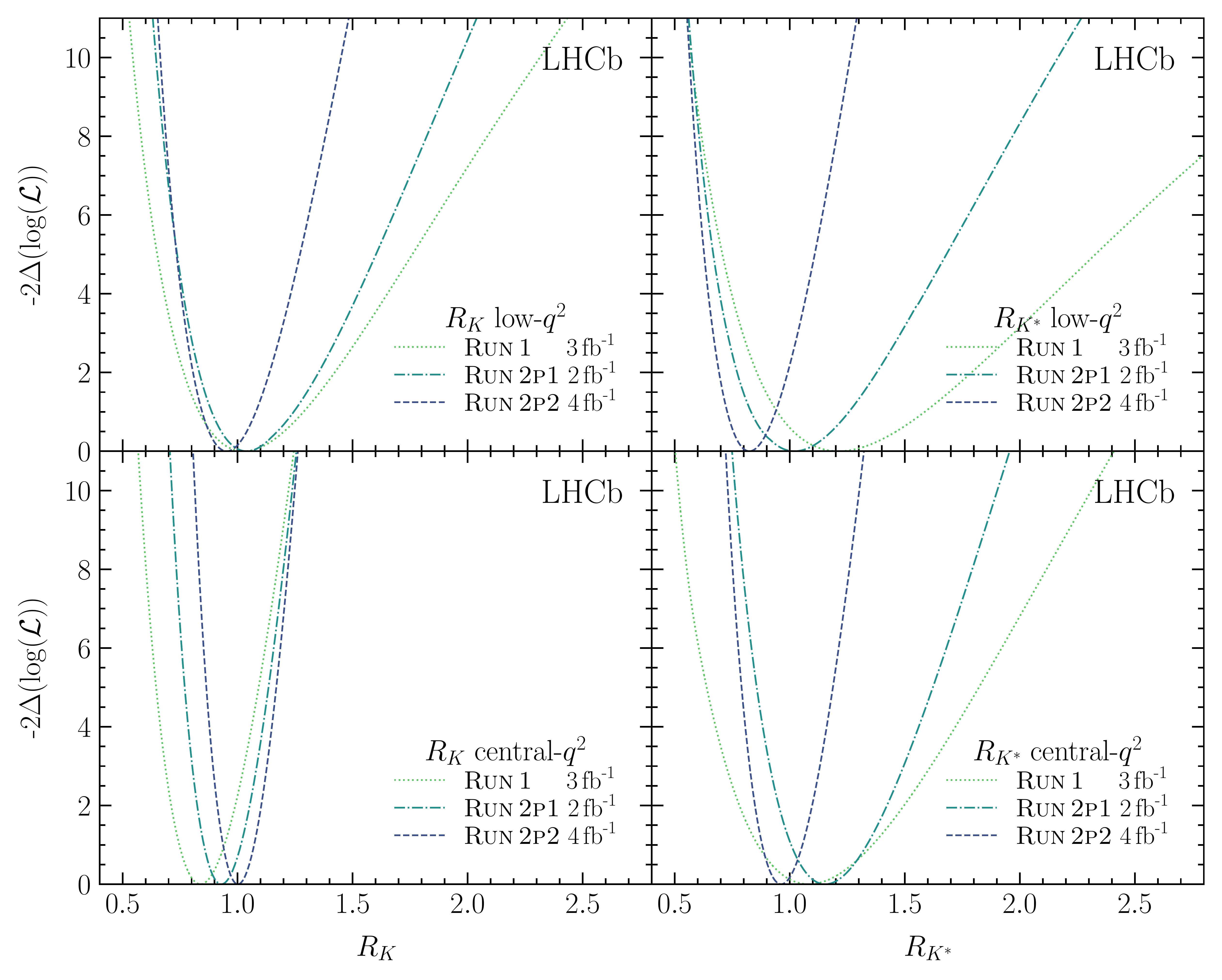}
\caption{One-dimensional likelihood scans for \RK and \RKst in the low- and \cqsq regions,  performing the measurements in each data-taking period separately. The scan shown includes  both systematic and statistical uncertainties. \label{fig:appendixA:resultsSplitRunPeriod}}
\end{figure}

The results obtained using the \runone, \runtwopo, \runtwopt datasets alone for \RK and \RKst in the low- and \cqsq regions are shown together with their likelihood profiles including all systematic uncertainties. The results obtained from the best fit in individual run periods are shown in Table~\ref{tab:appendixA:resultsSplitRunPeriod} and the corresponding one-dimensional likelihood scans are shown in Fig.~\ref{fig:appendixA:resultsSplitRunPeriod}.

\begin{table}[!h]
\centering 
\renewcommand{\arraystretch}{1.3} 
\caption{Measured values of the LU observables obtained from the separate run periods. Uncertainties are split into statistical and systematic components and have been extracted from the one-dimensional likelihood scans.}  \label{tab:appendixA:resultsSplitRunPeriod}
\begin{tabular}{lccc}
LU observable & \runone                                   & \runtwopo                                & \runtwopt                                \\ \hline
 \RK \lqsq   & $1.027\,^{\,+\,0.243\,+\,0.092}_{\,-\,0.180\,-\,0.073}$ & $1.039\,^{\,+\,0.203\,+\,0.027}_{\,-\,0.149\,-\,0.027}$ & $0.953\,^{\,+\,0.123\,+\,0.029}_{\,-\,0.104\,-\,0.026}$ \\
 \RKst \lqsq & $1.212\,^{\,+\,0.344\,+\,0.149}_{\,-\,0.240\,-\,0.114}$ & $1.021\,^{\,+\,0.234\,+\,0.036}_{\,-\,0.187\,-\,0.027}$ & $0.825\,^{\,+\,0.108\,+\,0.036}_{\,-\,0.091\,-\,0.031}$ \\
 \RK \cqsq   & $0.839\,^{\,+\,0.083\,+\,0.062}_{\,-\,0.073\,-\,0.056}$ & $0.929\,^{\,+\,0.082\,+\,0.023}_{\,-\,0.073\,-\,0.020}$ & $1.001\,^{\,+\,0.066\,+\,0.024}_{\,-\,0.061\,-\,0.022}$ \\
 \RKst \cqsq & $1.082\,^{\,+\,0.214\,+\,0.176}_{\,-\,0.165\,-\,0.148}$ & $1.154\,^{\,+\,0.179\,+\,0.027}_{\,-\,0.147\,-\,0.023}$ & $0.962\,^{\,+\,0.091\,+\,0.020}_{\,-\,0.080\,-\,0.018}$ \\
\hline
\end{tabular}
\end{table}

\end{appendices}

\end{document}